\documentclass[12pt, a4paper]{book}
\usepackage[utf8]{inputenc}

\usepackage{vub}

\usepackage[T1]{fontenc}
\usepackage{lmodern}
\usepackage[toc,page]{appendix}

\usepackage{amssymb}
\usepackage{amsmath}
\usepackage{fancyhdr}
\usepackage{graphicx}
\usepackage{caption}
\usepackage{microtype}
\usepackage{wasysym}
\usepackage{amsthm}
\usepackage{tabu}
\usepackage{esint}
\usepackage{tikz}
\usepackage{multicol} 
\usepackage{amssymb}
\usepackage{mdframed}
\usepackage{eufrak}
\usepackage{epigraph}
\usepackage{siunitx}
\usepackage{subcaption} 
\usepackage{url} 
\usepackage{mathtools}
\usepackage{mathrsfs}

\usepackage{bbm}
\usepackage{cancel}
\usepackage{mathtools}
\usepackage{accents}
\usepackage{tikz-feynman,contour}
\usepackage{slashed}
\usepackage{simpler-wick}
\usepackage{lipsum}

\usepackage{empheq}
\usepackage[most]{tcolorbox}

\newtcbox{\greybox}[1][]{%
    nobeforeafter, math upper, tcbox raise base,
    enhanced, colframe=black,
    colback=black!5, boxrule=1pt,
    #1}
    
\newtcbox{\hollowbox}[1][]{%
    nobeforeafter, math upper, tcbox raise base,
    enhanced, colframe=black,
    colback=white, boxrule=1pt,
    #1}

\usepackage{xcolor}
\usepackage[colorlinks]{hyperref}

\definecolor{VUB blue}{rgb}{0.0, 0.2, 0.6}
\definecolor{VUB orange}{rgb}{1, 0.4, 0.0}

\hypersetup{
 linkcolor= black
,citecolor= blue
,urlcolor= blue
}

\usepackage{sectsty}

\usepackage{subcaption}
\usepackage{gensymb}

\usepackage{halloweenmath}
\usepackage{marvosym}

\title{Localisation of the Third Way Theory}
\pretitle{\flushleft{Dissertation for the degree of Master in Physics and Astronomy}}
\author{Dimitri Kanakaris Decavel}
\date{June~2023}
\promotors{Promotors: \\prof.\ dr.\ Alexandre Sevrin 
\\prof.\ dr.\ Alexandros Spyridon Arvanitakis}
\faculty{sciences and bioengineering sciences} 

\begin{document}
\frontmatter
\maketitle%




\newcommand{\extd}{{\mathrm{d}}}
\newcommand{\extdd}{\extd^\dagger}
\newcommand{\del}{{\partial}}
\newcommand{\dels}{{\slashed{\del}}}
\newcommand{\delt}{{\accentset{\sim}{\del}}}
\newcommand{\delsl}{{\accentset{\leftarrow}{\dels}}}
\newcommand{\Dt}{{\accentset{\sim}{D}}}
\newcommand{\nablat}{{\accentset{\sim}{\nabla}}}
\newcommand{\nablas}{{\slashed{\nabla}}}

\newcommand{\der}[2]{\frac{\extd #1}{\extd #2}}
\newcommand{\pder}[2]{\frac{\del #1}{\del #2}}
\newcommand{\fder}[2]{\frac{\delta #1}{\delta #2}}
\newcommand{\EV}[1]{\left\langle #1\right\rangle}
\newcommand{\bra}[1]{{\left\langle #1\right|}}
\newcommand{\ket}[1]{{\left| #1\right\rangle}}
\newcommand{\braket}[2]{\left\langle #1\middle| #2\right\rangle}
\newcommand{\boobket}[2]{\left( #1\middle| #2\right)}
\newcommand{\overeq}[1]{\overset{\text{#1}}{=}}

\newcommand{\gLie}{{\mathfrak{g}}}
\newcommand{\isoLie}{{\mathfrak{iso}}}
\newcommand{\gintLie}{{\gLie_{\text{int}}}}

\newcommand{\NSUSY}{{\mathcal{N}}}
\newcommand{\Og}{{\mathcal{O}}}
\newcommand{\Lie}{{\mathcal{L}}}
\newcommand{\Qloc}{{\mathcal{Q}}}
\newcommand{\Bg}{{\mathcal{B}}}
\newcommand{\Dg}{{\mathcal{D}}}
\newcommand{\Dgd}{\Dg^\dagger}
\newcommand{\Dglr}{{\accentset{\leftrightarrow}{\Dg}}}
\newcommand{\Dgr}{{\accentset{\leftarrow}{\Dg}}}
\newcommand{\Dgb}{{\bar \Dg}}
\newcommand{\Dgs}{{\slashed{\Dg}}}
\newcommand{\Dgbs}{{\bar{\Dgs}}}
\newcommand{\Fg}{{\mathcal{F}}}
\newcommand{\Fgt}{{\accentset{\sim}{\mathcal{F}}}}
\newcommand{\Fgo}{{\accentset{\circ}{\mathcal{F}}}}
\newcommand{\Fgs}{{\cancel{\mathcal{F}}}}
\newcommand{\Lg}{{\mathcal{L}}}
\newcommand{\Jg}{{\mathcal{J}}}
\newcommand{\Mg}{{\mathcal{M}}}
\newcommand{\Ag}{{\mathcal{A}}}
\newcommand{\Agt}{{\accentset{\sim}{\mathcal{A}}}}
\newcommand{\Tg}{{\mathcal{T}}}
\newcommand{\Tgo}{{\overset{\circ}{\mathcal{T}}}}
\newcommand{\Xg}{{\mathcal{X}}}
\newcommand{\Xgt}{{\accentset{\sim}{\mathcal{X}}}}
\newcommand{\Vg}{{\mathcal{V}}}
\newcommand{\Vgt}{{\accentset{\sim}{\mathcal{V}}}}
\newcommand{\Vgtt}{{\tilde{\mathcal{V}}}}
\newcommand{\Vgb}{{\bar{\mathcal{V}}}}

\newcommand{\Fscr}{{\mathscr{F}}}
\newcommand{\Gscr}{{\mathscr{G}}}

\newcommand{\diag}{\operatorname{diag}}
\newcommand{\tr}{\operatorname{tr}}
\newcommand{\Ker}{\operatorname{Ker}}
\newcommand{\Imm}{\operatorname{Im}}
\newcommand{\Pf}{\operatorname{Pf}}
\newcommand{\Vol}{\operatorname{Vol}}
\newcommand{\Ber}{\operatorname{Ber}}
\newcommand{\Det}{\operatorname{Det}}
\newcommand{\sign}{\operatorname{sign}}
\newcommand{\ch}{\operatorname{ch}}
\newcommand{\ad}{\operatorname{ad}}
\newcommand{\inn}{\operatorname{\lrcorner}}
\newcommand{\innr}{\operatorname{\llcorner}}
\newcommand{\BPS}{{\text{BPS}}}
\newcommand{\bos}{{\text{bos}}}
\newcommand{\fer}{{\text{fer}}}
\newcommand{\aux}{{\text{aux}}}
\newcommand{\loc}{{\text{loc}}}
\newcommand{\YM}{{\text{YM}}}
\newcommand{\SYM}{{\text{SYM}}}
\newcommand{\CS}{{\text{CS}}}
\newcommand{\SCS}{{\text{SCS}}}
\newcommand{\TMYM}{{\text{TMYM}}}
\newcommand{\ThirdWay}{^{\text{3rd}}_{\text{Way}}}
\newcommand{\SuperThirdWay}{^{\text{super}}_{\text{3\textsuperscript{rd}Way}}}
\newcommand{\PCS}{{\text{PCS}}}
\newcommand{\SPCS}{{\text{SPCS}}}
\newcommand{\mass}{{\text{mass}}}
\newcommand{\bi}{{\text{bi}}}
\newcommand{\res}{{\text{res}}}
\newcommand{\new}{{\textcolor{blue}{\text{new}}}}
\newcommand{\other}{{\text{other}}}

\newcommand{\idop}{{\mathbbm{1}}}
\newcommand{\Dbb}{{\mathbb{D}}}
\newcommand{\Dbbd}{{\Dbb^\dagger}}

\newcommand{\Qt}{{\accentset{\sim}{Q}}}
\newcommand{\zetat}{{\accentset{\sim}{\zeta}}}
\newcommand{\zetatt}{{\tilde{\zeta}}}
\newcommand{\zetab}{{\bar{\zeta}}}
\newcommand{\zetabt}{{\accentset{\sim}{\zetab}}}
\newcommand{\zetabtt}{{\tilde{\zetab}}}
\newcommand{\phit}{{\accentset{\sim}{\phi}}}
\newcommand{\psit}{{\accentset{\sim}{\psi}}}
\newcommand{\Ft}{{\accentset{\sim}{F}}}
\newcommand{\thetat}{{\accentset{\sim}{\theta}}}
\newcommand{\thetatt}{{\tilde{\theta}}}
\newcommand{\chit}{{\accentset{\sim}{\chi}}}
\newcommand{\St}{{\accentset{\sim}{S}}}
\newcommand{\lambdat}{{\accentset{\sim}{\lambda}}}
\newcommand{\Phit}{{\accentset{\sim}{\Phi}}}
\newcommand{\Lambdat}{{\accentset{\sim}{\Lambda}}}
\newcommand{\Lambdatt}{{\tilde{\Lambda}}}
\newcommand{\deltat}{{\accentset{\sim}{\delta}}}
\newcommand{\deltab}{{\bar{\delta}}}
\newcommand{\deltabt}{{\accentset{\sim}{\deltab}}}
\newcommand{\etat}{{\accentset{\sim}{\eta}}}
\newcommand{\etatt}{{\tilde{\eta}}}
\newcommand{\etab}{{\bar\eta}}
\newcommand{\etabt}{{\accentset{\sim}{\etab}}}
\newcommand{\etabtt}{{\tilde{\etab}}}

\newcommand{\ass}{{\slashed{a}}}
\newcommand{\As}{{\slashed{A}}}
\newcommand{\Ks}{{\slashed{K}}}

\newcommand{\Phih}{{\hat\Phi}}
\newcommand{\Phith}{{\hat\Phit}}

\newcommand{\Lgd}{{\Lg^\dagger}}
\newcommand{\zetad}{{\zeta^\dagger}}

\newcommand{\invast}{{\ast^{-1}}}
\newcommand{\asst}{{\ast}}
\newcommand{\invasst}{\ast^{-1}}
\newcommand{\deltal}{{\accentset{\leftarrow}{\delta}}}

\newcommand{\Ab}{{\bar A}}
\newcommand{\Abs}{{\bar{\slashed A}}}
\newcommand{\Fb}{{\bar F}}
\newcommand{\mb}{{\bar m}}
\newcommand{\kb}{{\bar k}}
\newcommand{\thetab}{{\bar\theta}}
\newcommand{\sigmab}{{\bar\sigma}}
\newcommand{\lambdab}{{\bar\lambda}}
\newcommand{\lambdabt}{{\accentset{\sim}{\lambdab}}}
\newcommand{\Db}{{\bar D}}
\newcommand{\Kb}{{\bar K}}

\newcommand{\intau}{{\oint\extd\tau}}
\newcommand{\xdot}{{\dot x}}
\newcommand{\psid}{{\psi^\dagger}}
\newcommand{\psidot}{{\dot{\psi}}}
\newcommand{\psidotd}{{\dot{\psi}{^\dagger}}}
\newcommand{\ydot}{{\dot{y}}}
\newcommand{\chid}{{\chi^\dagger}}
\newcommand{\chidot}{{\dot{\chi}}}
\newcommand{\chidotd}{{\dot{\chi}{^\dagger}}}
\newcommand{\Qop}{{\mathcal{Q}}}
\newcommand{\Qopd}{{\mathcal{Q}{^\dagger}}}
\newcommand{\WI}{{\mathcal{I}}}
\newcommand{\Tr}{\operatorname{Tr}}
\newcommand{\Son}{S^{\text{on}}}
\newcommand{\Soff}{S^{\text{off}}}
\newcommand{\thetad}{{\theta^\dagger}}
\newcommand{\Lagr}{{\mathbb{L}}}
\newcommand{\Dd}{{D^\dagger}}
\newcommand{\Qd}{{Q^\dagger}}
\newcommand{\lambdad}{{\lambda^\dagger}}
\newcommand{\Xf}{{\mathcal{X}}}


\section*{Acknowledgements}

I'd like to express gratitude towards my two promotors: Alex S. Arvanitakis and Alexandre Sevrin. I'm very grateful to Prof. Sevrin for being given the opportunity to work on such an interesting problem in mathematical physics, and I'm especially grateful to Alex S. Arvanitakis for directly supervising me throughout the process of writing this thesis. This thesis has presented to me an opportunity to grow as a person and as a researcher. I'd also like to thank my family for their continued support throughout my bachelor and master studies.

\section*{Disclaimer}

This version of my thesis has been reviewed for typos and errors for the purpose of uploading it on arXiv. However, the content itself hasn't been meaningfully altered. If you'd happen to find another typo or error feel free to contact me.

\chapter{Abstract}

The following is a master thesis centered around the concept of localisation and the Third Way Theory. 
This thesis discusses various aspects of supersymmetric localisation in one and three dimensions, and contains original results with regards to the Third Way Theory. 
It starts off with the Witten index for a one-dimensional supersymmetric system and derives various aspects through localisation. 
After this, the thesis moves on to the Third Way Theory. First, it offers a review of the Third Way Theory, a deformation of topologically massive Yang-Mills theory in three dimensions. Then it moves on to original results. These include a supersymmetrisation of the Third Way Theory and consequently a localisation of the Third Way Theory, which is to say, a method of deriving non-perturbative results.

\tableofcontents%

\hypersetup{
 linkcolor= blue
,citecolor= blue
,urlcolor= blue
}

\mainmatter%




\chapter{Introduction to Localisation}

\section{Introduction}

The main goal of this master thesis is to develop a method of doing non-perturbative computations in the Third Way Theory. This is a quantum field theory developed by Arvanitakis, Sevrin and Townsend \cite{arvanitakis2015yang}. 
The technique which will be used for this ---and one of the main themes of this thesis--- will be localisation.
Localisation is a technique in quantum field theory which borrows from the concept of equivariant localisation in differential geometry. 
Equivariant localisation is a technique for performing integration over Riemannian manifolds with isometries. One deforms the integrant such that its contributions localise to a lower dimensional ---if not discrete--- locus, all the while leaving the integral unchanged. This happens when the integrant shares its symmetries with the underlying geometry, and will localise to the points left unchanged by this symmetry.
As it turns out, this procedure can be generalised to infinite-dimensional field space, the space over which one integrates in quantum field theory when working in the path integral formalism. 
This is interesting because generally path integrals are extremely difficult to compute exactly and one is typically forced to use a perturbative approach, which for example only works for weak couplings. Particularly, we consider supersymmetric quantum field theories. In this context a localisation technique exists in which the path integral can be deformed in such a way that it will only receive contributions from a lower dimensional ---if not finite-dimensional or even discrete--- subspace of field space. This space will consist of BPS configurations. These are configurations in field space which one could consider to be supersymmetric. The observables for which one could then do exact quantum computations will be the space of BPS operators. That is, operators which are invariant under supersymmetry. A way to think of this in more physical terms is that the 1-loop approximation becomes exact \cite{cremonesi2013introduction}.

In this chapter we introduce the necessary background concepts for doing localisations. These will include some basic facts about supersymmetry as well a general description of the localisation argument. 

In chapter 2 we will consider as a warm up exercise the so called Witten index. This is an integer defined for a superparticle moving in a superpotential which is closely related to the partition function. We will show that this object can be computed through a localisation. The standard approach to this topic involves the Hilbert space formalism as well as the path integral formalism. However, in this thesis previously known results which have to our knowledge only been derived in the Hilbert space formalism will be derived from the path integral formalism. These constitute new computations in that they weren't performed directly following a different source and could be new altogether.

In chapter 3 we will move on to describing the Third Way Theory, the theory we want to localise in this thesis. Particularly, we present a detailed description of its original conception by Arvanitakis, Sevrin and Townsend in \cite{arvanitakis2015yang}.

Finally, in chapter 4 we will try to localise the Third Way Theory. We will start off by giving a brief review of $d=3+0$, $\NSUSY = 2$ gauge theory, starting from a superconnection in superspace and arriving at the vector multiplet in Wess-Zumino gauge with its gauge covariant supersymmetry transformation. This supermultiplet will then be used to construct various standard supersymmetric actions and their supersymmetric localisations. After this we will move on to a toy model version of the Third Way Theory, which we coined Proca-Chern-Simons theory, which is simply Chern-Simons theory with a Proca mass term. A new kind of supersymmetry was then constructed, which we named massive supersymmetry, which is a supersymmetry of super-Proca-Chern-Simons theory. Interestingly this supersymmetry forces the Proca mass term upon us if one were to try and construct a super-Chern-Simons theory with this supersymmetry. After this we will move on to describe localisation in this theory. Then we will move on to supersymmetrisations of the Third Way Theory. We start off by presenting previous results by Arvanitakis, resulting in a so-called $\delta$-localisation of the Third Way Theory. After this, the reader will be presented with our own results, consisting of a generalisation of Proca-Chern-Simons theory to the Third Way Theory. This then leads to its superanalogue coined the Super-Third Way Theory, which is invariant under so-called Third Way supersymmetries. We will then use these supersymmetries to localise the Third Way Theory.

\newpage 

\section{Some Basics of Supersymmetry}
\label{Some Basics of Supersymmetry}

\subsection{A Brief Motivation}

A supersymmetry is simply defined as a symmetry between bosons and fermions. Bosons are spin-integer particles which make up the forces and Higgs boson in the Standard Model where on the other hand, fermions are spin-half-odd particles which make up matter in the Standard Model \cite{sohnius1985introducing}\cite{peskin2018introduction}.

There are a couple of motivations for supersymmetry. Firstly, there's the concept of renormalisation in quantum field theory. Renormalisation is a way to make sense of divergent integrals which appear in quantum field theory. In this regard supersymmetry is promising because bosons and fermions tend to contribute in opposite signs to transition amplitudes and for some supersymmetric theories these concellations could be so strong that the theory is finite. Another motivation is that it would explain various unanswered questions in the Standard Model, such as the Higgs mass. Furthermore, it is also a predictive hypothesis, namely predicting that there exist yet to be detected superpartners to the current known field content of the Standard Model, which we're looking for at the LHC. On the theoretical front, a great motivation for supersymmetry is the so called Coleman-Mandula theorem. This theorem states that ---under mild assumptions--- the symmetry algebra of a quantum field theory is generally the direct sum of the Poincaré algebra and an internal symmetry algebra. This would mean that gauge theory is restricted to Yang-Mills type gauge fields and gravity. However, there exists a loophole in this theorem, being that the symmetry algebra is assumed to be a Lie algebra. If one assumes that fermionic symmetries could exist, this no-go theorem can be bypassed. It's quite possible that supersymmetry is the only extension to the symmetry algebras of quantum field theory. Supersymmetry also plays a key role in string theory and M-theory \cite{sohnius1985introducing}\cite{wess1992supersymmetry}\cite{blumenhagen2013basic}.

Finally, there are applications of supersymmetry in other domains of physics. This is where for example localisation comes in. As a matter of fact localisation can be used to do non-perturbative computations in quantum field theories which aren't necessarily supersymmetric. Namely, if the supersymmetric extension of a theory is physically equivalent to it's non-supersymmetric version, one could use the techniques of localisation to do non-perturbative computations in such theories. One example of this is Chern-Simons theory, and it will in fact be in the spirit of this example that we will attempt to localise the Third Way Theory without altering its physics \cite{willett2017localization}\cite{kapustin2010exact}.

\newpage

\subsection{Algebraic Aspects of Supersymmetry}

\subsubsection{Non-supersymmetric symmetry algebra}

We will introduce supersymmetries through their algebra. We start off by considering the general form of a supersymmetry algebra. For non-supersymmetric quantum field theories one has a symmetry algebra given by

\begin{equation}
\gLie = \isoLie(d_-,d_+) \oplus \gintLie
\end{equation}

\noindent where ---in accordance with the Coleman-Mandula theorem--- $\isoLie$ denotes the Poincaré algebra for some metric $\eta_{ab}$ of signature $(d_-,d_+)$ and $\gLie$ a Lie algebra corresponding to internal symmetries \cite{sohnius1985introducing}\cite{fuchs2003symmetries}. We denote the generators of $\isoLie$ by translations $P_a$ and rotations $M_{ab}$. These satisfy an algebra

\begin{align}
[P_a,P_b] &= 0
&
[P_a,M_{bc}] &= 2\eta_{a[b}P_{c]}
&
[M_{ab},M^{cd}] &= -4\eta_{[a}{^{[c}}M_{b]}{^{d]}}
\\
[T_I,T_J] &= f_{IJ}{^K}T_K
&
[T_I,P_a] &= 0
&
[T_I,M_{ab}] &= 0
\end{align}

\noindent where $f_{IJ}{^K}$ denote the structure constants of the Lie group $\gintLie$ and indices for the Poincaré generators are raised and lowered using $\eta_{ab}$ \cite{fuchs2003symmetries}. 

\subsubsection{Supersymmetry algebra for $d = 0+3$, $\NSUSY = 2$}

We now consider a supersymmetric extension of a symmetry algebra. Particularly, the case we will consider will be $\NSUSY = 2$, $d = 0 + 3$. $\NSUSY$ refers to the number of supercharges and $d$ refers to the spacetime dimension, in this case describing Euclidean space $\eta = \idop_{3\times 3}$. The reason for using Euclidean signature is that this thesis will mainly be concerned with quantum field theory and hence path integrals, which need to be defined on a Wick rotated space to be well-defined. In this case we introduce two spinorial supercharges $Q_\alpha$ and $\Qt_\alpha$, with $\alpha,\beta,\dots = 1,2$ referring to 2-spinor indices. For conventions I refer the reader to appendix \ref{Spinors and Gamma Matrices}. These introduce new (anti)commutators

\begin{align}
\{Q_\alpha,\Qt_\beta\} &= -2i\gamma^a{_{\alpha\beta}}P_a
&
\{Q_\alpha,Q_\beta\} &= \{\Qt_\alpha,\Qt_\beta\} = 0\\
[Q_\alpha,M_{ab}] &= \frac{i}{2}\epsilon_{abc}\gamma^c{_\alpha}{^\beta}Q_\beta
&
[Q_\alpha,T_I] &= [\Qt_\alpha,T_I] = 0
\\
[\Qt_\alpha,M_{ab}] &= \frac{i}{2}\epsilon_{abc}\gamma^c{_\alpha}{^\beta}\Qt_\beta
\end{align}

\noindent The most remarkable of these relations is the first, which states that supersymmetries square to translations. This algebra is an example of a Lie superalgebra. This is a generalisation of a Lie algebra, which may also contain fermionic generators \cite{freedman2012supergravity}.

\subsection{Representations: The Chiral and Antichiral Multiplets}

\subsubsection{Component fields}

In this section we consider an example of a representation of supersymmetry on field content. Field content in supersymmetric theories appear in supermultiplets. These are multiplets of fields on which a representation of supersymmetry is represented which closes on the fields. Supersymmetry transformations typically take on the form

\begin{align}
\delta (\text{boson}) &= \text{fermion}
\\
\delta (\text{fermion}) &= \del(\text{boson}) + \text{auxiliary}
\\
\delta (\text{auxiliary}) &= \del(\text{auxiliary})
\end{align}

\noindent  By `auxiliary', one means that the field in question won't be dynamical. As an example we consider a fairly simple representation of supersymmetry \cite{sohnius1985introducing}\cite{closset2013supersymmetric}. Namely, we consider the chiral and antichiral multiplet. These multiplets respectively have the following field content:

\begin{itemize}
\item a real scalar $\phi$, resp. $\phit$
\item a complex spinor $\psi$, resp. $\psit$
\item a real auxiliary scalar $F$, resp. $\Ft$
\end{itemize}

\noindent Translations will simply be represented as $P_\mu = \del_\mu$. Supersymmetry is represented slightly more complicatedly. The supersymmetry transformations are \textit{parametrised} by constant \textit{bosonic} spinors $\zeta$, $\zetat$. Note, however, that the \textit{spinorial field content} (e.g. $\psi$, $\psit$) is \textit{fermionic}. The supersymmetry transformations are represented on the multiplets $(\phi,\psi,F)$ and $(\phit,\psit,\Ft)$ as

\begin{align}
\label{chiral SUSY 1}
\delta_{\zeta,\zetatt}\phi &= 2\zeta\psi
&
\delta_{\zeta,\zetatt}\phit &= 2\zetat\psit
\Big.\\
\label{chiral SUSY 2}
\delta_{\zeta,\zetatt}\psi &= \zeta F + i\zetat\dels\phi
&
\delta_{\zeta,\zetatt}\psit &= \zetat\Ft + i\zeta\dels\phit
\Big.\\
\label{chiral SUSY 3}
\delta_{\zeta,\zetatt}F &= -2i\zetat\dels\psi
&
\delta_{\zeta,\zetatt}\Ft &= -2i\zeta\dels\psit
\Big.
\end{align}

\noindent A direct computation indeed yields

\begin{align}
\{\delta_\zeta,\delta_\zetatt\}
\begin{pmatrix}
\phi \\ \psi \\ F
\end{pmatrix}
&= 2i\zeta\dels\zetat
\begin{pmatrix}
\phi \\ \psi \\ F
\end{pmatrix}
&
\{\delta_\zeta,\delta_\zetatt\}
\begin{pmatrix}
\phit \\ \psit \\ \Ft
\end{pmatrix}
&= 2i\zeta\dels\zetat
\begin{pmatrix}
\phit \\ \psit \\ \Ft
\end{pmatrix}
\end{align}

\noindent verifying that the algebra is indeed represented in these fields. The reason for introducing bosonic spinors might be slightly counterintuitive, as for fields spinors are typically taken to be fermionic. However, in the case of parameters for supersymmetry the more fundamental choice turns out to be bosonic spinors. This is important because as we will later see the choice of bosonic spinors is fundamental to the workings of localisation. 
One reason for this is so that one could normalise by dividing by $|\zeta|^2 = \zetad{^\alpha}\zeta_\alpha$. It should further be noted then that supersymmetry variations are Grassmann-odd instead of Grassmann-even, that is to say,

\begin{equation}
\delta_{\zeta,\zetatt}\big\{AB\big\} = (\delta_{\zeta,\zetatt}A)B + (-)^{|A|}A(\delta_{\zeta,\zetatt}B).
\end{equation}

\noindent One can obtain Grassmann-even supersymmetry transformations parametrised by fermionic spinors simply by multiplying the variation by a Grassmann parameter $\epsilon$:
$\delta_{\zeta,\zetatt} \to \epsilon\delta_{\zeta,\zetatt} = \delta_{\epsilon\zeta,\epsilon\zetatt}$. However since the bosonic spinors are fundamental to the localisation arguments we'll keep working with bosonic supersymmetry parameters.

\subsubsection{Superspace approach}

This is a very interesting case of a supersymmetry representation. But one could wonder, \textit{isn't this a little ad hoc?} It would be better if we had a way of finding the supersymmetry transformations from first principles, instead of guessing them out of thin air. One way one could `geometrise' supersymmetry is through the construct of superspace. In the superspace approach one extends spacetime with Grassmann coordinates $\theta^\alpha$, $\thetat{^\alpha}$. A supermultiplet is then described through a superfield $\Phi(x,\theta,\thetat)$ on superspace. A superfield can be expanded in powers of its Grassmann coordinates and the corresponding field valued coefficients of each term will correspond to a component field \cite{sohnius1985introducing}\cite{closset2013supersymmetric}. 


Supersymmetry generators are represented through the differential operators

\begin{align}
Q_\alpha &= \del_\alpha + i\dels_\alpha\thetat := \del_\alpha + i\dels_\alpha{^\beta}\thetat_\beta
\\
\Qt_\alpha &= \delt_\alpha + i\dels_\alpha\theta := \delt_\alpha + i\dels_\alpha{^\beta}\theta_\beta
\end{align}

\noindent One notes that these satisfy an algebra

\begin{equation}
\{Q_\alpha,\Qt_\beta\} = i\dels_\beta{^\gamma}\{\del_\alpha,\theta_\gamma\} + i\dels_\alpha{^\gamma}\{\thetat_\gamma,\delt_\beta\} = -2i\dels_{\alpha\beta}
\end{equation}

\noindent Another important construct in superspace will be be a notion of differentiation which preserves supersymmetry. That is, a form of differentiation which (anti)commutes with supersymmetry. We introduce covariant derivatives as

\begin{align}
\label{covariant superspace derivatives}
D_\mu &= \del_\mu
&
D_\alpha &= \del_\alpha - i\dels_\alpha\thetat
&
\Dt_\alpha &= \delt_\alpha - i\dels_\alpha\theta
\end{align}

\noindent These indeed preserve supersymmetry as

\begin{align}
[Q_\alpha,\del_\mu] &= \{Q_\alpha,D_\beta\} = \{Q_\alpha,\Dt_\beta\} = 0
\\
[\Qt_\alpha,\del_\mu] &= \{\Qt_\alpha,D_\beta\} = \{\Qt_\alpha,\Dt_\beta\} = 0
\end{align}

\noindent We further also find anticommutators

\begin{align}
\label{covariant superspace derivatives algebra}
\{D_\alpha,\Dt_\beta\} &= 2i\dels_{\alpha\beta}
&
\{D_\alpha,D_\beta\} &= \{\Dt_\alpha,\Dt_\beta\} = 0.
\end{align}

\noindent These tools will now allow us to define the chiral and antichiral multiplets in superspace. We start off by noticing that an unconstrained superfield $\Phi(x,\theta,\thetat)$ will not define an irreducible representation of the supersymetry algebra. Indeed, since the covariant spinorial derivatives $D_\alpha$ and $\Dt_\alpha$ anticommute with the supersymmetry generators we can use them to contrain the chiral and antichiral superfields $\Phi$ and $\Phit$ as

\begin{align}
\Dt_\alpha\Phi(x,\theta,\thetat) &= 0
&
D_\alpha\Phit(x,\theta,\thetat) &= 0.
\end{align}

\noindent Solving this constraint for a most general expression of these fields might seem a little tricky at first glance. However, this is greatly simplified if one notices that

\begin{align}
[\del_\alpha,e^{-i\thetatt\dels\theta}] &= e^{-i\thetatt\dels\theta}[\del_\alpha,-i\thetat\dels\theta] = e^{-i\thetatt\dels\theta}i\dels_\alpha\thetat
\\
[\delt_\alpha,e^{+i\thetatt\dels\theta}] &= e^{+i\thetatt\dels\theta}[\delt_\alpha,+i\thetat\dels\theta] = e^{+i\thetatt\dels\theta}i\dels_\alpha\theta
\end{align}

\noindent so that

\begin{align}
D_\alpha e^{-i\thetatt\dels\theta} &= e^{-i\thetatt\dels\theta}\del_\alpha
&
D_\alpha e^{+i\thetatt\dels\theta} &= e^{+i\thetatt\dels\theta}\big(\del_\alpha - 2i\dels_\alpha\thetat\big)
\\
\Dt_\alpha e^{+i\thetatt\dels\theta} &= e^{+i\thetatt\dels\theta}\delt_\alpha
&
\Dt_\alpha e^{+i\thetatt\dels\theta} &= e^{+i\thetatt\dels\theta}\big(\delt_\alpha - 2i\dels_\alpha\theta\big)
\end{align}

\noindent The constraints for the chiral and antichiral superfields are now trivially solved as

\begin{align}
\Phi &= e^{+i\thetatt\dels\theta}\Big[\phi + 2\psi\theta + F\theta^2\Big]
\\
\Phit &= e^{-i\thetatt\dels\theta}\Big[\phit + 2\psit\thetat + F\thetat{^2}\Big]
\end{align}

\noindent Consequently one finds that

\begin{equation}
\begin{split}
\delta_{\zeta,\zetatt}\Phi 
&= (\zeta Q + \zetat\Qt)\Phi 
\\
&= e^{i\thetatt\dels\theta}\big(\zeta\del + 2i\zetat\dels\theta\big)\Big[\phi + 2\psi\theta + F\theta^2\Big]
\\
&= e^{i\thetatt\dels\theta}\Big[2\zeta\psi + 2\big(\zeta F + i\zetat\dels\phi\big)\theta - 2i\zetat\dels\psi\theta^2\Big]
\\
&=: e^{i\thetatt\dels\theta}\Big[\delta_{\zeta,\zetatt}\phi + 2\delta_{\zeta,\zetatt}\psi\theta + \delta_{\zeta,\zetatt}F\theta^2\Big]
\end{split}
\end{equation}

\noindent and similarly that

\begin{equation}
\begin{split}
\delta_{\zeta,\zetatt}\Phit 
&= e^{-i\thetatt\dels\theta}\Big[2\zeta\psi + 2\big(\zetat \Ft + i\zeta\dels\phit\big)\thetat - 2i\zeta\dels\psit\thetat{^2}\Big]
\\
&=: e^{i\thetatt\dels\theta}\Big[\delta_{\zeta,\zetatt}\phit + 2\delta_{\zeta,\zetatt}\psit\thetat + \delta_{\zeta,\zetatt}\Ft\thetat{^2}\Big]
\end{split}
\end{equation}

\noindent which reproduces the results from equations \ref{chiral SUSY 1}-\ref{chiral SUSY 3} \cite{sohnius1985introducing}.

\section{Equivariant Localisation}

In this section we introduce a finite-dimensional toy model for localisation. After doing so we move on next section to do this for supersymmetric theories. Doing so we will mostly follow Cremonesi's lecture notes on introducing localisation \cite{cremonesi2013introduction}.

\subsection{Stationary Phase Approximation}

We start off by introducing the localisation argument for the finite-dimensional case. To do so we consider an integral over some $2\ell$-dimensional Riemannian manifold $M$, given by

\begin{equation}
Z_f(t) = \int_M\extd^{2\ell}x\sqrt{g}e^{itf(x)}
\end{equation}

\noindent which functions as a toy model for a partition function. $f$ is here taken to be a Morse function, that is, it has a discrete set of stationary points $x_k$ such that $\extd f(x_k) = 0$. In this context one can regard $t$ as $t = 1/\hbar$. The stationary phase approximation can then simply be given by

\begin{equation}
Z_f(t) = \bigg(\frac{2\pi i}{t}\bigg)^\ell\sum_{x_k:\extd f(x_k) = 0} (-i)^{\lambda(x_k)}\frac{e^{itf(x_k)}}{\sqrt{\det g^{-1}H_f(x_k)}} + \Og(t^{-\ell-1})
\end{equation}

\noindent where we defined $\lambda(x_k) = \#(\text{negative eigenvalues of $g^{-1}H_f(x_k)$})$ known as the Morse index at $x_k$. This result ---which follows simply from solving Gaussian integrals--- is known as stationary phase approximation. The way it can be interpret is that as we take $t\to\infty$, or equivalently, $\hbar\to 0$ the contributions to this integral become localised to the stationary points of the Morse function $f$. This is analogous to the semi-classical approach in quantum field theory, in which one integrates over field configurations which minimise the action weighted by the 1-loop determinant of the quadratic fluctuations around these configurations \cite{cremonesi2013introduction}.

\subsubsection{Illustrating example}

As an illustrating example we consider the height function on the 2-sphere \cite{cremonesi2013introduction}. That is, we take $M = S^2$ with a metric

\begin{equation}
\extd s^2 = \extd\theta^2 + \sin^2\theta\extd\varphi^2
\end{equation}

\noindent and the Morse function to be the height function

\begin{equation}
f(\theta,\varphi) = \cos\theta.
\end{equation}

\noindent with stationary points $\theta = 0,\pi$. The partition function is then simply given by

\begin{equation}
\label{illustrating example}
Z_f(t) = \int_{S^2}\extd\theta\extd\varphi~\sin\theta e^{it\cos\theta} = \frac{2\pi i}{t}\big(-e^{it} + e^{-it}\big)
\end{equation}

\noindent which \textit{precisely} agrees with the stationary phase approximation! This now raises the question, \textit{how do we interpret this result?} It turns out that the correct interpretation is due to the work of Duistermaat and Heckman. The key things to notice is that $(1)$ the manifold $S^2$ has a U(1) isometry, $(2)$ this isometry is also a symmetry of the height function and that $(3)$ it has fixed points, in this case corresponding to the poles of the 2-sphere.

\subsection{Equivariant Cohomology}

We now move on to describe the above mentioned principles in a formal manner. We consider a $2\ell$-dimensional compact Riemannian manifold $M$ without boundary, which has an Abelian (for simplicity) isometry group $G$. 

On this manifold we would like to define the appropriate kind of cohomology suited for essentially reducing the integral from the manifold $M$ to the coset space $M/G$. We further consider a Killing vector $V$, that is, a vector satisfying

\begin{equation}
0 = \Lie_V g_{\mu\nu} = 2\nabla_{(\mu}V_{\nu)}
\end{equation}

\noindent assumed to generate a U(1) symmetry. Let us further denote $\Omega(M)$ the space of polyforms on $M$. On this space we define the $V$-equivariant exterior derivative $\extd_V$ as

\begin{equation}
\extd_V := \extd - \iota_V : \Omega(M) \to \Omega(M)
\end{equation}

\noindent where $\extd : \Omega(M)\to\Omega(M)$ the exterior derivative on polyforms and $\iota_V : \Omega(M)\to\Omega(M)$ the interior product by $V$. These respectively incerase and decrease the differential degree by one. We note that the $V$-equivariant exterior derivative has the interesting property

\begin{equation}
\extd^2_V = -\{\extd,\iota_V\} = -\Lie_V
\end{equation}

\noindent of squaring to the Lie derivative with respect to $-V$. We now wish to restrict to the space of polyforms on which $\extd_V$ is nilpotent. This motivates us to define the space of $V$-equivariant polyforms $\Omega_V(M) = \{\alpha\in\Omega(M)~|~\Lie_V\alpha = 0\}$. Let $\alpha \in \Omega_V(M)$, we define

\begin{align}
&\alpha \text{ $V$-equivariantly closed}
&
&\Longleftrightarrow
&
\alpha &\in \Ker\extd_V\big|_{\Omega_V(M)}
\\
&\alpha \text{ $V$-equivariantly exact}
&
&\Longleftrightarrow
&
\alpha &\in \Imm\extd_V\big|_{\Omega_V(M)}
\end{align}

\noindent This in turn allows us to define the $V$-equivariant cohomology ring

\begin{equation}
H_V(M) = \frac{\Ker\extd_V|_{\Omega_V(M)}}{\Imm\extd_V|_{\Omega_V(M)}}
\end{equation}

\noindent consisting of $V$-equivariantly closed polyforms modulo $V$-equivariantly exact polyforms. We further recall that integration over polyforms is defined as integrating over the top form:

\begin{equation}
\int_M\alpha := \int_M\alpha_{2\ell}
\end{equation}

\noindent where $\alpha_r$ denotes the $r$-form component of the polyform $\alpha$. Noting that $(\extd_V\alpha)_{2\ell} = \extd\alpha_{2\ell-1}$ we find that

\begin{equation}
\int_M\extd_V\alpha = \int_M\extd\alpha_{2\ell-1} = \int_{\del M}\alpha_{2\ell-1} = 0.
\end{equation}

\noindent Thus, we conclude that the integral of a $V$-equivariant polyform $\alpha$ only depends on its $V$-equivariant cohomology class $[\alpha]$ \cite{cremonesi2013introduction}!


\subsection{Localisation Arguments for Equivariant Integrals}

Now we introduce two localisation arguments for equivariant integrals \cite{cremonesi2013introduction}. Namely, the claim is that the integral's contributions are localised to the zero locus of $V$:

\begin{equation}
M_V = \{x \in M ~|~ V|_x = 0\}.
\end{equation}

\noindent We now give two arguments for this claim:

\subsubsection{First localisation argument}

This argument is centered around proving a version of Poincaré's lemma. Let $\alpha$ be $V$-equivariantly closed: $\extd_V\alpha = 0$. We wish to show that it is equivariantly exact on the complement $M\setminus M_V$ of the zero locus. We can define the 1-form dual to the Killing vector as

\begin{align}
\eta &= V_\mu\extd x^\mu \in \Omega^1(M)
&
\eta(X) &= g(V,X)
\end{align}

\noindent We note that $\eta$ is a $V$-equivariant polyform since $V$ is a Killing vector:

\begin{equation}
\Lie_V \eta_\mu = V^\nu\nabla_\nu V_\mu + \nabla_\mu V^\nu V_\nu = 2V^\nu\nabla_{(\mu}V_{\nu)} = 0.
\end{equation}

\noindent Hence we find that $\eta \in \Omega_V(M)$. We further note that

\begin{equation}
\extd_V\eta = \extd\eta - \iota_V\eta = -|V|^2 + \extd\eta.
\end{equation}

\noindent Interestingly, for polyforms there exists a notion of inversion, but this is only if the 0-form component is non-vanishing and is defined through the terminating Taylor expansion of the inverse in its form components. It thus follows that one can invert $\extd_V\eta$ on the complement $M\setminus M_V$ of the zero locus as

\begin{equation}
\frac{1}{\extd_V\eta} = -\frac{1}{|V|^2}\Big(1 - \frac{\extd\eta}{|V|^2}\Big)^{-1} = -\frac{1}{|V|^2}\sum_{k=0}^\ell\Big(\frac{\extd\eta}{|V|^2}\Big)^k.
\end{equation}

\noindent This polyform is well-defined and is furthermore $V$-equivariantly closed. Indeed, one finds that

\begin{equation}
\extd_V\frac{1}{\extd_V\eta} = -\frac{\extd^2_V\eta}{(\extd_V\eta)^2} = \frac{\Lie_V\eta}{(\extd_V\eta)^2} = 0.
\end{equation}

\noindent We now define the polyform

\begin{equation}
\Theta_V := \frac{\eta}{\extd_V\eta}.
\end{equation}

\noindent This polyform satisfies the interesting identity

\begin{equation}
\extd_V\Theta_V = \frac{\extd_V\eta}{\extd_V\eta} - \eta\extd_V\frac{1}{\extd_V\eta} = 1 + 0 = 1
\end{equation}

\noindent where we applied the graded Leibniz rule. We now come back to the $V$-equivariantly closed polyform $\alpha$. Using the aforementioned results it directly follows that $\alpha$ is $V$ equivariantly exact on $M\setminus M_V$:

\begin{equation}
\alpha = 1\cdot \alpha = \extd_V\Theta_V\cdot\alpha = \extd_V(\Theta_V\alpha).
\end{equation}

\noindent Hence, it follows that the contributions to the equivariant integral have to localise to the zero locus $M_V$ \cite{cremonesi2013introduction}.

\subsubsection{Second localisation argument}

This argument will be more direct than the previous localisation argument. It will also be more directly related to the generalisation to path integrals. This argument hinges on the previous observation that equivariant integrals only depend on the respective equivariant cohomology class. Let us for example consider an equivariant polyforms $\alpha,\beta \in \Omega_V(M)$. We note that we can deform $\alpha$ without changing its equivariant cohomology class as

\begin{equation}
\alpha_t := \alpha e^{t\extd_V\beta}
\end{equation}

\noindent To see this, we start off by noting that

\begin{equation}
\extd_V e^{t\extd_V\beta} = \extd_V(t\extd_V\beta)e^{t\extd_V\beta} = t\extd^2_V\beta e^{t\extd_V\beta} = -t\Lie_V\beta e^{t\extd_V\beta} = 0.
\end{equation}

\noindent From this it follows that

\begin{equation}
\alpha_t - \alpha = \int_0^t\extd t ~ \der{}{t}\alpha_t = \int_0^t\extd t ~ \extd_V\beta\alpha e^{t\extd_V\beta} = \extd_V\int_0^t\extd t ~ \beta\alpha e^{t\extd_V\beta}
\end{equation}

\noindent hence we conclude that this kind of deformation doesn't change the equivariant cohomology class of $\alpha$:

\begin{equation}
[\alpha] = [\alpha e^{t\extd_V\beta}]
\end{equation}

\noindent and thus,

\begin{equation}
\int_M\alpha = \int_M\alpha e^{t\extd_V\beta}.
\end{equation}

\noindent If we now take $\beta = \eta$ we find that

\begin{equation}
\int_M\alpha = \int_M\alpha e^{t\extd_V\eta} = \int_M\alpha e^{-t|V|^2}e^{t\extd\eta}.
\end{equation}

\noindent Taking the limit $t \to \infty$ we thus find that

\begin{equation}
\int_M\alpha = \lim_{t\to\infty}\int_M\alpha e^{-t|V|^2}e^{t\extd\eta}.
\end{equation}

\noindent Note that for all points on the manifold where $V$ is nonzero, the corresponding contributions will be suppressed by the deformation of the form. This leaves us with contributions from the infinitesimal regions around the zero locus $M_V$ \cite{cremonesi2013introduction}.

\subsubsection{Atiyah-Bott-Berline-Vergne localisation formula}

Let us now compute the equivariant integral in question. Around the loci we use the `inertial' Cartesian coordinate system in which the metric becomes

\begin{equation}
\extd s^2 
\approx \sum_{i = 1}^\ell \big[\extd x_i^2 + \extd y_i^2\big]
= \sum_{i = 0}^\ell \big[\extd r_i^2 + r_i^2\extd \varphi_i^2\big].
\end{equation}

\noindent Around these loci we can furthermore also write the Killing vector $V$ as

\begin{equation}
V \approx \sum_{i = 1}^\ell\omega_{k,i}\bigg[-y_i\pder{}{x_i} + x_i\pder{}{y_i}\bigg] = \sum_{i = 1}^\ell\omega_{k,i}\pder{}{\varphi_i}.
\end{equation}

\noindent Here the index $k$ enumerates the zero locus $M_V = \{x_k\}$. The Killing vector generates approximately a linear transformation on these `inertial' Cartesian coordinates given by the block diagonal matrix

\begin{equation}
\label{L_V}
L_{V,k} = \diag\Bigg(
\begin{pmatrix}
0 & \omega_{k,1}\\
-\omega_{k,1} & 0
\end{pmatrix},
\dots,
\begin{pmatrix}
0 & \omega_{k,\ell}\\
-\omega_{k,\ell} & 0
\end{pmatrix}
\Bigg).
\end{equation}

\noindent We further note that the dual 1-form $\eta$ is given around the $k^\text{th}$ locus by

\begin{equation}
\eta \approx \sum_{i=1}^\ell \omega_{k,i}\Big[-y_i\extd x_i + x_i\extd y_i\Big] = \sum_{i=1}^\ell\omega_{k,i}r_i^2\extd\varphi_i.
\end{equation}

\noindent from which one obtains an equivariant differential

\begin{equation}
\extd_V\eta 
\approx \sum_{i=1}^\ell\bigg[2\omega_{k,i}\extd x_i\extd y_i - \omega_{k,i}^2(x_i^2+y_i^2)\bigg]
= \sum_{i = 1}^\ell\bigg[\omega_{k,i}\extd(r_i^2)\extd\varphi_i - \omega_{k,i}^2r_i^2\bigg].
\end{equation}

\noindent We now find that the equivariant integral can be computed as

\begin{equation}
\begin{split}
\int_M\alpha 
&= \lim_{t\to\infty}\int_M\alpha e^{t\extd_V\eta} = \sum_{x_k}\lim_{t\to\infty}\int_{U_k}\alpha e^{t\extd_V\eta}
\\
&= \sum_{x_k}\lim_{t\to\infty}\alpha_0(x_k)\prod_{i=1}^\ell\int\extd(tr_i^2)\extd\varphi_i ~ e^{-t\omega_{k,i}^2r_i^2}
\\
&= \sum_{x_k}\alpha_0(x_k)\frac{(2\pi)^\ell}{\prod_{i=1}^\ell\omega_{k,i}}
\end{split}
\end{equation}

\noindent where $U_k$ are sufficiently small neighbourhoods around $x_k$. We note that we can rewrite the product in this expression as

\begin{equation}
\prod_{i=1}^\ell\omega_{k,i} = \Pf(-L_V(x_k)).
\end{equation}

\noindent Here the Pfaffian is defined on antisymmetric $2\ell\times 2\ell$ matrices as

\begin{equation}
\Pf M = \varepsilon^{i_1\dots i_{2\ell}}M_{i_1i_2}\dots M_{i_{2\ell-1}i_{2\ell}}.
\end{equation}

\noindent We thus find the following powerful result:

\begin{tcolorbox}

\subsubsection{Atiyah-Bott-Berline-Vergne localisation formula}

Let $M$ be a Riemannian manifold with Killing vector $V$ which generates a U(1) action with discrete zero locus $M_V$. Let furthermore $\alpha$ be a $V$-equivariantly closed polyform. Its integral over $M$ is then given by

\begin{equation}
\int_M\alpha = (2\pi)^\ell\sum_{x_k}\frac{\alpha_0(x_k)}{\Pf(-L_V(x_k))}
\end{equation}

\noindent where we defined $L_V(x_k)$ as in \ref{L_V} \cite{cremonesi2013introduction}.

\end{tcolorbox}

\subsubsection{Duistermaat-Heckman localisation formula}

As a corrolary to the Atiyah-Bott-Berline-Vergne localisation formula we will now also derive the Duistermaat-Heckman localisation formula. Let us consider a symplectic manifold $(M,\omega)$. For completeness, this is an even-dimensional manifold endowed with a \textit{closed} non-degenerate 2-form $\omega$, called the symplectic form \cite{cremonesi2013introduction}. 

Let us now consider a Hamiltonian $H$ on this symplectic manifold with $V$ its associated Hamiltonian vector field. That is, $V$ satisfies the equation

\begin{align}
\extd H &= \iota_V\omega
&
&\Longleftrightarrow
&
\extd_V(H+\omega) &= 0
\end{align}

\noindent since $H$ is a 0-form and $\omega$ is closed. We can apply the Atiyah-Bott-Berline-Vergne localisation formula ---assuming that $V$ is also a Killing vector to some underlying metric--- to solve oscillatory integrals of the form

\begin{equation}
Z_H(t) = \int_M\frac{\omega^\ell}{\ell!}e^{iHt} = \frac{1}{(it)^\ell}\int_M e^{it(H+\omega)}.
\end{equation}

\noindent Indeed, applying the Atiyah-Bott-Berline-Vergne localisation formula we find the following celebrated result \cite{cremonesi2013introduction}:

\begin{tcolorbox}

\subsubsection{Duistermaat-Heckman formula}

\begin{equation}
Z_H(t) = \int_M\frac{\omega^\ell}{\ell!}e^{iHt} = \bigg(\frac{2\pi i}{t}\bigg)^\ell\sum_{x_k:\extd H(x_k) = 0}\frac{e^{itH(x_k)}}{\Pf L_V(x_k)}
\end{equation}

\end{tcolorbox}

\subsubsection{Illustrating example}

Let us now go back to the example given in equation \ref{illustrating example} of a case where the stationary phase approximation is exact. That is, the integral

\begin{equation}
Z(t) = \int_{S^2}\extd\varphi\extd\theta~\sin\theta e^{it\cos\theta}.
\end{equation}

\noindent In this case the U(1) action is generated by the Killing vector $\del_\varphi$, which we note is also a symmetry of the integrand. We further note that this partition function can be rewritten as

\begin{equation}
Z(t) = \int_{S^2}\extd\varphi\extd\cos\theta ~ e^{it\cos\theta} = \frac{1}{it}\int_{S^2}e^{it(\cos\theta + \extd\varphi\extd\cos\theta)}.
\end{equation}

\noindent This lends the interesting interpretation of a Hamiltonian $H = \cos\theta$ with corresponding symplectic form $\omega = \extd\varphi\extd\cos\theta$ with the Killing vector $\del_\varphi$ generating the Hamiltonian flow. And indeed, this interpretation is correct since

\begin{equation}
\iota_{\del_\varphi}\omega = \pder{}{\extd\varphi}\big(\extd\varphi\extd\cos\theta\big) = \extd\cos\theta = \extd H.
\end{equation}

\noindent As such, application of the Duistermaat-Heckman localisation formula leads to

\begin{equation}
Z(t) = \frac{2\pi i}{t}\big(-e^{it} + e^{-it}\big).
\end{equation}

\noindent in agreement with equation \ref{illustrating example} \cite{cremonesi2013introduction}.


\section{Supersymmetric Localisation}

\subsection{Analogy between Equivariant Localisation and Supersymmetric Localisation}

Now we ready to move on to path integrals. Again, we will for the most part follow Cremonesi's lecture notes \cite{cremonesi2013introduction} in this section. In our context, we will assume that these path integrals are defined on compact, Riemannian manifolds. The reason for this is that we'd like to avoid infrared divergences for the path integral to be well-defined. Another more concrete reason is that we want the analogue of $-|V|^2$ to have a definite signature in the upcoming field space generalisations of the localisation argument. In our context we will denote the field content collectively by $\Phi$, with $\Phi|_\bos$ denoting the collective bosonic field content and $\Phi|_\fer$ denoting the collective fermionic field content \cite{cremonesi2013introduction}.

We start off by noticing that the previous approach was a toy model for what we're about to do now. In our previous example, even and odd forms were to be understood as corresponding to respectively bosonic and fermionic fields, with integration over a manifold being the toy version of integration over field space: 
\begin{align}
&\alpha_{\text{even}}/\alpha_{\text{odd}}
&
&\longleftrightarrow
&
&\text{bosons/fermions}
\bigg.\\
&\int_M\alpha
&
&\longleftrightarrow
&
&\int\Dg\Phi~\Og e^{-S[\Phi]}
\bigg.
\end{align}

\noindent Furthermore, the $V$-equivariant exterior derivative is to be understood as a supersymmetry transformation. Indeed, we recall that

\begin{equation}
(\extd_V\alpha)_p = \extd\alpha_{p-1} + \iota_V\alpha_{p+1}
\end{equation}

\noindent where if $p$ even/odd one has $p\pm 1$ odd/even, and likewise for supersymmetry transformations one typically has the form
\begin{align}
\Qloc(\text{boson}) &= \text{fermion}
\\
\Qloc(\text{fermion}) &= \del(\text{boson}) + \text{auxiliary}
\\
\Qloc(\text{auxiliary}) &= \del(\text{fermion})
\end{align}

\noindent Also the algebra shows a strong analogy. Indeed, we have

\begin{equation}
\extd_V^2 = -\Lie_V
\end{equation}

\noindent and likewise

\begin{equation}
\Qloc^2 = \Bg = \text{diffeo} + \text{gauge} + \text{field equations} + \dots
\end{equation}

\noindent where $\Bg$ stands for a bosonic operator. The objects we were integrating over previous were $V$-equivariantly closed polyforms $\alpha$: $\extd_V\alpha = 0$. It is clear that its analogues in supersymmetric field theory will be operators which are $\Qloc$-closed: $\Qloc\Og_\BPS = 0$, which we will refer to as BPS-operators. Similarly, we define $\Qloc$-exact operators as being of the form $\Qloc\Og$. In summary, we have the following analogies \cite{cremonesi2013introduction}:

\begin{align}
&\extd_V
&
&\longleftrightarrow
&
&\Qloc
\Big.\\
&\extd_V^2 = -\Lie_V
&
&\longleftrightarrow
&
&\Qloc^2 = \Bg
\Big.\\
&\extd_V\alpha
&
&\longleftrightarrow
&
&\Qloc\Og_\BPS = 0
\Big.\\
&\alpha = \extd_V\beta
&
&\longleftrightarrow
&
&\Og = \Qloc\Og'
\Big.
\end{align}

\subsection{$\Qloc$-cohomology and Path Integrals}

We now move on to describe how cohomology works on field space and what its implications are for path integrals. We recall that Stokes' theorem on manifolds is given by

\begin{equation}
\int_M\extd\alpha = \int_{\del M}\alpha.
\end{equation}

\noindent One could now wonder how this generalises to field space. On field space, the analogues of derivatives $\del_\mu$ will simply be functional derivatives $\delta/\delta\Phi(x)$, where indices are replaced by points in spacetime and the Kronecker $\delta$-symbol between various indices by the Dirac $\delta$-distribution. However, a generalisation which won't be straightforward is the field space analogue of a boundary. In our setting we will assume that a boundary is associated to fields which decay fast enough as they go to (a possible) infinity. In this case the field space analogue of Stokes' theorem will be given by

\begin{equation}
\int\Dg\Phi ~ \fder{}{\Phi}\Big\{\Og[\Phi]e^{-S[\Phi]}\Big\} = 0.
\end{equation}

\noindent To relate this result to its $V$-equivariant analogue, we further note that the supersymmetry operator $\Qloc$ can be written as

\begin{equation}
\Qloc = \int\extd^dx ~ \bigg[(\fer)\fder{}{(\bos)} + \Big(\del(\bos) + \aux\Big)\fder{}{(\fer)} + \del(\fer)\fder{}{(\aux)}\bigg]
\end{equation}

\noindent As such, we arrive at the analogy

\begin{align}
&\int_M\extd_V\alpha = 0
&
&\longleftrightarrow
&
&\int\Dg\Phi ~ \Qloc\Big\{\Og[\Phi]e^{-S[\Phi]}\Big\} = 0
\end{align}

\noindent It should be noted that this analogue is equivalent to the invariance of the path integral measure with respect to supersymmetry. That is, throughout the rest of the thesis we will assume this is the case but generally this doesn't have to be the case. Now, let's suppose that we consider a supersymmetric theory, that is a theory with a $\Qloc$-closed action:

\begin{equation}
\Qloc S[\Phi] = 0.
\end{equation}

\noindent We consider the expectation value of various operators now. We recall that the expectation value is given by

\begin{equation}
\EV{\Og} = \int\Dg\Phi ~ \Og[\Phi]e^{-S[\Phi]}.
\end{equation}

\noindent Interestingly, as it turns out, this expectation value will only depend on the $\Qloc$-cohomology class of the operator. Indeed, we find that

\begin{equation}
\EV{\Qloc\Og} = \int\Dg\Phi ~ \Qloc\Og[\Phi]e^{-S[\Phi]} = \int\Dg\Phi ~ \Qloc\Big\{\Og[\Phi]e^{-S[\Phi]}\Big\} = 0
\end{equation}

\noindent where the second step follows from the fact that the supersymmetry variation $\Qloc$ satisfies the Leibniz rule and that the action is supersymmetric. We thus conclude

\begin{tcolorbox}

\subsubsection*{$\Qloc$-cohomology}

\begin{equation}
\EV{\Og_\BPS} = \EV{\Og_\BPS + \Qloc\Og}
\end{equation}

In supersymmetric theories, the expectation value of a BPS operator is determined by its $\Qloc$-cohomology class.

\end{tcolorbox}

\noindent Note that nowhere we required the operator inserted to be a BPS operator for the above identity to hold. However, this is necessary firstly for cohomology to be well defined, and this will also become relevant for the localisation arguments \cite{cremonesi2013introduction}.

\subsection{Localisation Arguments for Supersymmetric Path Integrals}
\label{Localisation Arguments for Supersymmetric Path Integrals}

In this section we outline localisation arguments for supersymmetric path integrals. These will be analogous to the arguments made in previous section for equivariant localisation. 

\subsubsection{First localisation argument}

This argument was due to Witten \cite{witten1991mirror}. Suppose we have a fermionic symmetry group $G$ which acts on field space $\Fscr$, which has fixed points $\Fscr_\Qloc$. It then follows that the symmetry group acts freely on $\Fscr\setminus\Fscr_\Qloc$, and hence we can `integrate along the fibers' generated by the fermionic symmetry group $G$. This yields a path integral

\begin{equation}
\int_{\Fscr\setminus\Fscr_\Qloc}\Dg\Phi ~ \Og[\Phi]e^{-S[\Phi]} = \Vol(G)\int_{(\Fscr\setminus\Fscr_\Qloc)/G}\Dg\Phi ~ \Og[\Phi]e^{-S[\Phi]}.
\end{equation}

\noindent However, since we're dealing with a fermionic group, we find that the volume of this group will be of the form

\begin{equation}
\Vol(G) = \int\extd\theta \cdot 1 = 0
\end{equation}

\noindent for some Grassmann parameter $\theta$. This yields a result

\begin{equation}
\int_{\Fscr\setminus\Fscr_\Qloc}\Dg\Phi ~ \Og[\Phi]e^{-S[\Phi]} = 0
\end{equation}

\noindent hence showing that the path integral localises to the $\Qloc$-supersymmetric field configurations $\Fscr_\Qloc$ \cite{cremonesi2013introduction}.

\subsubsection{Second localisation argument}

The second localisation argument follows in strong analogy to that for equivariant integrals. We note that we can deform the path integral continuously using a fermionic functional $\Fg[\Phi]$ which we require to be annihilated by the square of the supercharge:

\begin{equation}
\Qloc^2\Fg[\Phi] = \Bg\Fg[\Phi] = 0.
\end{equation}

\noindent and for which we take its supersymmetry transformation to be \textit{positive semi-definite} in its bosonic components: 

\begin{equation}
\Qloc\Fg[\Phi]\Big|_\bos \geq 0.
\end{equation}

\noindent In this case it turns out that the path integral can be continuously deformed as

\begin{equation}
\begin{split}
\EV{\Og_\BPS} 
&= \int_\Fscr\Dg\Phi ~ \Og_\BPS[\Phi]e^{-S[\Phi]}
\\
&= \int_\Fscr\Dg\Phi ~ \Og_\BPS[\Phi]e^{-S[\Phi]-t\Qloc\Fg[\Phi]}
\end{split}
\end{equation}

\noindent for any value of $t$, without changing the expetation value, assuming that $\Og_\BPS$ is (as its name suggests) a \textit{BPS operator}. Indeed, we find that

\begin{multline}
\der{}{t}\int_\Fscr\Dg\Phi ~ \Og_\BPS[\Phi]e^{-S[\Phi]-t\Qloc\Fg[\Phi]} 
= -\int_\Fscr\Dg\Phi ~ \Og_\BPS[\Phi]\Qloc\Fg[\Phi]e^{-S[\Phi]-t\Qloc\Fg[\Phi]}
\\
= -\int_\Fscr\Dg\Phi~\Qloc\Big\{\Og_\BPS[\Phi]\Fg[\Phi]e^{-S[\Phi]-t\Qloc\Fg[\Phi]}\Big\}
= 0
\end{multline}

\noindent where in the third step we used the Leibniz rule for variations and in the final step we used Stokes' theorem on field space. Since the bosonic part of $\Qloc\Fg$ is positive semi-definite, we are free to take the limit $t\to\infty$. Taking this limit we find that the parts for which the bosonic part of the localising action

\begin{equation}
S_\loc[\Phi] = \Qloc\Fg[\Phi]
\end{equation}

\noindent is strictly positive get exponentially supressed in the limit $t\to\infty$. Taking this limit we find that the path integral localises to

\begin{equation}
\Fscr_\Qloc = \Big\{\Phi\in\Fscr~\Big|~S_\loc[\Phi]\big|_\bos = 0,~~\Phi\big|_\fer = 0\Big\}.
\end{equation}

\noindent The reason that we can localise around zero configurations of the fermionic component fields $\Phi|_\fer$ that due to their anticommutig nature we can expand them around any given point, since we don't have to worry about things such as divergence for series \cite{cremonesi2013introduction}. 

\subsubsection{An example}

A standard choice for localising actions is for example

\begin{equation}
S_\loc[\Phi] \overset{\text{e.g.}}{=} \Qloc\int\extd^dx \sum_{\psi\in\Phi|_\fer} (\Qloc\psi)^\dagger\psi
\end{equation}

\noindent This localising term has a bosonic part

\begin{equation}
S_\loc[\Phi]\Big|_\bos = \int\extd^dx \sum_{\psi\in\Phi|_\fer}|\Qloc\psi|^2
\end{equation}

\noindent which is indeed positive semi-definite. We thus find that for this example the path integral localises to

\begin{equation}
\Fscr_\Qloc = \big\{\Phi\in\Fscr~\big|~\forall\psi\in\Phi|_\fer:\Qloc\psi=0,~\psi = 0\big\}
\end{equation}

\noindent consisting of so-called BPS configurations \cite{cremonesi2013introduction}.

\subsubsection{Working out the localisation}

Let us now work out how this localisation gives us results for supersymmetric path integrals. We expand the field content around the localisation locus $\Phi_0 \in \Fscr_\Qloc$ as

\begin{equation}
\Phi = \Phi_0 + \frac{1}{\sqrt{t}}\delta\Phi.
\end{equation}

\noindent The localising action is then expanded as

\begin{multline}
tS_\loc[\Phi] = tS_\loc[\Phi_0] + \sqrt{t}\int\fder{S_\loc}{\Phi}[\Phi_0]\delta\Phi + \frac{1}{2}\iint\fder{^2S_\loc}{\Phi^2}[\Phi_0](\delta\Phi)^2 + \Og(t^{-1/2})
\\
= \frac{1}{2}\iint\fder{^2S_\loc}{\Phi^2}[\Phi_0](\delta\Phi)^2 + \Og(t^{-1/2})
\xrightarrow{t\to\infty}
\frac{1}{2}\iint\fder{^2S_\loc}{\Phi^2}[\Phi_0](\delta\Phi)^2
\end{multline}

\noindent The first in the expansion vanishes due to the fact the defining quality of the localisation locus is $S_\loc[\Phi_0] = 0$. The second term vanishes due to the fact that since the localising term is positive definite its first order contributions around a local extremum vanish. This leaves us after taking the limit with only the second order contributions to the path integral. As for the action of the theory as well as the BPS operator, we find that they are simply expanded as

\begin{alignat}{2}
S[\Phi] &= S[\Phi_0] &&+ \Og(t^{-1/2})
\xrightarrow{t\to\infty} 0
\\
\Og_\BPS[\Phi] &= \Og_\BPS[\Phi_0] &&+ \Og(t^{-1/2})
\xrightarrow{t\to\infty} 0
\end{alignat}

\noindent The aforementioned results allow us to solve the path integral as

\begin{equation}
\begin{split}
\EV{\Og_\BPS} 
&= \lim_{t\to\infty}\int_\Fscr\Dg\Phi ~ \Og_\BPS[\Phi]e^{-S[\Phi] - tS_\loc[\Phi]}
\\
&= \int_{N\Fscr_\Qloc}\Dg\Phi_0\Dg\delta\Phi ~ \Og_\BPS[\Phi_0]\exp\bigg\{-S[\Phi_0] - \frac{1}{2}\iint\fder{^2S_\loc}{\Phi^2}[\Phi_0](\delta\Phi)^2\bigg\}
\\
&= \int_{\Fscr_\Qloc}\Dg\Phi_0 ~ \frac{\Og_\BPS[\Phi_0]}{\Ber\left[\fder{^2S_\loc}{\Phi^2}[\Phi_0]\right]}e^{-S[\Phi_0]}.
\end{split}
\end{equation}

\noindent where $\Ber$ stands for the functional Berezinian, which we took of the kinetic operator of the localising action at the localisation locus. $N\Fscr_\Qloc$ stands for the normal bundle of the localisation locus, which has field space coordinates $([\Phi_0],[\delta\Phi])$. In conclusion, we find that the localisation of the supersymmetric path integral is given by

\begin{tcolorbox}[breakable, enhanced]

\subsubsection*{Supersymmetic localisation formula}

\begin{equation}
\EV{\Og_\BPS} = \int_{\Fscr_\Qloc}\Dg\Phi_0 ~ \frac{\Og_\BPS[\Phi_0]}{\Ber\left[\fder{^2S_\loc}{\Phi^2}[\Phi_0]\right]}e^{-S[\Phi_0]}
\end{equation}

the localisation formula for a BPS operator $\Og_\BPS$, with the localising action given by

\begin{equation}
S_\loc[\Phi] = \Qloc\Fg[\Phi]
\end{equation}

\noindent for some fermionic functional $\Fg[\Phi]$ satisfying
\begin{align}
\Qloc\Fg[\Phi] &\geq 0
&
\Qloc^2\Fg[\Phi] &= 0,
\end{align}

\noindent and the localisation locus given by

\begin{equation}
\Fscr_\Qloc = \Big\{\Phi_0 \in \Fscr ~\Big|~ S_\loc[\Phi_0]\big|_\bos = 0,~~\Phi_0\big|_\fer = 0\Big\}
\end{equation}

\noindent which may itself be another field space, or may be a finite-dimensional manifold or even a discrete set of points.

\end{tcolorbox}

\noindent In conclusion, we have given a method which allows one to systematically reduce the dimension of a path integral. This in turn would allow one to do non-perturbative computations \cite{cremonesi2013introduction}.

It should be noted that this is not the only way one can localise. There are more ways of continuously deforming path integrand in such a way that the corresponding expectation value of a BPS operator is left unchanged. One such example will be treated next chapter in the Witten index. In the case of the Witten index, rather than adding a localising term which we scale to infinity, we will rescale certain parameters of the theory, resulting in an eventual localisation.

\chapter{The Witten Index}

\section{Introduction}

In this chapter, we discuss the Witten index of a one-dimensional supersymmetric quantum mechanical system. This serves as a `warm up' before we move on to the more challenging cases of supersymmetric quantum field theories.

We start off by introducing the Witten index through the Hamiltonian formalism. We discuss a Hilbert space expression for this object which involves periodic Euclidean time and also show the independence of the Witten index on this Euclidean time periodicity. We then move on to describe the Witten index through a Euclidean action formalism, in the language of path integrals. It's in this context that we can connect this to localisation methods. This action can then also be described in superspace, allowing one to efficiently derive all desired properties of the Witten index. 
These notes start off by formulating a topic of discussion in David Tong's notes on supersymmetric quantum mechanics \cite{DavidTong} but will quickly start diverging from these notes. 
In fact, it is believed that we made a new contribution to this, because we described a procedure through the path intergal formalism which has previously only been described through the Hilbert space formalism.

\section{Witten Index in the Hilbert Space Formalism}

\subsection{On-Shell Supersymmetric Point Particle}

We start off by giving the on-shell description of the supersymmetric point particle. That is, we start off by considering the action of the supersymmetric point particle where potential auxiliary fields are taken to be on-shell. We take this description because 
it's this action that allows us to go to the Hilbert space formalism. We also immediately go to Euclidean time, since that's all we'll be making use of. This being said, the on-shell Euclidean action for $\beta$-periodic Euclidean time $\tau$ is given by

\begin{equation}
S_E^{\text{on}} = \underset{S^1(\beta)}{\intau}\left[\frac{1}{2}\xdot^2+\psid\psidot+\frac{1}{2}h'(x)^2-h''(x)\psid\psi\right].
\end{equation}

\noindent Let us take a look at the content of this action. $x(\tau)$ is taken to be the position of the point particle in one-dimensional space. $\psi(\tau)$ and $\psid(\tau)$ describe the
spin of this particle and are Grassmann-valued fields on the Euclidean time circle. $h(x)$ is a real-valued Morse function of $x$,
known as the superpotential of the system. By Morse, we mean that the stationary points of this function are taken to be isolated. $S^1(\beta)$ denotes the Euclidean time circle of period $\beta$ over which we're integrating. Finally, a dot denotes a time derivative with respect to Euclidean time.

This action is supersymmetric, with its supersymmetries given by

\begin{align}
\Qop  &= \intau\left[\psi\fder{}{x}+(-\xdot+h')\fder{}{\psid}\right],
&
\Qopd &= \intau\left[-\psid\fder{}{x}+(\xdot+h')\fder{}{\psi}\right].
\end{align}

\noindent Indeed, a straightforward computation yields

\begin{equation}
\Qop S_E^{\text{on}} = \Qopd S_E^{\text{on}} = 0.
\end{equation}

\noindent We can finally also introduce a `Eucidean Hamiltonian'

\begin{equation}
\label{Euclidean energy}
H_E = -L_E + \pder{L_E}{\xdot}\xdot + \pder{L_E}{\psidot}\psidot = \frac{1}{2}\xdot^2 - \frac{1}{2}h'^2 + h''\psid\psi = -H.
\end{equation}

\noindent In the final step, it's important to remember that $(\extd x/\extd\tau)^2 = -(\extd x/\extd t)^2$ \cite{DavidTong}.

\subsection{Hilbert Space Formalism}

Before moving on to the Witten index, we make the necessary commentary on the Hilbert space of this system. The system in consideration carries bosonic as well as fermionic degrees of freedom. Accordingly, its Hilbert space will be that of a one-dimensional point particle with a single spin, given by

\begin{equation}
\mathcal{H} = \mathcal{H}_B \otimes \mathcal{H}_F \cong L^2(\mathbb{R})\otimes \mathbb{C}^2.
\end{equation}

\noindent In accordance with general properties of supersymmetry generators, the Euclidean Hamiltonian will be given by

\begin{equation}
H_E = -\frac{1}{2}\{Q_E,Q_E{^\dagger}\} = \frac{1}{2}\left(\der{^2}{x^2}-h'^2\right)\otimes\mathbbm{1} + \frac{1}{2} h''\otimes\sigma^3
\end{equation}

\noindent where $\sigma^3=\operatorname{diag}(1,-1)$ the third Pauli matrix and $Q_E$, $Q_E{^\dagger}$ the supersymmetry operators given by

\begin{align}
Q_E &= \left(-\der{}{x}+h'\right)\otimes
\begin{pmatrix}
0 & 0\\
1 & 0
\end{pmatrix},
&
Q_E{^\dagger} &= \left(\der{}{x}+h'\right)\otimes
\begin{pmatrix}
0 & 1\\
0 & 0
\end{pmatrix}.
\end{align}

\noindent They also satisfy the nilpotency conditions

\begin{equation}
(Q_E)^2 = (Q_E{^\dagger})^2 = 0.
\end{equation}

\noindent These algebraic properties of the Hamiltonian and supersymmetry oparators have a number of consequences. Firstly, we note that the Euclidean energy is always negative, since for any state $\ket{\psi}$

\begin{equation}
\bra{\psi}H_E\ket{\psi} = -\frac{1}{2}|Q_E\ket{\psi}|^2-\frac{1}{2}|Q_E{^\dagger}\ket{\psi}|^2 \leq 0
\end{equation}

\noindent where one assumes the Hilbert space inner product to be positive definite, which is obviously the case for our system in consideration. 
Secondly, we note that for a supersymmetric system positive energy states are 1:1. This can be seen as follows: Recall firstly that the fermion number operator is given by

\begin{equation}
F = 
\begin{pmatrix}
0 & 0\\
0 & 1
\end{pmatrix}
\end{equation}

\noindent highlighting that in the spin Hilbert space the components refer to respectively bosonic and fermionic states. We now consider the subspace of negative Euclidean energy states, spanned by

\begin{align}
&\ket{E}, &&\text{s.t.} & H_E\ket{E} &= -E\ket{E}, & E > 0.
\end{align}

\noindent Note however that negative Euclidean energy corresponds to positive energy, see equation \ref{Euclidean energy}. This allows us now to define operators $c$ and $c^\dagger$ on this space by linearly extending

\begin{align}
c\ket{E} &= \frac{1}{\sqrt{2E}}Q_E\ket{E}, & c^\dagger\ket{E} &= \frac{1}{\sqrt{2E}}Q_E{^\dagger}\ket{E}.
\end{align}

\noindent These operators satisfy the algebra

\begin{align}
\{c,c^\dagger\} &= \mathbbm{1}, & c^2 &= (c^\dagger)^2 = 0
\end{align}

\noindent being the commutation relations characterizing a fermionic harmonic oscillator. This means that negative Euclidean energy states come in pairs which represent this algebra. We furthermore note that these operators preserve energy and change fermion number, due to the algebraic relations

\begin{align}
[c,H_E] &= [c^\dagger,H_E] = 0, & [F,c] &= -c, & [F,c^\dagger] &= +c^\dagger.
\end{align}

\noindent This confirms that bosonic and fermionic states negative Euclidean energy states come 1:1 with the same energy. 
However, this needn't be the case for zero energy states \cite{DavidTong}.

\subsection{Hilbert Space Witten Index Expression}

We now introduce the Witten index as expressed in the Hilbert space formalism. In this formalism, the Witten index is given by

\begin{equation}
\WI = \Tr(-1)^Fe^{-\beta H} = \Tr(-1)^Fe^{\beta H_E}.
\end{equation}

\noindent where we assume a discrete energy spectrum. Let us think about how to compute this: Since bosonic and fermionic same positive energy states are $1:1$, their contributions will cancel in the above trace, due to the $(-1)^F$ factor. We hence find that the Witten index in the Hilbert space formalism is given by

\begin{equation}
\WI = \Tr(-1)^Fe^{-\beta H} = \dim \ker H|_{\text{bos}} - \dim \ker H|_{\text{fer}}.
\end{equation}

\noindent This can be regarded as a statement about the kernel of a differential operator ---the Hamiltonian operator--- on a manifold. This will later link this result to more general index theorems. 
We furthermore note that the Witten index is independent of our choice of $\beta$. Indeed, differentiating with respect to $\beta$ we find that

\begin{equation}
\der{\WI}{\beta} = \Tr(-1)^FH_Ee^{H_E} = 0.
\end{equation}

\noindent The vanishing in the final step is due to the fact that positive energy fermion-boson doublets again cancel, as well as the fact that for this trace zero energy states don't contribute altogether.
From this we conclude that

\begin{tcolorbox}

\begin{align}
\WI &= \Tr(-1)^Fe^{-\beta H} = \dim \ker H|_{\text{bos}} - \dim \ker H|_{\text{fer}}, & \der{\WI}{\beta} &= 0
\end{align}

the Hilbert space expression and $\beta$-independence of the Witten index, derived from the Hilbert space formalism \cite{DavidTong}.

\end{tcolorbox}

\section{The Witten index and the Path Integral}

In this section, we will discuss the Witten index through the path integral formalism. We will start off by obtaining a path integral expression for the Witten index from its Hilbert space expression. Then, we will discuss the off-shell expression for the Witten index. To do so, we will start off by reintroducing a missing auxiliary field which in the on-shell expression is taken to be on-shell. We will then further generalize our description of the Witten index to superspace. After this, we will discuss symmetries of the Witten index. To do this, we will start off by motivating the systematics through geometrical arguments generalized to function spaces. We then combine this with our superspace description of the system to obtain manifest symmetries of the Witten index. When this is finished we will use these symmetries to show the $\beta$-independence of the Witten index through the path integral formalism, as well as obtain a simple expression for the Witten index also through the path integral formalism.

\subsection{Path Integral Expression of the Witten Index}

We start off this section by deriving a path integral expression of the Witten index. To do so, we note that the trace of some operator $\mathcal{O}$ acting on our Hilbert space is given by

\begin{equation}
\begin{split}
\Tr\mathcal{O} 
&= \int\extd x\extd\psid\extd\psi e^{-\psid\psi}~\bra{-\psi,x}\mathcal{O}\ket{\psi,x}
\\
&= \int\extd x ~ \bigg[\bra{0,x}\Og\ket{0,x} + \bra{1,x}\Og\ket{1,x}\bigg]
\end{split}
\end{equation}

\noindent where we defined the states

\begin{align}
\ket{0,x} &:= \ket{x}\otimes\ket{0}
&
\ket{\psi,x} &:= \ket{x}\otimes\ket{0} + \ket{x}\otimes\ket{1}\psi, 
\\
\ket{1,x} &:= \ket{x}\otimes\ket{1}
&
\bra{\psi,x} &:= \bra{x}\otimes\bra{0} + \bra{x}\otimes\psid\bra{1}
\end{align}

\noindent where $\ket{0}$, $\ket{1}$ denote the bra-ket basis of the fermionic Hilbert space. The exponential factor in this trace might seem a little confusing but is in fact completely justified. The reason for this being that only terms in the integrand proportional to $\psid\psi$ survive Berezinian integration.

From this it follows that

\begin{equation}
(-1)^F\ket{\psi,x} = \ket{x}\otimes \Big((-1)^0\ket{0}+(-1)^1\ket{1}\psi\Big) = \ket{-\psi,x}.
\end{equation}

\noindent Using this we now find that

\begin{equation}
\label{3.1comp}
\begin{split}
\WI = \Tr(-1)^Fe^{-\beta H} 
&= \int\extd x\extd\psid\extd\psi e^{-\psid\psi}~\bra{-\psi,x}(-1)^Fe^{-\beta H}\ket{\psi,x}\\
&= \int\extd x\extd\psid\extd\psi e^{-\psid\psi}~\bra{\psi,x}e^{-\beta H}\ket{\psi,x}.
\end{split}
\end{equation}

\noindent Furthermore noting the path integral identity

\begin{equation}
\bra{\psi,x}e^{-\beta H}\ket{\eta,y} = \overset{x(\beta)=y,~\psi(\beta)=\eta,~\text{h.c.}}{\underset{x(0)=x,~\psi(0)=\psi,~\text{h.c.}}{\int\Dg x\Dg\psid\Dg\psi}}~e^{-S_E^\text{on}}
\end{equation}

\noindent we conclude that

\begin{tcolorbox}

\begin{equation}
\WI = \Tr(-1)^Fe^{-\beta H} = \int\underbrace{\Dg x\Dg\psid\Dg\psi}_{\text{$\beta$-periodic}}~e^{-\Son}
\end{equation}

\noindent the on-shell path integral expression of the Witten index. We dropped the subscript $E$ in the Euclidean action and will keep doing so from now on.

\end{tcolorbox}

\noindent Note that the additional factor $e^{-\psid\psi}$ in equation \ref{3.1comp} gets absorbed into the path integral acting as an object which glues the two ends of the Euclidean time $[0,\beta]$ into a circle $S^1(\beta)$ of circumference $\beta$ \cite{nakahara2018geometry}.

\subsection{The Off-Shell Action}

So far we have only considered the on-shell action $\Son$ of the supersymmetric point particle. 
This can be seen from the fact that while the bosonic and fermionic states seem to show matching behaviour, the field content doesn't. Namely, there is one real bosonic degree of freedom but two real (i.e. one complex) fermionic degrees of freedom.
However, to properly study the Witten index from the path integral formalism we will find it to be necessary to include the remaining the off-shell degree of freedom in our discussion.
From this point on we diverge completely from Tong's notes \cite{DavidTong}, as the off-shell description isn't mentioned in Tong's notes whatsoever.

\subsubsection{Reintroducing the auxiliary field}

To match the degrees of freedom of the field content, we need one more bosonic degree of freedom, matching the bosonic and fermionic field content degrees of freedom two-to-two. This remaining degree of freedom will be a real bosonic auxiliary field $F(\tau)$, which is taken to be non-dynamical. We now want to find an action such that upon eliminating this non-dynamical field we obtain the on-shell action again. This action is given by

\begin{tcolorbox}

\begin{equation}
\label{Soff}
\Soff = \intau\left[\frac{1}{2}\xdot^2+\psid\psidot+\frac{1}{2}F^2+ih'F-h''\psid\psi\right]
\end{equation}

\noindent the (Euclidean) off-shell action.

\end{tcolorbox}

\noindent We note that this action can be regarded as a WZ model with interactions up to arbitrary order. Let us verify that on-shell this action indeed gives the us the on-shell action. The equation of motion of the auxiliary field is

\begin{equation}
\begin{aligned}
0 &= \fder{\Soff}{F} = F + ih' & &\Leftrightarrow & F &= -ih'.
\end{aligned}
\end{equation}

\noindent Substitution in the off-shell action yields

\begin{equation}
\Soff[x,-ih',\psi,\psid] = \intau\left[\frac{1}{2}\xdot^2+\psid\psidot+\frac{1}{2}h'^2-h''\psid\psi\right] = \Son[x,\psi,\psid]
\end{equation}

\noindent giving us the on-shell action. This is all good and well, but this doesn't guarantee that the physics will we the same for this system or that its supersymmetries are preserved. We hence have two goals: Show the existence of supersymmetries of the off-shell action and relate them to the on-shell action, as well as show that introducing these auxiliary fields doesn't change the Witten index.

To tackle the first goal, we let us be inspired by the typical form of SUSY variations, being

\begin{equation}
\begin{aligned}
\delta(\text{boson}) &= \text{fermion}, 
\\
\delta(\text{fermion}) &= \partial(\text{boson}) + \text{auxiliary},
\\
\delta(\text{auxiliary}) &= \partial(\text{fermion}).
\end{aligned}
\end{equation}

\noindent Following this form we consider

\begin{tcolorbox}

\begin{align}
\Qop &= \intau\left[\psi\fder{}{x}+(-\xdot+iF)\fder{}{\psid}-i\psidot\fder{}{F}\right]
\\
\Qopd &= \intau\left[-\psid\fder{}{x}+(\xdot+iF)\fder{}{\psi}-i\psidotd\fder{}{F}\right]
\end{align}

\noindent the off-shell supersymmetry variations. We typically supress the superscript `off' unless ambiguities arise.

\end{tcolorbox}

\noindent Indeed, we note that

\begin{equation}
\Qop^{\text{off}}|_{F=-ih'} = \intau\left[\psi\fder{}{x}+(-\xdot+h')\fder{}{\psid}\right] = \Qop^{\text{on}}
\end{equation}

\noindent and similarly $\Qop^{\text{off}\dagger}|_{F=-ih'} = \Qop^{\text{on}\dagger}$. Furthermore, we find that it also indeed is a symmetry of the off-shell action since

\begin{multline}
\Qop\Soff = \intau\bigg[\psi(-\ddot{x}+ih''F)+(-\xdot+iF)(\psidot-h''\psi)-i\psidot(F+ih')\bigg]
\\
= 
\intau\bigg[
(-\ddot{x}\psi-\xdot\psidot)
+(\xdot h''\psi+h'\psidot)\bigg] 
= 0
\end{multline}

\noindent and similarly $\Qopd\Soff = 0$.

Now we move on to our second goal of relating it to the on-shell Witten index. We claim that the on-shell Witten index agrees with the off-shell expression

\begin{equation}
\WI = \int\Dg x\Dg F\Dg\psid\Dg\psi ~e^{-\Soff}.
\end{equation}

\noindent This is indeed the case and can be shown by completing the square in the off-shell action as

\begin{equation}
\Soff = \intau\left[\frac{1}{2}\xdot^2+\psid\psidot+\frac{1}{2}h'^2-h''\psid\psi+\frac{1}{2}(F+ih')^2\right] = \Son + \intau\frac{1}{2}(F+ih')^2.
\end{equation}

\noindent This hence yields

\begin{multline}
\int\Dg x\Dg F\Dg\psid\Dg\psi ~e^{-\Soff} 
= \int\Dg x\Dg F\Dg\psid\Dg\psi~\exp\left\{-\Son-\frac{1}{2}\intau (F+ih')^2\right\}\\
= \Det^{-\frac{1}{2}}\mathbbm{1}\int\Dg x\Dg\psid\Dg\psi~e^{-\Son} = \int\Dg x\Dg\psid\Dg\psi~e^{-\Son}.
\end{multline}

\noindent This confirms that

\begin{tcolorbox}

\begin{equation}
\WI = \int\Dg x\Dg F\Dg\psid\Dg\psi~e^{-\Soff} = \int\Dg x\Dg\psid\Dg\psi~e^{-\Son}
\end{equation}

\noindent the off-shell and the on-shell expressions of the Witten index agree.

\end{tcolorbox}

\subsubsection{Going to superspace}

This won't fully conclude us extending the way we describe this system. We won't introduce any more new fields, rather we will reformulate our theory in the superspace formalism, as this will make a lot of properties which are important later on for obtaining the right results become manifest.
We will promote the Euclidean time circle to a superspace. That is, instead of just having a coordinate $\tau$, superspace will have coordinates $(\tau,\theta,\thetad)$ with the latter two being Grassmann variables. In accordance with the periodicity of the fermionic fields we take these coordinates to be periodic instead of antiperiodic on the Euclidean time circle. 
We will now interpret the fields involved on this circle to correspond to a scalar superfield $X$. That is, we take our superfield to be of the form

\begin{equation}
X(\tau,\theta,\thetad) = x(\tau) + \theta\psid(\tau) + \thetad\psi(\tau) + i\theta\thetad F(\tau).
\end{equation}

\noindent Accordingly, we now introduce a `superspace Lie derivative' representation of the supersymmetries $\Qop$, $\Qopd$. We will denote these representations by $Q$, $\Qd$ respectively. These representations have to satisfy

\begin{equation}
\begin{aligned}
Q X 
&= Q x - \theta\Qop\psid - \thetad\Qop\psi + i\theta\thetad \Qop F 
& 
\Qd X
&= \Qopd x - \theta\Qopd\psid - \thetad\Qopd\psi + i\theta\thetad\Qopd F
\\
&= \psi - \theta(-\xdot+iF) + i\theta\thetad (-i\psidot)
&
&= -\psid - \thetad(\xdot+iF) +  i\theta\thetad (-i\psidotd)
\end{aligned}
\end{equation}

\noindent We can satisfy these conditions, namely through choosing

\begin{tcolorbox}

\subsubsection*{Superspace supersymmetry representation}

\begin{align}
Q &= \pder{}{\thetad} + \theta\der{}{\tau},
&
\Qd &= -\pder{}{\theta} - \thetad\der{}{\tau}
\end{align}


\end{tcolorbox}

\noindent One of the advantages of using this representation is that it becomes much easier to study general algebraic properties of supersymmetry. For example, we find that

\begin{equation}
\{Q,\Qd\} = -\{\pder{}{\thetad},\pder{}{\theta}\}-\{\pder{}{\thetad},\thetad\der{}{\tau}\} - \{\theta\der{}{\tau},\pder{}{\theta}\} - \{\theta\der{}{\tau},\thetad\der{}{\tau}\} = -2\der{}{\tau}
\end{equation}

\noindent as well as $Q^2=(\Qd)^2=0$ which is in alignment with general properties of supersymmetric theories. 

We now move on to constructing the Lagrangian in superspace. This Lagrangian $\Lagr$ has to be chosen such that the action is given by

\begin{equation}
S = \intau\extd\thetad\extd\theta~\Lagr(X,DX,\Dd X)
\end{equation}

\noindent where we demand that the superspace Lagrangian transforms as a density under supersymmetry transformations. The derivatives $D$ and $\Dd$ are covariant spinor derivatives. They're covariant as in they are required to anticommute with the supersymmetries $Q$ and $\Qd$, i.e.

\begin{align}
\{Q,D\} = \{\Qd,D\} = \{Q,D^\dagger\} = \{\Qd,D^\dagger\} = 0.
\end{align}

\noindent These relations are satisfied if we choose

\begin{tcolorbox}

\subsubsection*{Covariant spinor derivatives}

\begin{align}
D &= \pder{}{\thetad} - \theta\der{}{\tau},
&
\Dd &= -\pder{}{\theta} + \thetad\der{}{\tau}
\end{align}


\end{tcolorbox}

\noindent We note that these also satisfy 

\begin{align}
\{D,\Dd\} &= 2\der{}{\tau}, & D^2 &= (\Dd)^2 = 0.
\end{align}

\noindent These now allow us to construct the free part of the action. We compute

\begin{align}
DX &= \psi - \theta(\xdot+iF) - \theta\thetad \psidot,
&
\Dd X &= -\psid + \thetad(\xdot-iF) - \theta\thetad \psidotd.
\end{align}

\noindent We hence find that

\begin{equation}
\Dd XDX = -\psid\psi - \theta\psid(\xdot+iF) + \thetad\psi(\xdot-iF) + \theta\thetad(\xdot^2+F^2+\psid\psidot-\psidotd\psi)
\end{equation}

\noindent of which the highest Grassmann order term is proportional to the free sector of the off-shell action $\Soff$ given in equation \ref{Soff}. As for the interaction part of the action, doing some combinatorics we note that

\begin{equation}
\begin{split}
X^n 
&= x^n + nx^{n-1}(\theta\psid+\thetad\psi) + \theta\thetad\Big(inx^{n-1}F - n(n-1)x^{n-2}\psid\psi\Big)\\
&= \left[1 + (\theta\psid+\thetad\psi)\der{}{x} + \theta\thetad\left(iF\der{}{x}-\psid\psi\der{^2}{x^2}\right)\right]x^n.
\end{split}
\end{equation}

\noindent Hence, formally Taylor expanding the superpotential $h$ in powers of the superfields $X$ instead of $x$ we find that

\begin{equation}
\begin{split}
h(X) 
&\equiv \exp\left(X\der{}{x}\right)h(0)\\
&= \left[1 + (\theta\psid+\thetad\psi)\der{}{x} + \theta\thetad\left(iF\der{}{x}-\psid\psi\der{^2}{x^2}\right)\right]\exp\left(x\der{}{x}\right)h(0)\\
&= h(x) + h'(x)(\theta\psid+\thetad\psi) + \theta\thetad\Big(ih'(x)F-h''(x)\psid\psi\Big)
\end{split}
\end{equation}

\noindent of which the highest Grassmann order term is the interaction part of the action. From this we conclude that

\begin{tcolorbox}

\subsubsection*{Superspace Lagrangian}

\begin{equation}
\label{Lsuperspace}
\begin{split}
\Lagr 
&= \frac{1}{2}\Dd XDX + h(X)\\
&= 
\!\begin{multlined}[t]
\left(-\frac{1}{2}\psid\psi+h\right) + \theta\psid\left(-\frac{1}{2}(\xdot+iF)+h'\right) + \thetad\psi\left(\frac{1}{2}(\xdot-iF)+h'\right)
\\
+ \theta\thetad\bigg(\frac{1}{2}\Big(\xdot^2
+\psid\overleftrightarrow{\der{}{\tau}}\psi
+F^2\Big)+ih'F-h''\psid\psi\bigg)
\end{multlined}
\end{split}
\end{equation}


\end{tcolorbox}

\noindent The reason we went ahead and wrote down the full expansion is because this expansion will become important later on.

\subsection{Symmetries of the Witten Index}

\subsubsection{Systematics}

In this section we will discuss symmetries of the Witten index. What we mean by symmetries is that we consider transformations of the off-shell action and verify whether or not they leave the Witten index unchanged. It is here that a connection is made with the previous chapter. Particularly, as explained in Cremonesi's lecture notes \cite{cremonesi2013introduction}, the expectation value of a BPS operator depends only on its $Q$-cohomology class. In the current context, this is expressed as

\begin{equation}
\EV{\Og_\BPS} = \EV{\Og_\BPS + \Qop\mathcal{G} + \Qopd\mathcal{H}}.
\end{equation}

\noindent The BPS operator in consideration here is the $(-1)^F$ operator. As such, if the Witten index path integrand is deformed by a $Q$-exact term the Witten index is left unchanged.
The fact that we can freely deform a path integral integrand by $Q$-exact functionals will be used deform the integrand in such a way that the path integral will in some limit only receive contributions from lower dimensional subspaces of the function space, if not finite-dimensional or even discrete contributions \cite{cremonesi2013introduction}.

\subsubsection{$\hbar$-rescalings}

We will consider three symmetries in this section. The first symmetry we will consider is an 
$\hbar$-rescaling.
This is a symmetry given by $\Soff\to\mu\Soff$ with $\mu > 0$
, essentially corresponding to a rescaling $\hbar\to\mu^{-1}\hbar$. 
The Witten index then becomes

\begin{equation}
\WI(\mu) = \int\Dg x\Dg F\Dg\psid\Dg\psi~e^{-\mu\Soff}.
\end{equation}

\noindent This means that under changes of $\mu$ the Witten index changes as

\begin{equation}
\pder{\WI}{\mu} = \int\Dg x\Dg F\Dg\psid\Dg\psi~(-\Soff)e^{-\mu\Soff}.
\end{equation}

\noindent Note that these rescalings aren't the same as on-shell 
$\hbar$-rescalings
, as you can't integrate out the auxiliary fields without obtaining an additional functional detereminant factor! Following our discussion earlier this section it now suffices to show that $\Soff$ is $Q$-exact. This can be most easily seen in the superspace formalism. This is because the superspace Lagrangian represents the supersymmetry algebra as a scalar density. Expanding the superspace Lagrangian as

\begin{equation}
\Lagr = \ell + \theta\Lg + \thetad\Lgd + \theta\thetad L
\end{equation}

\noindent this means that under $Q$ and $\Qd$ it transforms as

\begin{align}
Q\Lagr &= \Qop\ell - \theta\Qop\Lg - \thetad\Qop\Lgd + \theta\thetad \Qop L + \der{}{\tau}(\dots),\\
\Qd\Lagr &= \Qopd\ell - \theta\Qopd\Lg - \thetad\Qopd\Lgd + \theta\thetad \Qopd L + \der{}{\tau}(\dots).
\end{align}

\noindent Further noting that at least for this system we're allowed to write

\begin{equation}
\intau\extd\thetad\extd\theta(\dots) = \intau\Qd Q(\dots)
\end{equation}

\noindent we find that

\begin{equation}
\Soff = \intau\extd\thetad\extd\theta ~\Lagr = \intau~\Qd Q\Lagr = \intau ~\Qopd\Qop\ell = \Qopd\Qop\left\{\intau~\ell\right\}.
\end{equation}

\noindent Comparing this to equation \ref{Lsuperspace} we then find

\begin{equation}
\Soff = \Qopd\Qop\left\{\intau\left(-\frac{1}{2}\psid\psi+h\right)\right\}.
\end{equation}

\noindent By a direct computation one could verify that this indeed gives the right result. Using the $Q$-closure of the action as well as the graded Leibniz rule we now find that

\begin{equation}
\pder{\WI}{\mu} = \int\Dg x\Dg F\Dg\psid\Dg\psi~\Qopd\Qop\left\{-\intau\left(-\frac{1}{2}\psid\psi+h\right)e^{-\mu\Soff}\right\} = 0
\end{equation}

\noindent showing the invariance of the Witten index under 
$\hbar$-rescalings. 
We conclude

\begin{tcolorbox}

\subsubsection*{$\hbar$-rescaling symmetry}

\begin{align}
\pder{\WI}{\mu} &= 0, & \Soff \to \mu \Soff
\end{align}

the Witten index is invariant under 
$\hbar$-rescalings.

\end{tcolorbox}

\subsubsection{Kinetic and superpotential rescalings}

We now consider the other two symmetries. These are kinetic and superpotential rescalings. Denoting $\Xf=(x,\psi,\psid,F)$ the collective fields this gives a transformation

\begin{equation}
\Soff[\Xf] \to \Soff_{\kappa,\lambda}[\Xf] = \intau~L(\Xf,\kappa\dot{\Xf})|_{h\to\lambda h}
\end{equation}

\noindent with $\kappa,\lambda>0$. In $x$-space this gives

\begin{equation}
\Soff_{\kappa,\lambda} = \intau\left[\frac{1}{2}\kappa^2\xdot^2+\kappa\psid\psidot+\frac{1}{2}F^2+i\lambda h'F - \lambda h''\psid\psi\right]
\end{equation}

\noindent with augmented supersymmetries

\begin{align}
\Qop_\kappa &= \intau\left[\psi\fder{}{x}+(-\kappa\xdot+iF)\fder{}{\psid}-i\kappa\psidot\fder{}{F}\right],\\
\Qop_\kappa{^\dagger} &= \intau\left[-\psid\fder{}{x}+(\kappa\xdot+iF)\fder{}{\psi}-i\kappa\psidotd\fder{}{F}\right].
\end{align}

\noindent Indeed, we find for example that

\begin{multline}
\Qop_\kappa \Soff_{\kappa,\lambda} = \intau\bigg[\psi(-\kappa^2\ddot{x}+i\lambda h''F)+(-\kappa\xdot+iF)(\kappa\psidot-\lambda h''\psi)-i\kappa\psidot(F+i\lambda h')\bigg]
\\
= \intau\bigg[-\kappa^2(\ddot{x}\psi+\xdot\psidot)+\lambda(1-1)ih''F\psi+\kappa(1-1)iF\psidot+\kappa\lambda(\xdot h''\psi+h'\psidot)\bigg] = 0
\end{multline}

\noindent and analogously $\Qop_{\kappa}{^\dagger}\Soff_{\kappa,\lambda} = 0$. It hence follows that to show the invariance of the Witten index under these rescalings we have to show that the corresponding deformation of the integrand is exact with respect to $\Qop_\kappa$ or $\Qop_\kappa{^\dagger}$. 

However, let us first out of practical considerations also formulate these transformations in superspace. In superspace we redefine the supersymmetry representation and covariant spinor derivatives as

\begin{align}
Q_\kappa &= \pder{}{\thetad} + \kappa\theta\der{}{\tau}
&
D_\kappa &= \pder{}{\thetad} - \kappa\theta\der{}{\tau}
\\
Q_\kappa{^\dagger} &= -\pder{}{\theta} - \kappa\thetad\der{}{\tau}
&
D_\kappa{^\dagger} &= -\pder{}{\theta} + \kappa\thetad\der{}{\tau}
\end{align}

\noindent The transformed superspace Lagrangian $\Lagr_{\kappa,\lambda}$ is now given by

\begin{equation}
\Lagr_{\kappa,\lambda} = \frac{1}{2}D_\kappa{^\dagger}XD_\kappa X + \lambda h(X).
\end{equation}

We are now set to show that kinetic and superpotential rescales are indeed symmetries of the Witten index. Starting off with superpotential rescalings we find that the Witten index changes as

\begin{equation}
\pder{\WI}{\lambda} = \int\Dg x\Dg F\Dg\psid\Dg\psi~\left(-\pder{}{\lambda}\Soff_{\kappa,\lambda}\right)e^{-\Soff_{\kappa,\lambda}}
\end{equation}

\noindent where

\begin{equation}
\pder{}{\lambda}\Soff_{\kappa,\lambda} = \intau\extd\thetad\extd\theta~h(X).
\end{equation}

\noindent Noticing that $h(X)$ transforms as a scalar under supersymmetry transformations we find in analogy to the 
$\hbar$-rescalings
that

\begin{equation}
\pder{}{\lambda}\Soff_{\kappa,\lambda} = \Qop_\kappa{^\dagger}\Qop_\kappa\left\{\intau~h(X)|_{\theta=\thetad=0}\right\} = \Qop_\kappa{^\dagger}\Qop_\kappa\left\{\intau~h(x)\right\}
\end{equation}

\noindent by the same reasoning as the 
$\hbar$-rescalings
it now follows that the Witten index is invariant under superpotential rescalings.

We now move on to kinetic rescalings. These will prove to be slightly more tricky to deal with, but can nevertheless be shown to give $Q$-exact changes to the path integrand of the Witten index. As can be expected by now we find that the Witten index changes as

\begin{equation}
\pder{\WI}{\kappa} = \int\Dg x\Dg F\Dg\psid\Dg\psi~\left(-\pder{}{\kappa}\Soff_{\kappa,\lambda}\right)e^{-\Soff_{\kappa,\lambda}}
\end{equation}

\noindent under kinetic rescalings. In this case we find

\begin{equation}
\begin{split}
\pder{}{\kappa}\Soff_{\kappa,\lambda} 
&= \intau\extd\thetad\extd\theta~\frac{1}{2}\left[\left(\pder{}{\kappa}D_\kappa{^\dagger}X\right)D_\kappa X+D_\kappa{^\dagger}X\left(\pder{}{\kappa}D_\kappa X\right)\right]\\
&=\intau\extd\thetad\extd\theta~\frac{1}{2}\bigg[\thetad\dot{X}D_\kappa X+\theta D_\kappa{^\dagger}X\dot{X}\bigg].
\end{split}
\end{equation}

\noindent To show that these are $Q$-exact, we'd have to show that the integrand of the above expression corresponds to a supersymmetry multiplet. To see this, we note that the Euclidean time derivative is covariant with respect to supersymmetry. We further also note that

\begin{align}
\intau\extd\thetad\extd\theta~\theta(\dots) &= \intau\extd\thetad\left[1-\theta\pder{}{\theta}\right](\dots) = \intau\extd\thetad(\dots)\Big|_{\theta=0},\\
\intau\extd\thetad\extd\theta~\thetad(\dots) &= -\intau\extd\theta\left[1-\thetad\pder{}{\thetad}\right](\dots) = -\intau\extd\theta(\dots)\Big|_{\thetad=0}.
\end{align}

\noindent This follows from the fact that the integrals on the left hand side can't be $\theta$,$\thetad$-graded. We hence find that

\begin{equation}
\begin{split}
\pder{}{\kappa}\Soff_{\kappa,\lambda}
&= -\frac{1}{2}\intau\extd\theta~\dot{X}D_\kappa X + \frac{1}{2}\intau\extd\thetad~D_\kappa{^\dagger}X\dot{X}\\
&= \frac{1}{2}\intau Q_\kappa{^\dagger}\Big\{\dot{X}D_\kappa X\Big\}\Big|_{\thetad=0} + \frac{1}{2}\intau Q_\kappa\Big\{D_\kappa{^\dagger}X\dot{X}\Big\}\Big|_{\theta=0}\\
&= \frac{1}{2}\Qop_\kappa{^\dagger}\left\{\intau~\dot{X}D_\kappa X\Big|_{\theta=\thetad=0}\right\} + \frac{1}{2}\Qop_\kappa\left\{\intau~D_\kappa{^\dagger}X\dot{X}\Big|_{\theta=\thetad=0}\right\}.
\end{split}
\end{equation}

\noindent The lowest Grassmann order components are given by

\begin{align}
\dot{X}D_\kappa X\Big|_{\theta=\thetad=0} &= \xdot\psi, 
&
D_\kappa{^\dagger}X\dot{X}\Big|_{\theta=\thetad=0} &= -\xdot\psid.
\end{align}

\noindent We thus find that

\begin{equation}
\pder{}{\kappa}\Soff_{\kappa,\lambda} = \frac{1}{2}\Qop_\kappa{^\dagger}\left\{\intau~\xdot\psi\right\} - \frac{1}{2}\Qop_\kappa\left\{\intau~\xdot\psid\right\}.
\end{equation}

\noindent Again, by the same reasoning as we had for the 
$\hbar$-rescalings
we now find that the Witten index is invariant under kinetic rescalings. We hence conclude

\begin{tcolorbox}

\subsubsection*{Kinetic and superpotential rescaling symmetry}

\begin{align}
\pder{\WI}{\kappa} &= \pder{\WI}{\lambda} = 0,
&
\Soff \to \Soff_{\kappa,\lambda} = \intau~L(\Xf,\kappa\dot{\Xf})|_{h\to\lambda h}
\end{align}

the Witten index is invariant under kinetic and superpotential rescalings.

\end{tcolorbox}

\subsection{$\beta$-independence of the Witten Index}

We now use the previously derived symmetries to show the $\beta$-independence of the Witten index through the path integral formalism. We do this in the typical fashion of the previous section by taking a derivative with respect to $\beta$ and showing that it vanishes. Taking a $\beta$-derivative of the Witten index isn't entirely straightforward though, since the $\beta$-dependence is carried in the path integral measure and the integration domain of the action. We can get around this by setting the circumference of the circle to unity, but in return rescaling the Euclidean time parameter as $\tau=\beta\sigma$. Again denoting $\Xf = (x,\psi,\psid,F)$ the collective fields, we then find that

\begin{equation}
\begin{split}
\WI(\beta) 
&= \int\underbrace{\Dg\Xf(\tau)}_{\text{per. $\beta$}}\exp\bigg\{-\underset{S^1(\beta)}{\intau}~L\left(\Xf,\der{\Xf}{\tau}\right)\bigg\}
\\
&= \int\underbrace{\Dg\Xf(\sigma)}_{\text{per. $1$}}\exp\bigg\{-\underset{S^1(1)}{\oint\extd\sigma}~\beta L\left(\Xf,\beta^{-1}\der{\Xf}{\sigma}\right)\bigg\}
\end{split}
\end{equation}

\noindent which is nothing but a combination of an 
$\hbar$-rescaling
and a kinetic rescaling! Indeed, we find that

\begin{equation}
\der{\WI}{\beta} = \int\Dg\Xf\left(-\frac{1}{\beta}\Soff+\oint\frac{\extd\tau}{\beta}\pder{L}{\dot{\Xf}}\dot{\Xf}\right)e^{-\Soff} = \frac{1}{\beta}\left.\pder{\WI}{\mu}\right|_{\mu=1} - \frac{1}{\beta}\left.\pder{\WI}{\kappa}\right|_{\kappa=1} = 0.
\end{equation}

\noindent We hence conclude

\begin{tcolorbox}

\subsubsection*{$\beta$-independence}

\begin{equation}
\der{\WI}{\beta} = 0
\end{equation}

the Witten index is also $\beta$-independent in the path integral formalism.

\end{tcolorbox}

\subsection{Localising the Witten Index}

Finally, we compute the Witten index through the path integral formalism. This will be achieved through a localisation. This goes as follows: We make use of the freedom to deform the path integrand of the Witten index by a $Q$-exact functional in such a way that the contributions to the path integral get localised to a subspace of the function space. This can still be an infinite-dimensional space, but can also reduce to finite-dimensional subspaces or even discrete critical points.

The deformation we consider consists of diagonal kinetic-superpotential rescalings we recall are given by

\begin{equation}
\begin{split}
\Soff \to \Soff_{\lambda,\lambda} &
= \intau \left[\frac{1}{2}\lambda^2\xdot^2+\lambda\psid\psidot+\frac{1}{2}F^2+i\lambda h'F-\lambda h''\psid\psi\right]\\
&= \intau \left[\frac{1}{2}\lambda^2\xdot^2+\lambda\psid\psidot+\frac{1}{2}\lambda^2h'^2-\lambda h''\psid\psi\right]+\intau\frac{1}{2}(F+i\lambda h')^2.
\end{split}
\end{equation}

\noindent From this it is clear that in the $\lambda\to\infty$ limit the path integral localises to the critical points $x_0\in\ker\extd h$. These points are discrete since we assumed the superpotential $h$ to be Morse. Accordingly, we expand around these points as 

\begin{align}
x &= x_0 + \frac{y}{\lambda}, & \psi &= \frac{\chi}{\sqrt{\lambda}}, & \psid &= \frac{\chid}{\sqrt{\lambda}}
\end{align}

\noindent yielding an expanded action

\begin{equation}
\Soff_{\lambda,\lambda} =
\intau
\!\begin{multlined}[t] 
\bigg[
\overbrace{\bigg.
\frac{1}{2}\ydot^2+\chid\chidot+\frac{1}{2}F^2
}^\text{free terms}
+
\overbrace{\bigg.
h''(x_0)\Big(iFy-\chid\chi\Big)
}^{\text{harmonic terms}}
\\
+
\underbrace{
\sum_{n\geq 3}\frac{1}{n!}\lambda^{2-n}h^{(n)}(x_0)\Big[iF\der{}{y}-\chid\chi\der{^2}{y^2}\Big]y^n
}_{\text{$\Og(\lambda^{-1})$ non-linear interactions}}
\bigg]
\equiv \Soff_0 + \Og(\lambda^{-1}).
\end{multlined}
\end{equation}

\noindent Under this reparametrisation the path integral measure transforms as

\begin{equation}
\Dg x\Dg F\Dg\psid\Dg\psi = \Dg y\Dg F\Dg\chid\Dg\chi ~ \Ber\fder{(x,\psi,\psid,F)}{(y,\chi,\chid,F)}
\end{equation}

\noindent where the functional Berezinian is given by

\begin{equation}
\Ber\fder{(x,\psi,\psid,F)}{(y,\chi,\chid,F)}
= \Det\fder{x}{y}\Det^{-1}\fder{\psid}{\chid}\Det^{-1}\fder{\psi}{\chi}
= \frac{\Det \lambda^{-1}}{\Det\lambda^{-\frac{1}{2}}\Det\lambda^{-\frac{1}{2}}}
= 1
\end{equation}

\noindent where `$\Det$' denotes the formal functional determinant of the field rescalings, hence yielding

\begin{equation}
\Dg x\Dg F\Dg\psid\Dg\psi = \Dg y\Dg F\Dg\chid\Dg\chi.
\end{equation}

\noindent We also note that these are local reparametrisations on the regions around the critical points in field space. 
The Witten index will be computed through a loop expansion around these critical points. Accordingly, we need to define expectation values. We define the expectation value of some functional $G$ as

\begin{equation}
\EV{G} = \mathcal{N}^{-1}\int\Dg y\Dg F\Dg\chid\Dg\chi~G[y,\chi,\chid,F]e^{-\Soff_0}
\end{equation}

\noindent where we defined the normalisation to be

\begin{equation}
\mathcal{N} = \int\Dg y\Dg F\Dg\chid\Dg\chi ~e^{-\Soff_0} = \Ber^{-1}\fder{^2\Soff_0}{(y,\chi,\chid,F)^2}.
\end{equation}

\noindent This functional Berezinian is somewhat nontrivial, since the free action $\Soff_0$ contains nondiagonal terms. One way to get past this is to eliminate the auxiliary field first and then to compute the functional Berezinian of the on-shell free action. Alternatively, it can also be computed as

\begin{equation}
\begin{split}
\Ber\fder{^2\Soff_0}{(y,\chi,\chid,F)^2} 
&= 
\Det^{\frac{1}{2}}
\begin{pmatrix}
-\extd^2/\extd\tau^2 & ih''(x_0)\\
ih''(x_0) & \mathbbm{1}
\end{pmatrix}
\Det^{-1}\left(\der{}{\tau}+h''(x_0)\right)\\
&= \Det^{\frac{1}{2}}\left(-\der{^2}{\tau^2}+h''(x_0)^2\right)\Det^{-1}\left(\der{}{\tau}+h''(x_0)\right)\\
&= \frac{\sinh(\beta|h''(x_0)|/2)}{\sinh(\beta h''(x_0)/2)}
= \sign ~h''(x_0)
\end{split}
\end{equation}

\noindent thus yielding a normalisation

\begin{equation}
\mathcal{N} = \sign ~h''(x_0).
\end{equation}

\noindent Combining these results we are now ready to compute the Witten index. We find

\begin{multline}
\WI = \int\Dg x\Dg F\Dg\psid\Dg\psi~e^{-\Soff_{\lambda,\lambda}} = \sum_{x_0}\int\Dg y\Dg F\Dg\chid\Dg\chi~e^{-\Soff+\Og(\lambda^{-1})}
\\
= \sum_{x_0}\sign~h''(x_0)\EV{e^{\Og(\lambda^{-1})}}
\xrightarrow{\lambda\to\infty} \sum_{x_0}\sign~h''(x_0).
\end{multline}

\noindent We hence conclude that

\begin{tcolorbox}

\subsubsection*{Localisation of the Witten index}

\begin{equation}
\WI = \int\underbrace{\Dg x\Dg F\Dg\psid\Dg\psi}_{\text{period $\beta$}}~e^{-\Soff} = \sum_{x_0}\sign~h''(x_0)
\end{equation}

the Witten index in the path integral formalism.

\end{tcolorbox}

\newpage

\section{Conclusion}

In conclusion, we have

\begin{tcolorbox}

\begin{equation}
\begin{aligned}
\WI &= \Tr(-1)^Fe^{-\beta H} &&= \dim \ker H|_{\text{bos}} - \dim \ker H|_{\text{fer}}\\
&= \int\underbrace{\Dg x\Dg F\Dg\psid\Dg\psi}_{\text{period $\beta$}}~e^{-\Soff} &&= \sum_{x_0}\sign~h''(x_0)
\end{aligned}
\end{equation}

the Witten index given in both the Hilbert space formalism and the path integral formalism. The Witten index further also satisfies

\begin{equation}
\der{\WI}{\beta} = 0.
\end{equation}

\noindent That is, it is indepedent of the Euclidean periodicity $\beta$.

\end{tcolorbox}

\noindent Let us comment on this. We consider an analytic interpretation of the Hilbert space. As was mentioned in the second section, the Hilbert space can be regarded as $\mathcal{H}\cong L^2(\mathbb{R})\otimes\mathbb{C}^2$. One way to look at this is as position space constituting a manifold, with the wavefunctions being $\mathbb{C}^2$-valued fields living on this manifold. The supercharges then constitute an operator complex and the Hamiltonian its associated Laplacian. With this interpretation, the Hilbert space expression of the Witten index becomes an analytical statement about the Hamiltonian operator. On the other hand, the expression of the Wi tten index in terms of a sum over signs of Hessians of the superpotential $h$ can be regarded as a statement of a topological nature \cite{DavidTong}. We hence find that our computations show a relation between analytical properties of an operator acting on some space and its topology. More general cases could include for example the Atiyah-Singer index theorem, which generally relates analytical properties of complexes on manifolds to topological properties of a corresponding principal bundle \cite{DavidTong}\cite{nakahara2018geometry}.

\chapter{The Third Way Theory}

\section{Introduction}

In this chapter, we introduce the Third Way Theory, originally studied by Arvanitakis, Sevrin and Townsend in \cite{arvanitakis2015yang}. The Third Way Theory is an example of what's called a third way consistent field theory. These are theories in which `auxiliary' field content is added which however can't be consistently integrated out of the action. This results in field equations which can't be the result of any gauge invariant action in only one set of gauge fields, hence constituting an interesting and new contribution to classical field theory. These were originally developed in the context of `minimally massive gravity' (MMG) \cite{Bergshoeff_2014}\cite{Arvanitakis_2014}\cite{Arvanitakis_2015} by ---among others--- Arvanitakis and Townsend, which is a modification of topologically massive gravity \cite{deser2000topologically}. The field theory which is central to this thesis is a third way consistent analogue of topologically massive Yang-Mills theory (TMYM). Its field equation was originally postulated by Sevrin, and the corresponding action was constructed by Arvanitakis \cite{arvanitakis2015yang}. In this chapter we will go in close detail over the contents of their original paper, as to set the ground for the goal of this thesis, namely to localise this theory. 

\section{Topologically Massive Yang-Mills Theory}

We start off by describing the theory on which the Third Way theory is based: Topologically Massive Yang-Mills theory. Throughout these notes I will work in the form-formalism for differential geometry. For conventions and basic identities, see appendix \ref{Differential Geometry}.

\subsection{Yang-Mills Theory}

Let us start off by describing Yang-Mills theory. The field content consists of a $\gLie$-valued spin-1 gauge vector $A_\mu$. The corresponding action is given by
\begin{equation}
S_\YM[A] = \frac{1}{2g^2}\int\tr\big[F\ast F\big] = \frac{1}{4g^2}\int\extd^3x~e\kappa_{IJ} F^I_{\mu\nu}F^{J\mu\nu} =: \frac{1}{2g^2}(F|F)
\end{equation}

\noindent Its field equations are computed to give

\begin{equation}
\delta S_\YM[A] = \frac{1}{g^2}(\delta F|F) = \frac{1}{g^2}(\Dg\delta A|F) = \frac{1}{g^2}(\delta A|\Dgd F)
\end{equation}

\noindent hence by the non-degeneracy of the inner product $(\bullet|\bullet)$ yielding field equations
\begin{align}
\delta S_\YM[A] &= 0
&
&\Longleftrightarrow
&
\Dgd F &= 0.
\end{align}

\noindent These field equations are the $d=3$ non-Abelian analogue of sourced Maxwell equations, whereas the Bianchi identity $\Dg F \equiv 0$ is the analogue of the unsourced Maxwell equations \cite{nakahara2018geometry}.

\subsection{Chern-Simons Theory}

\subsubsection{The classical theory}

Let us now move on to the second component of topologically massive Yang-Mills theory: The Chern-Simons action. This action warrants a little more detail since there are a lot of interesting things to say about this slightly more obscure action. The action of Chern-Simons theory is given by

\begin{equation}
\begin{split}
S_\CS[A] 
&= \int\Omega_\CS(A,F) = \frac{1}{2}\int\tr\bigg[A\extd A + \frac{2}{3}A^3\bigg]
\\
&= \int\extd^3x ~ \frac{1}{2}\varepsilon^{\mu\nu\rho}\kappa_{IJ}\Big(A^I_\mu\del_\nu A^J_\rho + \frac{1}{3}f_{KL}{^J}A^I_\mu A^K_\nu A^L_\rho\Big)
\end{split}
\end{equation}

\noindent with the Chern-Simons 3-form $\Omega_\CS(A,F)$ given by

\begin{equation}
\Omega_\CS(A,F) = \frac{1}{2}\tr\bigg[AF - \frac{1}{3}A^3\bigg] = \frac{1}{2}\tr\bigg[A\extd A + \frac{2}{3}A^3\bigg].
\end{equation}

\noindent This theory has the special property of not being coupled to gravity. Indeed, the action contains no dependence on the metric whatsoever. Theories like this are called topological field theories. The geometric origins of the Chern-Simons form are rooted in characteristic classes. These are polyforms which contain topological information about the principal bundle to which a particular connection belongs \cite{nakahara2018geometry}. Particularly, the characteristic class connected to the Chern-Simons form is the 4\textsuperscript{th} Chern character

\begin{equation}
\ch_4(F) = \frac{1}{2!}\tr\bigg(\frac{iF}{2\pi}\bigg)^2 = -\frac{1}{8\pi^2}\tr F^2
\end{equation}

\noindent being the 4-form component of the Chern character

\begin{equation}
\ch(F) = \tr\exp\frac{iF}{2\pi} 
=: \ch_0(F) + \ch_2(F) + \ch_4(F) + \dots
\end{equation}

\noindent The Chern character ---and in fact any characteristic class--- has the property of being closed, that is,

\begin{equation}
\extd \ch(F) = 0.
\end{equation}

\noindent From Poincaré's lemma \cite{nakahara2018geometry} it therefore follows that the Chern character is \textit{locally} exact. Particularly, one finds that locally the 4\textsuperscript{th} Chern character can be written as

\begin{equation}
\ch_4(A) = \bigg(\frac{i}{2\pi}\bigg)^2\extd \Omega_\CS(A,F)
\end{equation}

\noindent where $\Omega_\CS$ the aforementioned Chern-Simons form. Since the Chern character is gauge invariant, it follows then that \textit{locally} the Chern-Simons form changes by a total derivative under gauge transformations. This is very interesting, because it means that while the Chern-Simons action may not actually be gauge invariant the way it changes under gauge transformations doesn't actually change the classical dynamics! Varying the Chern-Simons action we find that

\begin{equation}
\delta S_\CS[A] = \frac{1}{2}\int\tr\bigg[\delta A\extd A + A\extd\delta A + 2\delta AA^2\bigg] = \int\tr\Big[\delta AF - \extd(A\delta A)\Big].
\end{equation}

\noindent Assuming we work on boundariless spaces we then find field equations
\begin{align}
\delta S_\CS[A] &= 0
&
&\Longleftrightarrow
&
F &= 0.
\end{align}

\noindent That is, the classical solutions consist of flat connections. We note that under an infinitesimal gauge transformation $\delta_\theta A = -\Dg\theta$ with parameter $\theta$ the Chern-Simons action transforms as

\begin{equation}
\delta_\theta S_\CS[A] = \int \tr\big[\delta_\theta AF\big] = \int\tr\big[-\Dg(\theta F)\big] = \int\extd\tr\big[-\theta F\big]
\end{equation}

\noindent where we used the Bianchi identity $\Dg F \equiv 0$. This is indeed a total derivative as we expected \cite{nakahara2018geometry}.

\subsubsection{The quantum theory}

Let us now move on to make a few comments on the quantum theory. Since in the quantum theory we are interested in computing expectation values

\begin{equation}
\EV{\Og} = \int\Dg A ~ \Og[A]e^{-S[A]}
\end{equation}

\noindent it becomes now important that the Chern-Simons action isn't gauge invariant! Let us sketch the solution to this problem without getting into too much technical details. 
In the quantum theory, the parameters defining the theory are 
\begin{itemize}
\item a simple gauge group $G$,
\item a bilinear form of $G$ (the Killing form) of level $k$.
\end{itemize}
The level of the bilinear form manifests itself in the trace as $\tr_k = k\tr$ where $\tr$ the trace of the fundamental representation. The action is then given by
\begin{equation}
S_\CS^k[A] := \frac{1}{4\pi}\int\tr_k\bigg[A\extd A + \frac{2}{3}A^3\bigg] = \frac{k}{4\pi}\int\tr\bigg[A\extd A + \frac{2}{3}A^3\bigg] = \frac{k}{2\pi}S_\CS[A].
\end{equation}

\noindent This action is still not gauge invariant. However, now it has the interesting property of being \textit{gauge invariant modulo $2\pi i$}, hence, leaving the expectation values unchanged and thus making the quantum theory gauge invariant. These integer multiples of $2\pi$ by which the action changes can be related to the fact that the third homotopy group for simple gauge groups is given by $\pi_3(G) = \mathbb{Z}$ \cite{closset2013supersymmetric}. 

\subsection{Topologically Massive Yang-Mills Theory}

Now that we've treated both the Yang-Mills and Chern-Simons actions we are ready to move on to the topologically massive Yang-Mills (TMYM) action which combines both these theories. The action is given by

\begin{equation}
S_\TMYM[A] = \frac{1}{g^2}S_\YM[A] + \frac{\mu}{g^2}S_\CS[A]
\end{equation}

\noindent where $g^2$ and $\mu$ are couplings which have the dimension of mass. The name of this action will become clear when we work out the corresponding field equations. Varying the action one finds

\begin{equation}
g^2\delta S_\TMYM[A] 
= \int\tr\bigg[\delta A\ast\Dgd F + \mu\delta AF\bigg] = \boobket{\delta A}{\Dgd F + \mu\invast F}
\end{equation}

\noindent We thus arrive at field equations

\begin{align}
\delta S_\TMYM &= 0
&
&\Longleftrightarrow
&
\Dg\ast F + \mu F &= 0
\\
&
&
&\Longleftrightarrow
&
\epsilon_{\mu\rho\sigma}\Dg^\rho(\asst F)^\sigma + \mu (\asst F)_\mu &= 0
\end{align}

\noindent Using the field equations we then find that

\begin{align}
(\Dg\ast)^2F - \mu^2F &= 0
&
&\Longleftrightarrow
&
\Dg^\nu\Dg_\nu(\ast F)_\mu + [F_{\mu\nu},(\ast F)^\nu] + \eta\mu^2(\ast F)_\mu &= 0
\end{align}

\noindent which are nothing but the field equations for a vector field $\ast F_\mu$ of mass $\mu$. Here we defined $\eta = \det(\eta_{ab})$ the determinant of the flat index metric, as is outlined in the appendix section \ref{Riemannian Geometry}.

\section{Third Way to Source TMYM Theory}

In this section we will discuss several ways in which to add a source TMYM theory. We will start off by going over the standard two ways one could add a source to TMYM theory, and then finally we will discuss the so called `third way' to add a source to this theory, with interesting consequences.

\subsection{Consistency Condition}

Adding a source current $j$ to TMYM theory is simply achieved by adding it to the field equation as
\begin{align}
\Dg\ast F + \mu F &= \invast j
&
&\Longleftrightarrow
&
\epsilon_{\mu\rho\sigma}\Dg^\rho(\asst F)^\sigma + \mu (\asst F)_\mu &= j_\mu.
\end{align}

\noindent However, we can't just pick any 1-form to be a source for these equations. Namely, these currents have to satisfy consistency conditions. Particularly, we find that

\begin{equation}
\label{consistency condition 1}
\Dg\big(\Dg\ast F + \mu F\big) = [F,\ast F] + \mu \Dg F \equiv 0
\end{equation}

\noindent where we used the Bianchi identity as well as $D^2 = \ad F$. The commutator in the above expression vanishes because it's antisymmetric in its Lie algebra components but symmetric in its form components under an interchange of the two field strengths.\footnote{This result may seem a bit surprising but in fact it's a generic consistency condition for gauge invariant theories, as we will see shortly.} Because of this, the source current $j$ has to satisfy the consistency condition
\begin{align}
\label{consistency condition 2}
\Dg\ast j &= 0
&
&\Longleftrightarrow
&
\Dg_\mu j^\mu &= 0.
\end{align}

\noindent That is, it has to be covariantly conserved. There are two standard ways of enforcing this result:

\begin{itemize}
\item One could have the source emerge through additional interaction terms in the gauge fields. The source would then be of the form $j^\mu = e^{-1}\delta I[A]/\delta A_\mu$, for $I[A]$ some gauge invariant functional. This source would be conserved off-shell.
\item As a Noether current for some lower spin matter content $\Phi$. In this case the source $j(\Phi)$ is conserved is the matter field equations are enforced. However, the gauge fields may be off-shell.
\end{itemize}

\subsection{Noether's Second Theorem}
\label{Noether's Second Theorem}

Let us now take a look at Noether's second theorem. Not only will this be important for understanding how the two aforementioned currents are conserved, but it will also highlight just how special the third way of getting a consistent source current is! We consider a functional $S[A,\Phi]$ of the gauge fields and matter content which is both Lorentz and gauge invariant.

From the gauge invariance of this action it follows that under infinitesimal gauge transformations with local parameter $\theta^I$ we have
\begin{multline}
0 \equiv \delta_\theta S[A,\Phi] 
= \int\ast\bigg[\delta_\theta A_\mu^I\frac{1}{e}\fder{S}{A_\mu^I} + \delta_\theta\Phi^i\frac{1}{e}\fder{S}{\Phi^i}\bigg]
= \int\ast\bigg[-\Dg_\mu\theta^I\frac{1}{e}\fder{S}{A_\mu^I} + \delta_\theta\Phi^i\frac{1}{e}\fder{S}{\Phi^i}\bigg]
\\
= \int\ast\bigg[\theta^I\Dg_\mu\bigg(\frac{1}{e}\fder{S}{A_\mu^I}\bigg) + \delta_\theta\Phi^i\frac{1}{e}\fder{S}{\Phi^i}\bigg].
\end{multline}

\noindent Since this holds for any local parameter $\theta^I$ we find the much celebrated

\begin{tcolorbox}

\subsubsection*{Noether's second theorem}

Let $S[A,\Phi]$ be a functional which is gauge and Lorentz invariant. It then satisfies the conservation law
\begin{align}
\Dg_\mu\bigg(\frac{1}{e}\fder{S}{A_\mu}\bigg) &= 0
&
&
\begin{cases}
\Phi & \text{on-shell}
\\
A_\mu & \text{off-shell}
\end{cases}
\end{align}

\end{tcolorbox}

\noindent For example, if we take $S = S_\TMYM$ we arrive at the consistency condition \refeq{consistency condition 1}. The two source currents can be achieved as follows: For a source current through coupling to lower spin matter we consider a total action

\begin{equation}
S[A,\Phi] = S_\TMYM[A] + S_{0,\frac{1}{2}}[\Phi,A]
\end{equation}

\noindent where $S_{0,\frac{1}{2}}$ a matter Lagrangian. In this case we find a conserved current

\begin{align}
j^\mu(\Phi) &:= -\frac{1}{e}\fder{S_{0,\frac{1}{2}}}{A_\mu}
&
\Dg_\mu j^\mu &= 0
&
\begin{cases}
\Phi & \text{on-shell}
\\
A_\mu & \text{off-shell}
\end{cases}
\end{align}

\noindent yielding the first kind of conserved current. The second can be found by taking an action

\begin{equation}
S[A] = S_\TMYM[A] - I[A]
\end{equation}

\noindent yielding an identically conserved current

\begin{align}
j^\mu &:= \frac{1}{e}\fder{I}{A_\mu}
&
\Dg_\mu j^\mu &\equiv 0.
\end{align}

\noindent Note, however, that the most general gauge invariant action which is (up to) second order derivatives in the gauge fields is exactly the TMYM action \cite{Arvanitakis_2015}. As such, sourcing the action this way would lead to higher order derivative terms.

\subsection{Third Way Source Term}

We now move on to discuss a third way to source TMYM theory. This way was due to Sevrin and borrows from the methodologies of MMG, instead now applied to TMYM theory. In this approach, the conserved current will be of the form
\begin{align}
j &\propto \ast[\ast F,\ast F]
&
&\Longleftrightarrow
&
j_\mu &\propto [F_{\mu\nu},(\asst F)^\nu].
\end{align}

\noindent And it is here that comes the catch to the story: The reason that this current isn't produced by the aforementioned methods of obtaining a source, \textit{is that it makes use of the field equations for the gauge fields!} We find that using the TMYM field equations
\begin{equation}
\label{vanishing 0}
\Dg\ast j \propto [\Dg\ast F,\ast F] \propto [F,\ast F] = 0.
\end{equation}

\noindent However, this by itself isn't enough because we used unsourced TMYM field equations. If we use the sourced TMYM field equations with this source we find
\begin{equation}
\Dg\ast j \propto [\Dg\ast F,\ast F] = -\mu[F,\ast F] + [\invast j,\ast F] \propto [[\ast F,\ast F],\ast F]
\end{equation}

\noindent Using the $\mathbb{Z}_2$-graded Jacobi identity we can then show that the RHS vanishes:
\begin{multline}
\label{vanishing 1}
[[\ast F,\ast F],\ast F] = \frac{1}{3}[[\ast F,\ast F],\ast F] - \frac{1}{3}\Big([\ast F,[\ast F,\ast F]]-[[\ast F,\ast F],\ast F]\Big)
\\
\overset{\text{\ref{Z2 graded Jacobi identity}}}{=} \frac{1}{3}[[\ast F,\ast F],\ast F] - \frac{1}{3}[[\ast F,\ast F],\ast F] = 0.
\end{multline}

\noindent From this, it follows that the aformentioned current can be consistently used to source the TMYM field equations. For dimensional reasons we now normalise the current as

\begin{equation}
j = -\frac{1}{2m}\ast[\ast F,\ast F]
\end{equation}

\noindent introducing a new dimensionful parameter $m$ with the dimension of mass. In conclusion we now have

\begin{tcolorbox}

\subsubsection*{Third Way sourced TMYM field equations}

\begin{align}
\Dg\ast F + \mu F &= \invast j
&
j &= -\frac{1}{2m}\ast[\ast F,\ast F]
\end{align}

\noindent satisfying the consistency condition
\begin{align}
\Dg_\mu j^\mu &= 0
&
&\text{$A_\mu$ \textbf{on}-shell}
\end{align}

\noindent which ---unlike Noether charges--- only holds on-shell.

\end{tcolorbox}

\section{The Third Way Theory}

We have now constructed a new way to source TMYM theory. However, as far as we have explained it isn't yet obvious how to translate this into an action formalism, mainly due to the fact that this approach is distinct from the usual ways one sources TMYM theory. Before moving on to an action description, we take a closer look at the field equations of the presumed Third Way Theory:

\begin{tcolorbox}

\subsubsection*{Third Way field equations}

\begin{gather}
\Dg\ast F + \frac{1}{2m}[\ast F,\ast F] + \mu F = 0
\\
\mathcal{m}\nonumber
\\
\epsilon_{\mu\rho\sigma}\Big(\Dg^\rho\asst F^\sigma + \frac{1}{2m}[\asst F^\rho,\asst F^\sigma]\Big) + \mu\asst F_\mu = 0
\end{gather}

\noindent the field equations of the presumed Third Way Theory, with dimensionful constants $m$ and $\mu$ of mass dimension 1.

\end{tcolorbox}

\noindent We will now go over various aspects of the conjectured Third Way Theory: It's parity, the way it couples to 3D gravity and the way it couples to lower spin matter fields. In all of these regards we will see that the Third Way Theory exhibits quite remarkable and very unorthodox properties.

\subsection{Parity}

Let us start off by studying whether or not the on-shell Third Way theory is parity invariant. By note that if we take $A$ to be parity-even, the field equations are not parity invariant, even if $\mu = 0$. Indeed, if $A$ is parity-even then so is $F$ from which it follows that $\ast F$ is parity-odd. This is in conflict with the $[\ast F,\ast F]$ term which has to be parity-even. To solve this issue one introduces the ad hoc parity transformations

\begin{align}
\label{parity on-shell}
A &\xrightarrow{P} A + \frac{1}{m}\ast F
&
&\Longrightarrow
&
F &\xrightarrow{P} F + 
\underbrace{
\frac{1}{m}\bigg[\Dg\ast F + \frac{1}{2m}[\ast F,\ast F]\bigg]
}_{\text{$\propto$ $\mu = 0$ field equations}}
\end{align}

\noindent One thus finds that for $\mu = 0$ on-shell, $F$ is parity-even, $\ast F$ is parity-odd and $[\ast F,\ast F]$. However, in contrast to the previous approach $\Dg\ast F$ will be neither. The way it transforms non-trivially will make it so that the field equations are parity invariant on-shell. Indeed, one finds that

\begin{multline}
\Dg\ast F + \frac{1}{2m}[\ast F,\ast F] \xrightarrow{\text{on-shell $P$}} \big(\Dg + \frac{1}{m}\ad\asst F\big)\big(-\asst F\big) + \frac{1}{2m}[-\asst F,-\asst F]
\\
= -\bigg(\Dg\ast F + \frac{1}{2m}[\ast F,\ast F]\bigg).
\end{multline}

\noindent For these transformations, parity is thus conserved on-shell. This feature which we now have introduced ad hoc will become manifest through the off-shell formulation of the Third Way Theory.



\subsection{Coupling to Lower Spin Fields}

Coupling the Third Way Theory to lower spin matter will in the on-shell formalism also turn out to be a non-trivial task. Let us start off by trying to naively source the Third Way field equations as

\begin{equation}
\label{naive sourcing}
\Dg\ast F + \frac{1}{2m}[\ast F,\ast F] + \mu F = \invast j
\end{equation}

\noindent where we assume $j^\mu$ some Noether current for lower spin content. We then find using previous results that

\begin{multline}
\Dg\invast j
\overeq{\ref{consistency condition 1}}
\frac{1}{m}[\Dg\ast F,\ast F] 
\overeq{\ref{naive sourcing}} 
-\frac{1}{2m^2}[[\ast F,\ast F],\ast F] - \frac{\mu}{m}[F,\ast F] +  \frac{1}{m}[\invast j,\ast F]
\\
\overeq{\ref{vanishing 1}} \frac{1}{m}[\invast j,\ast F].
\end{multline}

\noindent We hence arrive at the on-shell sourcing consistency condition

\begin{align}
\label{Third Way sourcing}
\Big(\Dg + \frac{1}{m}\ad\asst F\Big)\invast j &= 0
&
&\Leftrightarrow
&
\Dg^\mu j_\mu + \frac{1}{m}[\asst F^\mu,j_\mu] &= 0
\end{align}

\noindent which is clearly in conflict with ordinary sourcing, since $j$ can only be consistently interpret as a Noether current in the limit $m^{-1}\to 0$. Before offering a solution to this problem let us first simplify the notation. We note that we can simplify expression \ref{Third Way sourcing} as

\begin{align}
\Dbbd j &= 0
&
\Dbb &:= \Dg + \frac{1}{m}\ad\asst F
\end{align}

\noindent It turns out, however, that there is a way to solve this problem. For this we assume that $m \neq \mu$ and that we're given a Noether current $j$, i.e. $\Dgd j = 0$. Then we can construct a consistent sourcing of the on-shell Third Way field equation by using a current

\begin{equation}
\label{Jg}
\Jg := j - \frac{1}{m-\mu}\ast\bigg[\Dbb j + \frac{1}{2m(m-\mu)}[j,j]\bigg].
\end{equation}

\noindent Indeed, we start off by noting that

\begin{equation}
\label{Dbb^2}
\begin{split}
\Dbb^2 = \frac{1}{2}[\Dbb,\Dbb]
&= \frac{1}{2}[\Dg,\Dg] + \frac{1}{m}[\Dg,\ad\ast F] + \frac{1}{2m^2}[\ad\ast F,\ad\ast F]
\\
&= \ad\bigg[F + \frac{1}{m}\Dg\ast F + \frac{1}{2m^2}[\ast F,\ast F]\bigg]
\\
&= \ad\bigg[\frac{m-\mu}{m}F + \frac{1}{m}\bigg(\Dg\ast F + \frac{1}{2m}[\ast F,\ast F] + \mu F\bigg)\bigg]
\\
&\overeq{on-shell}
\ad\bigg[\frac{m-\mu}{m}F + \frac{1}{m}\invast \Jg\bigg]
\end{split}
\end{equation}

\noindent It thus follows that

\begin{equation}
\begin{split}
\invast\Dbbd\Jg 
&= -\Dbb\invast\Jg
= -\Dbb\invast j + \frac{1}{m-\mu}\ad\bigg[\Dbb^2 + \frac{1}{m(m-\mu)}\Dbb j\bigg]j
\\
&\overeq{\ref{Dbb^2}} \ad\bigg[-\frac{1}{m}F + \frac{1}{m-\mu}\bigg(\frac{m-\mu}{m}F + \frac{1}{m}\invast\Jg + \frac{1}{m(m-\mu)}\Dbb j\bigg)\bigg]j
\\
&= \frac{1}{m(m-\mu)}\ad\bigg[\invast\Jg + \frac{1}{m-\mu}\Dbb j\bigg]j
\\
&\overeq{\ref{Jg}} \frac{1}{m(m-\mu)}\bigg[[\invast j,j] + \frac{1}{2m(m-\mu)}[[j,j],j]\bigg]
\\
&\equiv 0
\end{split}
\end{equation}

\noindent where the final step follows in analogy to equations \ref{vanishing 0} and \ref{vanishing 1}. We thus conclude that

\begin{tcolorbox}

\subsubsection{Sourced Third Way field equations}

The Third Way Theory is consistently coupled to lower spin fields as

\begin{equation}
\label{Third Way sourced}
\Dg\ast F + \frac{1}{2m}[\ast F,\ast F] + \mu F = \invast\Jg
\end{equation}

\noindent where we defined the Third Way current $\Jg$ in terms of a Noether current $j$, satisfying $\Dg_\mu j^\mu = 0$ as

\begin{equation}
\Jg := j - \frac{1}{m-\mu}\ast\bigg[\Dg j + \frac{1}{m}[\ast F,j] + \frac{1}{2m(m-\mu)}[j,j]\bigg].
\end{equation}

\smallskip

\end{tcolorbox}

\section{The Third Way Theory: Off-Shell}

\subsection{Third Way Action}

\subsubsection{Noether's second theorem}

We now move on to describing the off-shell formalism of the Third Way theory. We start off by making the important observation that the Third Way field equations \textit{cannot} come from a local gauge invariant action. Indeed, we recall from section \ref{Noether's Second Theorem} that a consequence of Noether's second theorem is that local gauge invariant theories have field equations which identically satisfy

\begin{equation}
\Dg_\mu\fder{S}{A_\mu} \equiv 0.
\end{equation}

\noindent However, in the case of the Third Way field equation we find that

\begin{equation}
\Dg\bigg(\Dg\ast F + \frac{1}{2m}[\ast F,\ast F] + \mu F\bigg) 
= \frac{1}{m}[\Dg\ast F,\ast F]
~\slashed{\equiv}~ 0.
\end{equation}

\noindent Indeed, the whole point of the Third Way being a new consistent way of doing gauge theory in three dimensions was that the consistency of the field equations invoked the on-shell conditions for the gauge fields.

\subsubsection{Third Way Action}

As it turns out, the way to construct an action by introducing an `auxiliary' 1-form $G$ of mass dimension 2 and taking the dimensionful parameters to satisfy $m\neq \mu$. The action which gives the Third Way field equations is then given by

\begin{equation}
S\ThirdWay[A,G] = 
\begin{multlined}[t]
\frac{1}{g^2}\int\tr\bigg[FG - \frac{m-\mu}{2m}G\invast G + \frac{1}{2m}\bigg(G\Dg G + \frac{2}{3m}G^3\bigg)\bigg]
\\
+ \frac{\mu}{g^2}\int\tr\bigg[A\extd A + \frac{2}{3}A^3\bigg].
\end{multlined}
\end{equation}

\noindent We note that the `auxilary' field $G$ is not auxiliary in the typical sense we mean in the context of supersymmetry, since it appears with derivatives in the action. 
Because of this, we can't consistently eliminate it from the action.
To see this, let us work out the field equations: Varying the action yields

\begin{equation}
\begin{aligned}
\delta S\ThirdWay
= 
\frac{1}{g^2}\int\tr\bigg[\big(\delta A + \frac{1}{m}\delta G\big)\bigg(\Dg G + \frac{1}{2m}[G,G] + \mu F\bigg) 
\\
+ \frac{m-\mu}{m}\delta G\big(F - \invast G\big)\bigg]
\end{aligned}
\end{equation}

\noindent resulting in field equations

\begin{align}
\Dg G + \frac{1}{2m}[G,G] + \mu F &= 0,
&
G &= \ast F.
\end{align}

\noindent If we substitute the latter in the former we arrive at the Third Way field equation. 

\subsubsection{Parameter constraints}

We can constrain the parameters $\mu$ and $m$ by noting that the stress tensor will be that of Yang-Mills theory (at least using $G = \ast F$), multiplied by a factor $(m-\mu)/m$. By demanding positive energy we can constrain the parameters by the condition

\begin{equation}
m(m-\mu) > 0.
\end{equation}

\subsubsection{Sourcing the Third Way theory}

Sourcing the Third Way theory now becomes a straightforward procedure, and will also connect to the on-shell formalism. One simply adds the standard source term

\begin{equation}
S[A,G] = S\ThirdWay[A,G] - \frac{1}{g^2}\int\tr\big[A\invasst j\big]
\end{equation}

\noindent where $j$ is taken to be some Noether current, satisfying $\Dgd j = 0$. In this case varying the action yields

\begin{equation}
\begin{aligned}
\delta S
= 
\frac{1}{g^2}\int\tr\bigg[\big(\delta A + \frac{1}{m}\delta G\big)\bigg(\Dg G + \frac{1}{2m}[G,G] + \mu F - \invast j\bigg) 
\\
+ \frac{m-\mu}{m}\delta G\bigg(F + \frac{1}{m-\mu}\invast j - \invast G\bigg)\bigg]
\end{aligned}
\end{equation}

\noindent resulting in field equations

\begin{align}
\Dg G + \frac{1}{2m}[G,G] + \mu F &= \invast j,
&
G &= \ast F + \frac{1}{m-\mu}j.
\end{align}

\noindent One notes now that on-shell $G$ gains $j$ dependence. It is exactly this which after substituting the latter equation into the former produces the unusual source $\Jg$ introduced into the on-shell Third Way field equations \ref{Third Way sourced}.

\subsection{Manifest Parity}

So far, we have discussed all aspects of the Third Way theory discussed in the on-shell approach, except for the way parity is preserved for $\mu = 0$. To see how parity is preserved we will have to consider an alternate way to parametrise the Third Way action. We introduce a second connection $\Ab$ for the gauge group which transforms \textit{diagonally} with $A$ under gauge transformations. That is, their gauge transformations are parametrised by the same local parameter. The 1-form $G$ which is an adjoint tensor of the gauge group will then be related to the connections $A$ and $\Ab$ as

\begin{equation}
\label{G(A,Ab)}
G = m(\Ab-A)
\end{equation}

\noindent which is still an adjoint tensor of the gauge group since differences of connections are tensors. In this case the action will take on the much more elegant form

\begin{tcolorbox}

\subsubsection{Parity manifest Third Way action}

\medskip

\begin{equation}
\begin{aligned}[t]
S\ThirdWay[A,\Ab] 
={}& \frac{m}{g^2}S_\CS[\Ab] - \frac{\mb}{g^2}S_\CS[A] 
- \frac{m\mb}{2g^2}\int\tr\Big[(\Ab-A)\invast(\Ab-A)\Big]
\end{aligned}
\end{equation}

\noindent the Third Way action in terms of $A$ and $\Ab$, with field equations 
\begin{align}
\label{field equation F}
0 &= F - m\invast(\Ab-A)
\\
\label{field equation Fb}
0 &= \Fb - \mb\invast(\Ab-A)
&
\mb &:= m - \mu
\end{align}

\noindent and parity transformations 

\begin{equation}
A \overset{P}{\longleftrightarrow} \Ab
\end{equation}

\end{tcolorbox}

\noindent Let us verify that this indeed agrees with previous results. We start off by noting that the barred field strength can be rewritten as

\begin{equation}
\label{tussenstap Fb}
\Fb 
= \extd \Ab + \Ab{^2} 
\overeq{\ref{G(A,Ab)}}
F + \frac{1}{m}\Dg G + \frac{1}{m^2}G^2.
\end{equation}

\noindent With this the Chern-Simons term in $\Ab$ becomes
\begin{equation}
\begin{split}
\frac{m}{g^2}S_\CS[\Ab]
&=
\frac{m}{2g^2}\int\tr\bigg[\Ab\Fb - \frac{1}{3}\Ab{^3}\bigg]
\\
&\overeq{\ref{tussenstap Fb}} \frac{m}{g^2}S_\CS[A] + \frac{1}{2g^2}\int\tr\bigg[
\big(A\Dg G + FG -A^2G\big)
+\frac{1}{m}G\Dg G + \frac{2}{3m^2}G^3\bigg]
\\
&= \frac{m}{g^2}S_\CS[A] + \frac{1}{g^2}\int\tr\bigg[FG + \frac{1}{2m}\bigg(G\Dg G + \frac{2}{3m}G^3\bigg)\bigg]
\end{split}
\end{equation}

\noindent where in the final step we integrated by parts and used the fact that $\extd A = F - A^2$.

\subsubsection{Connection to previous parity transformations}

We now show that these parity transformations also yield the previous parity transformations as in equation \ref{parity on-shell}. This follows from field equation \ref{field equation F} as

\begin{equation}
A \overset{P}{\longrightarrow} \Ab \overeq{\ref{field equation F}} A + \frac{1}{m}\ast F
\end{equation}

\noindent indeed agreeing with the on-shell results, when the fields are taken to be on-shell.

\subsubsection{Further constraints on the parameters in the quantum theory}

Let us also briefly discuss how the parameters of the Third Way are constrained in the quantum theory. In accordance with the results of quantum Chern-Simons theory we find constraints for gauge invariance given by
\begin{align}
\frac{m}{g^2} &=: \frac{k}{2\pi},
&
\frac{\mb}{g^2} &=: \frac{\kb}{2\pi}
&
&\Rightarrow
&
\frac{m\mb}{2g^2} &= g^2\frac{k\kb}{8\pi^2}.
\end{align}

\noindent with $k,\kb\in\mathbb{Z}_0$. This in turn allows us to rewrite the Third Way action as
\begin{equation}
S\ThirdWay = \frac{k}{2\pi}S_\CS[\Ab] - \frac{\kb}{2\pi}S_\CS[A] - \frac{k\kb}{4\pi^2}\int\tr\bigg[\frac{1}{2\ell}(\Ab-A)\invasst(\Ab-A)\bigg]
\end{equation}

\noindent where we defined the parameter $\ell := g^{-2}$ with the dimension of length.

\subsection{The Bifundamental Scalar and the Brout-Englert-Higgs Mechanism}

To close off this chapter we discuss a context in which the Third Way Theory emerges through the BEH mechanism. This was originally discovered in M-theory, particularly for multi M2-brane dynamics in \cite{Mukhi_2008} by Mukhi and Papageorgakis in 2008. Later in 2011 Mukhi studied the symmetry breaking of the bifundamental scalar in greater detail in \cite{mukhi2011unravelling}. Here we will briefly outline how the Third Way Theory (together with interactions with a BEH scalar) emerges in this context.

\subsubsection{The bifundamental scalar and connection}

We start off by describing the gauge group and representations of this field theory. The gauge group of this theory will be of the form $G\times G$, where $G$ a simple Lie group. A connection $A^\bi$ of the gauge group $G\times G$ will then be of the form

\begin{equation}
A^\bi = A \otimes \idop + \idop \otimes \Ab
\end{equation} 

\noindent transforming under gauge transformations as

\begin{align}
A^\bi &\to e^{\theta^\bi}\Big(A^\bi + \extd\Big)e^{-\theta^\bi}
&
&\Leftrightarrow
&
&
\begin{cases}
A \to e^\theta(A+\extd)e^{-\theta}
\\
\Ab \to e^\thetab(\Ab+\extd)e^{-\thetab}
\end{cases}
\\
\delta_{\theta^\bi}A^\bi &= -\Dg^\bi \theta^\bi
&
&\Leftrightarrow
&
&
\begin{cases}
\delta_\theta A = -\Dg\theta
\\
\delta_\thetab\Ab = -\Dgb\thetab
\end{cases}
\end{align}

\noindent where we defined the local gauge parameter 

\begin{align}
\theta^\bi &:= \theta\otimes\idop + \idop\otimes\thetab
&
e^{\theta^\bi} &= e^\theta \otimes e^\thetab
\end{align}

\noindent It is thus clear that $A$ and $\Ab$ transform like \textit{indepedent} gauge connections of $G$. The scalar field we consider lies in the so-called bifundamental representation. That is, the bifundamental scalar $\Phi$ is a matrix of the fundamental representation of $G$ (which for all clarity doesn't have to lie in the Lie algebra) which transforms under $G\times G$ as

\begin{align}
\Phi &\to e^\theta\Phi e^{-\thetab},
&
\delta_{\theta,\thetab}\Phi &= \theta\Phi - \Phi\thetab.
\end{align}

\noindent In this representation the bifundamental covariant derivative thus acts on it as

\begin{equation}
\Dg^\bi\Phi = \extd\Phi + A\Phi - \Phi\Ab.
\end{equation}

\subsubsection{Symmetry breaking}

Let us now move on to the theory of interest. This theory is given by an action
\begin{equation}
S[A^\bi,\Phi]
=
\frac{k}{2\pi}S_\CS[\Ab] - \frac{\kb}{2\pi}S_\CS[A] - \int\tr\bigg[\frac{1}{2}\Dg^\bi\Phi\invasst\Dg^\bi\Phi\bigg] + \int\invasst V(\Phi).
\end{equation}

\noindent Here $V(\Phi)$ is a potential such that $\Phi$ has a vacuum expectation value $\EV{\Phi}$ of the form

\begin{align}
\EV{\Phi} &= v\idop
&
\Phi &= v\idop + \Sigma
\end{align}

\noindent where by $\Sigma$ we denote the fluctuations around the vacuum expectation value $\EV{\Phi}$. We then find that 

\begin{equation}
\Dg^\bi\Phi = v(\Ab-A) + \Dg^\bi\Sigma
\end{equation}

\noindent so that the kinetic term for $\Phi$ becomes
\begin{multline}
\tr\bigg[\frac{1}{2}\Dg^\bi\Phi\invasst\Dg^\bi\Phi\bigg]
=
\tr\bigg[\frac{1}{2}v^2(\Ab-A)\invasst(\Ab-A) + v(\Ab-A)\invasst\Dg^\bi\Sigma 
\\
+ \frac{1}{2}\Dg^\bi\Sigma\invasst\Dg^\bi\Sigma\bigg].
\end{multline}

\noindent If we now choose to redefine the vacuum expectation value $v$ to be

\begin{equation}
v^2 =: \frac{k\kb}{4\pi^2}\frac{1}{\ell}
\end{equation}

\noindent we find that the action reduces to

\begin{equation}
\begin{aligned}
S[A,\Ab,\Sigma]
=
\frac{k}{2\pi}S_\CS[\Ab] - \frac{\kb}{2\pi}S_\CS[A] - \frac{k\kb}{4\pi^2}\int\tr\bigg[\frac{1}{2\ell}(\Ab-A)\invasst(\Ab-A)\bigg]
\\
- \int\tr\bigg[\frac{1}{2}\Dg^\bi\Sigma\invasst\Dg^\bi\Sigma - \frac{1}{2\pi}\left(\frac{k\kb}{\ell}\right)^{\frac{1}{2}}(\Ab-A)\invasst\Dg^\bi\Sigma - \hat{V}(\Sigma)\bigg]
\end{aligned}
\end{equation}

\noindent which indeed yields the Third Way Theory along with some interaction terms. Let us now more closely look at the way the gauge group $G\times G$ reduces under this symmetry breaking. We find that our choice of the vacuum expectation value $\EV{\Phi}$ transforms under gauge transformations as

\begin{align}
\delta_{\theta,\thetab}\EV{\Phi} &= v(\thetab-\theta) = 0
&
&\Leftrightarrow
&
\theta &= \thetab.
\end{align}

\noindent We thus find that the gauge group $G\times G$ breaks to its diagonal subgroup as

\begin{equation}
G\times G \to \big(G\times G\big)\big|_{\text{diag}} \cong G.
\end{equation}

\noindent We hence arrive back at the case of the Third Way Theory where we demanded the gauge fields to transform diagonally under gauge transformations.

\chapter{Localisation of the Third Way Theory}

\section{Introduction}

In this chapter we move on to the main goal of this thesis: The localisation of the Third Way Theory. We will start off by getting deeper into the specifics of $\NSUSY = 2$ supersymmetry in Euclidean $d = 3$, continuing the discussion we started in section \ref{Some Basics of Supersymmetry}. Particularly, we will start off by taking a closer look at the gauge vector multiplet. We will discuss this object starting from superconnections in superspace and derive how the vector multiplet and its gauge covariant supersymmetry transformations naturally arise from this formalism, as well as how gauge transformations as we know and love them are to be understood in this formalism, that is, to relate `supergauge transformations' to `gauge transformations'. We then move on to describe some theories in this formalism. We will in particular take a closer look at super-Chern-Simons theory and super-Yang-Mills theory. After this, we move on to describe the way to localise these theories.

Having done these things, we have finally set the stage to describe progress on the localisation of the Third Way Theory. We start off by treating a toy model of the Third Way Theory which we coined Proca-Chern-Simons theory, which contains only a single gauge field. To localise it, we have to supersymmetrise this theory. Interestingly, this will be in complete analogy to super-Chern-Simons theory, but where we deform the supersymmetry transformations of the gauginos to accomodate for the mass term, yielding new supersymmetry transformations which we will coin `massive supersymmetry transformations'. These are somewhat analogous to a central charge but not quite the same. As we will see, these will localise Proca-Chern-Simons theory not on its field equations, but rather on sourced field equations. After having dealt with Proca-Chern-Simons theory we move on to the Third Way Theory. We start off by describing a first attempt by Arvanitakis to localise this theory. This will successfully localise the Third Way Theory on one of its \textit{diagonal} field equations, again sourced by the auxiliary fields. This achieves half a localisation of the Third Way Theory. Instead of trying to localise on the antidiagonal field equations, we take a slightly different approach and try to localise on the $A$ and $\Ab$ field equations \ref{field equation F} and \ref{field equation Fb}. This will be achieved by extending the results of Proca-Chern-Simons theory to the Third Way Theory, resulting in Third Way supersymmetries, which are deformations of standard supersymmetry transformations. These will then be used to localise the Third Way Theory on its sourced field equations.


\section{Superspace Gauge Theory}

In this section, we will for the most part follow Sohnius' ``Introducing supersymmetry'' \cite{sohnius1985introducing}, which instead of dealing with $d = 3+0$, $\NSUSY = 2$ supersymmetry deals with $d = 3+1$, $\NSUSY = 1$ supersymmetry. However, this isn't a bad thing as the former can actually be regarded as a dimensional reduction of the latter. From this viewpoint one of the directions is compactified to a circle and the dependence along this direction is ignored. The four dimensional Dirac spinor Grassmann coordinate in $d = 3+1$ then breaks up into two 2-spinors in $d = 3+0$ resulting in $\NSUSY = 2$ supersymmetry. If one were to consider modes along this circle it would manifest itself as a central charge \cite{sohnius1985introducing}\cite{closset2013supersymmetric}. 

\subsection{The Superconnection}

\subsubsection{Defining the superconnection}

We give a superspace geometric derivation of the field content and supersymmetries of supersymmetric gauge theories. This is achieved by considering the superspace analogue of a connection 1-form. Before proceiding through this section we encourage the reader to take another look at section \ref{Some Basics of Supersymmetry} and appendix \ref{Spinors and Gamma Matrices}. The superfield in consideration here is the superconnection

\begin{equation}
\Ag_A(x,\theta,\thetat) = \big\{\Ag_\mu, ~ \Ag_\alpha, ~ \Agt_\alpha\big\}
\end{equation}

\noindent for which each component is a $\gLie$-valued general complex superfield with overall Lorentz index $\mu$ or $\alpha$.\footnote{Since we're working in flat space here we don't bother to distinguish between flat and curved indices in our notation.} $\Ag_\mu$ is taken to be Grassmann-even and $\Ag_\alpha$, $\Agt_\alpha$ are taken to be Grassmann-odd. Under so-called \textit{super}gauge transformations this superconnection transforms as

\begin{equation}
\Ag_A \to e^X\big(\Ag_A + D_A\big)e^{-X},
\end{equation}

\noindent with $X$ a general $\gLie$-valued superfield and $D_A$ the covariant superspace derivatives introduced in equation \ref{covariant superspace derivatives}. Infinitesimally this gives

\begin{equation}
\delta_X\Ag_A = -\nabla_AX := -D_AX - [\Ag_A,X\}
\end{equation}

\noindent in analogy to connection 1-forms as introduced for non-supersymmetric theories. The bracket $[\bullet,\bullet\}$ denotes the $\mathbb{Z}_2$-graded commutator given by
\begin{align}
[A,B\} &:= AB - (-)^{|A||B|}BA
&
&\Leftrightarrow
&
[A,B\}^I &:= f_{JK}{^I}A^JB^K
\end{align}

\noindent which satisfies $\mathbb{Z}_2$-graded Jacobi identities

\begin{equation}
[A,[B,C\}\} = [[A,B\},C\} + (-)^{|A||B|}[B,[A,C\}\}.
\end{equation}

\noindent Here $|\bullet|$ is used to denote the Grassmann parity \cite{sohnius1985introducing}.

\subsubsection{Defining the superfield strength}

We now move on to describe the superanalogue of the field strength tensor. We define the superfield strength tensor $\Fg_{AB}$ along with the supertorsion tensor $\Tg_{AB}{^C}$ through the relations

\begin{align}
\label{super Bianchi identities}
[D_A,D_B\} &=: \Tg_{AB}{^C}D_C
\\
[\nabla_A,\nabla_B\} &=: \Tg_{AB}{^C}\nabla_C + \Fg_{AB}
\end{align}

\noindent From this definition and equation \ref{covariant superspace derivatives algebra} one can immediately derive that the supertorsion tensor is given by its only non-vanishing component

\begin{align}
\Tgo_{\alpha\beta}{^\mu} &= 2i\gamma^\mu{_{\alpha\beta}},
&
\{D_\alpha,\Dt_\beta\} &= \Tgo_{\alpha\beta}{^\mu}\del_\mu.
\end{align}

\noindent We further note that an explicit expression for the superfield strength tensor can be derived from working out
\begin{multline}
[\nabla_A,\nabla_B\}
= \big[D_A + \Ag_A, ~ D_B + \Ag_B\big\}
= \Tg_{AB}{^C}D_C + 2D_{[A}\Ag_{B\}} + [\Ag_A,\Ag_B\}
\\
= \Tg_{AB}{^C}\nabla_C + \Big(2D_{[A}\Ag_{B\}} + [\Ag_A,\Ag_B\} - \Tg_{AB}{^C}\Ag_C\Big).
\end{multline}

\noindent where $_{[\dots\}}$ denotes a $\mathbb{Z}_2$-graded antisymmetrisation, with vector indices regarded as `even' and spinor indices regarded as `odd'.\footnote{$\mu\leftrightarrow\nu$ and $\mu\leftrightarrow\alpha$ are antisymmetric and $\alpha\leftrightarrow\beta$ is symmetric} This gives us an explicit expression

\begin{equation}
\Fg_{AB} = 2D_{[A}\Ag_{B\}} + [\Ag_A,\Ag_B\} - \Tg_{AB}{^C}\Ag_C
\end{equation}

\noindent for the superfield strength tensor \cite{sohnius1985introducing}. The superconnection satisfies Bianchi identities
\begin{align}
[[\nabla_{[A},[\nabla_B,\nabla_{C\}}\}\} &\equiv 0
&
&\Leftrightarrow
&
\nabla_{[A}\Fg_{BC\}} - \Tg_{[AB|}{^D}\Fg_{D|C\}} &\equiv 0.
\end{align}

\subsection{Constraints}

We recall from the case of the chiral and antichiral multiplets that a general superfield is not an irreducible representation of supersymmetry. What we do then is to impose constraints on these superfields which preferably (anti)commute with supersymmetry to reduce the field content. For the vector multiplet we will introduce two kinds of constraints: The so-called conventional constraints and the representation preserving constraints \cite{sohnius1985introducing}.
Before moving on to describe these two kinds of constraints we quickly introduce the necessary notation for the field strength components:

\begin{align}
[\nabla_\mu,\nabla_\nu] &= \Fg_{\mu\nu}
&
\{\nabla_\alpha,\nabla_\beta\} &= \Fg_{\alpha\beta}
\Big.\\
[\nabla_\mu,\nabla_\alpha] &= \Fg_{\mu\alpha}
&
\{\nabla_\alpha,\nablat_\beta\} &= 2i\nablas_{\alpha\beta} + \Fgo_{\alpha\beta}
\Big.\\
[\nabla_\mu,\nablat_\alpha] &= \Fgt_{\mu\alpha}
&
\{\nablat_\alpha,\nablat_\beta\} &= \Fgt_{\alpha\beta}
\Big.
\end{align}

\subsubsection{Conventional constraints}

The conventional constraint arises from the fact that a connection 1-form can be arbitrarily shifted by an adjoint tensor of the gauge group without ceasing to be a connection 1-form. We thus have the freedom to redefine the superconnection as

\begin{align}
\Ag_A &\xrightarrow{\text{redef}} \Ag_A + \Xg_A
&
&
\begin{cases}
\Ag_A \to e^X(\Ag_A + D_A)e^{-X}
\\
\Xg_A \to e^X\Xg_Ae^{-X}
\end{cases}
\end{align}

\noindent From this it follows that the superfield strength transforms as

\begin{equation}
\Fg_{AB}
\xrightarrow{\text{redef}} \Fg_{AB} + 2\nabla_{[A}\Xg_{B\}} + [\Xg_A,\Xg_B\} - \Tg_{AB}{^C}\Xg_C
\end{equation}

\noindent If we now choose
\begin{align}
\Xg_\mu &= \frac{i}{4}\gamma_\mu{^{\alpha\beta}}\Fgo_{\alpha\beta}
&
\Xg_\alpha &= \Xgt_\alpha = 0
\end{align}

\noindent we find that

\begin{equation}
\Fgo_{\alpha\beta}
\xrightarrow{\text{redef}}
\Fgo_{\alpha\beta}
+ \frac{1}{2}\gamma^\mu{_{\alpha\beta}}\gamma_\mu{^{\gamma\delta}}\Fgo_{\gamma\delta}
\overeq{\ref{Lemma symmetric}}
\Fgo_{\alpha\beta} - \Fgo_{(\alpha\beta)}
=
\Fgo_{[\alpha\beta]}.
\end{equation}

\noindent That is, we can generally impose the constraint

\begin{align}
\Fgo_{(\alpha\beta)} &= 0
&
&\Leftrightarrow
&
\Fgo_{\alpha\beta} &= \Fgo_{[\alpha\beta]} =: \varepsilon_{\alpha\beta}\Fgo.
\end{align}

\noindent This constraint which can always be imposed is referred to as the \textit{conventional constraint} \cite{sohnius1985introducing}. A direct consequence of this is that we can express $\Ag_\mu$ in terms of $\Ag_\alpha$ and $\Agt_\alpha$. Indeed, we see that

\begin{equation}
0 = \Fgo_{(\alpha\beta)} = D_\alpha\Agt_\beta + \Dt_\beta\Ag_\alpha + \{\Ag_\alpha,\Agt_\beta\} - 2i\gamma^\mu{_{\alpha\beta}}\Ag_\mu
\end{equation}

\noindent Again making use of equation \ref{Lemma symmetric} we find that

\begin{equation}
\Ag_\mu = \frac{i}{4}\gamma_\mu{^{\alpha\beta}}\Big(D_\alpha\Agt_\beta + \Dt_\beta\Ag_\alpha + \{\Ag_\alpha,\Agt_\beta\}\Big).
\end{equation}

\subsubsection{Representation preserving constraint}

Now we move on to the second kind of constraint. These constraints are motivated by integrability conditions for so-called gauge-chiral and gauge-antichiral superfields $\Phi$ and $\Phit$ which are constrained by
\begin{align}
\nabla_\alpha\Phi &= 0,
&
\nablat_\alpha\Phit &= 0.
\end{align}

\noindent The representation preserving constraints are now given by
\begin{align}
\Fg_{\alpha\beta} &= 0,
&
\Fgt_{\alpha\beta} &= 0.
\end{align}

\noindent Indeed, if these weren't vanishing and one would take covariant derivatives of the equations above arive at conditions which overconstrain the gauge-chiral and gauge-antichiral multiplets $\Phi$ and $\Phit$ \cite{sohnius1985introducing}.



\subsection{Bianchi Identities}

We will now go over Bianchi identities as they were described in equation \ref{super Bianchi identities}. These will result in some new information because the constraints we imposed weren't taking account in any way with the superconnection origins of the superfield strengths \cite{nishino1993chern}.

\begin{itemize}
\item $\nabla_\alpha\nabla_\beta\nablat_\gamma$ and $\nabla_\alpha\nablat_\beta\nablat_\gamma$ Bianchi identities: These respectively yield
\begin{align}
\label{Bianchi sst}
\nabla_{(\alpha}\Fgo\varepsilon_{\beta)\gamma} &= +2i\Fg_{\mu(\alpha}\gamma^\mu{_{\beta)\gamma}}
&
&\Longrightarrow
&
\nabla_\alpha\Fgo &= +\frac{2i}{3}\gamma^\mu{_\alpha}{^\beta}\Fg_{\mu\beta}
\\
\label{Bianchi stt}
\nablat_{(\alpha}\Fgo\varepsilon_{\beta)\gamma} &= -2i\Fgt_{\mu(\alpha}\gamma^\mu{_{\beta)\gamma}}
&
&\Longrightarrow
&
\nablat_\alpha\Fgo &= -\frac{2i}{3}\gamma^\mu{_\alpha}{^\beta}\Fgt_{\mu\beta}
\end{align}
Using these equations as well as the Fierz identity \ref{Fierz identity} one then finds explicit expressions
\begin{align}
\label{F sv}
\Fg_{\alpha,\beta\gamma} 
&:=
\Fg_{\alpha\mu}\gamma^\mu{_{\beta\gamma}}
=
+\frac{3i}{4}\varepsilon_{\alpha\beta}\nabla_\gamma\Fgo + \frac{i}{2}\nabla_{(\alpha}\Fgo\varepsilon_{\beta)\gamma}
\\
\label{Ft sv}
\Fgt_{\alpha,\beta\gamma} 
&:=
\Fgt_{\alpha\mu}\gamma^\mu{_{\beta\gamma}}
=
-\frac{3i}{4}\varepsilon_{\alpha\beta}\nablat_\gamma\Fgo - \frac{i}{2}\nablat_{(\alpha}\Fgo\varepsilon_{\beta)\gamma}
\end{align}
\item $\nabla_\mu\nabla_\alpha\nabla_\beta$ and $\nabla_\mu\nablat_\alpha\nablat_\beta$ Bianchi identities: These respectively yield
\begin{align}
\nabla_{(\alpha}\Fg_{\beta)\mu} &= 0
&
&\overset{\ref{Bianchi sst}}{\Longrightarrow}
&
\nabla^\alpha\nabla_\alpha\Fgo &= 0
\\
\nablat_{(\alpha}\Fgt_{\beta)\mu} &= 0
&
&\overset{\ref{Bianchi stt}}{\Longrightarrow}
&
\nablat{^\alpha}\nablat_\alpha\Fgo &= 0
\end{align}
\item $\nabla_\mu\nabla_\alpha\nablat_\beta$ Bianchi identity: This identity yields
\begin{equation}
2i\gamma^\nu{_{\alpha\beta}}\Fg_{\mu\nu} + \varepsilon_{\alpha\beta}\nabla_\mu\Fgo
=
\nablat_\beta\Fg_{\mu\alpha} + \nabla_\alpha\Fgt_{\mu\beta}
\end{equation}
Tracing over $_{\alpha\beta}$ and contracting with the Levi-Civita tensor then yields
\begin{gather}
\label{tussenstap 1}
4i\Fg_{\mu\nu} = \nablat{^\alpha}\gamma_{\nu\alpha}{^\beta}\Fg_{\mu\beta} + \nabla{^\alpha}\gamma_{\nu\alpha}{^\beta}\Fgt_{\mu\beta}
=: \nablat\gamma_\nu\Fg_\mu + \nabla\gamma_\nu\Fgt_\mu
\Big.\\
\Updownarrow\nonumber
\Big.\\
\label{tussenstap 2}
8i{\ast}\Fg^\mu = \epsilon^{\mu\rho\sigma}\Big(\nablat\gamma_\sigma\Fg_\rho + \nabla\gamma_\sigma\Fgt_\rho\Big).
\Big.
\end{gather}
\noindent Contracting \ref{tussenstap 2} with $\gamma^\mu{_{\alpha\beta}}$ and using equation \ref{Lemma antisymmetric} as well as expressions \ref{F sv} and \ref{Ft sv} finally results in
\begin{align}
{\ast}\Fgs_{\alpha\beta} &= \frac{i}{16}[\nablat_{(\alpha},\nabla_{\beta)}]\Fgo
&
&\overset{\text{\ref{Lemma symmetric}}}{\Leftrightarrow}
&
{\ast}\Fg^\mu &= -\frac{i}{32}\gamma^{\mu\alpha\beta}[\nablat_\alpha,\nabla_\beta]\Fgo.
\end{align}
On the other hand, contracting $_{\mu\nu}$ in \ref{tussenstap 1} and using previous Bianchi identities \ref{Bianchi sst} and \ref{Bianchi stt} results in
\begin{equation}
\big(\nabla^\alpha\nablat_\alpha - \nablat{^\alpha}\nabla_\alpha\big)\Fgo = 0.
\end{equation}
\end{itemize}

\noindent These sum up the most interesting constraints placed on $\Fgo$ by using the Bianchi identities.

\subsection{Solving the Constraints}

We now wish to formulate an explicit solution to the constraints placed on the connection. To this end we recall that in the case of gauge theories a flat connection ---that is, a vanishing field strength--- corresponds to what's called a `pure gauge' connection 1-form
\begin{align}
F &= 0
&
&\Longleftrightarrow
&
\exists \theta : A &= e^{-\theta}\extd e^\theta
\end{align}

\noindent which can obviously always be gauged away by choosing a gauge transformation with parameter $\theta$ \cite{nakahara2018geometry}. In the spirit of this fact we choose to write the spinorial superconnection components $\Ag_\alpha$ and $\Agt_\alpha$ as

\begin{align}
\Fg_{\alpha\beta} &= 0
&
&\Longrightarrow
&
\Ag_\alpha &= e^{-2\Vg}D_\alpha e^{2\Vg}
\\
\Fgt_{\alpha\beta} &= 0
&
&\Longrightarrow
&
\Agt_\alpha &= e^{-2\Vgtt}\Dt_\alpha e^{2\Vgtt}
\end{align}

\noindent which is referred to as a \textit{spinorially flat connection}. $\Vg$ and $\Vgt$ are $\gLie$-valued superfields which we refer to as the \textit{prepotentials} \cite{sohnius1985introducing}\cite{nishino1993chern}. One could now wonder whether this means that the superconnection is flat and thus contains no physical degrees of freedom. However it isn't since $\Vg$ and $\Vgt$ aren't generally the same and thus can't be generally gauged away simultaneously.

\subsubsection{Pregauge and supergauge transformations}

Now we move on to describe how gauge symmetries work in this propotential formalism, and particularly how the prepotentials transform under gauge transformations. Due to the graded Leibniz property of the spinorial covariant derivatives we find that the prepotentials transform under supergauge transformations as
\begin{align}
e^{2\Vg} &\to e^{2\Vg}e^{-X}
&
e^{2\Vgtt} &\to e^{2\Vgtt}e^{-X}
\end{align}

\noindent More concretely using the BCH formulas outlined in the appendix section \ref{Matrix Exponentiation and BCH Formula} we find that
\begin{align}
\label{chiral residual gauge symmetries}
\Vg &\to \frac{1}{2}(2\Vg)\star(-X)
&
\Vgt &\to \frac{1}{2}(2\Vgt)\star(-X)
\end{align}

\noindent as defined in equation \ref{BCH formula} of the appendix.\footnote{The five-pointed star $\star$ of the BCH formula is not to be confused with the six-pointed star $\ast$ of the Hodge operator.} Following equation \ref{BCH formula infinitesimal} we find infinitesimal supergauge transformations

\begin{align}
\delta_X\Vg &= \frac{-\ad\Vg}{1-e^{-2\ad\Vg}}X,
&
\delta_X\Vgt &= \frac{-\ad\Vgt}{1-e^{-2\ad\Vgtt}}X.
\end{align}

\noindent Another kind of `gauge' symmetry is one which arises as a redundancy of the way we write down the spinorially flat solution to the constraints we imposed. Namely, we note that the spinorial superconnections $\Ag_\alpha$ and $\Agt_\alpha$ are left invariant under transformations
\begin{align}
e^{2\Vg} &\to e^\Lambdatt e^{2\Vg}
&
&\Leftrightarrow
&
\Vg &\to \frac{1}{2}\big(\Lambdat\star(2\Vg)\big)
\\
e^{2\Vgtt} &\to e^\Lambda e^{2\Vgtt}
&
&\Leftrightarrow
&
\Vg &\to \frac{1}{2}\big(\Lambda\star(2\Vgt)\big)
\end{align}

\noindent where $\Lambda$ and $\Lambdat$ are respectively taken to be $\gLie$-valued chiral and antichiral superfields. These transformations are referred to as \textit{pregauge transformations}. Similarly one finds following the results of equation \ref{BCH formula infinitesimal} that the pregauge transformations are infinitesimally given by
\begin{align}
\delta_\Lambdatt\Vg &= \frac{-\ad\Vg}{1-e^{2\ad\Vg}}\Lambdat,
&
\delta_\Lambda\Vgt &= \frac{-\ad\Vgt}{1-e^{2\ad\Vgtt}}\Lambda.
\end{align}

\noindent In summary, we find that:

\begin{tcolorbox}

\subsubsection{Gauge transformations in the prepotential formalism}

The prepotentials $\Vg$ and $\Vgt$ contain gauge transformations
\begin{align}
e^{2\Vg} &\to e^\Lambdatt e^{2\Vg} e^{-X}
&
&\Longleftrightarrow
&
\Vg &\to \frac{1}{2}\big(\Lambdat\star(2\Vg)\star (-X)\big)
\\
e^{2\Vgtt} &\to e^\Lambda e^{2\Vgt} e^{-X}
&
&\Longleftrightarrow
&
\Vgt &\to \frac{1}{2}\big(\Lambda\star(2\Vgt)\star(-X)\big)
\end{align}

\noindent infinitesimally given by
\begin{align}
\delta_{X,\Lambdatt}\Vg &= \frac{-\ad\Vg}{1-e^{-2\ad\Vg}}X + \frac{-\ad\Vg}{1-e^{2\ad\Vg}}\Lambdat
\\
\delta_{X,\Lambda}\Vgt &= \frac{-\ad\Vgt}{1-e^{-2\ad\Vgtt}}X + \frac{-\ad\Vgt}{1-e^{2\ad\Vgtt}}\Lambda
\end{align}

\noindent where $X$ is a $\gLie$-valued general superfield parameter for the supergauge transformations and $\Lambdat$ and $\Lambda$ are respectively $\gLie$-valued antichiral and chiral superfield parameters of the pregauge transformations.

\end{tcolorbox}

\subsection{Wess-Zumino Gauge}

We now move on to make a supergauge choice. This kind of gauge fixing is different from the kind we're used to in gauge theory. Usually when we think of gauge fixing this corresponds to imposing constraints on the field components of the connection 1-form. However, in the case of supergauge fixing the components which will be constrained will be the component fields of the supermultiplet.

\subsubsection{The chiral representation}

\noindent We start off with a gauge consition which is generally preserved by supersymmetry: We note that by supergauge transformations we have enough freedom to set one of the spinorial gauge superfields to zero. Choosing $X = 2\Vgt$ we set
\begin{align}
\Agt_\alpha &= 0
&
&\Longrightarrow
&
\nablat_\alpha &= \Dt_\alpha,
&
\Ag_\mu &= -\frac{i}{4}\Dt\gamma_\mu\Ag.
\end{align}
\noindent However taking into account the pregauge transformation freedom of $\Vgt$ we see that this doesn't fully use all the freedom of $X$. We're left with the freedom to take $X$ to be chiral. That is, the residual gauge symmetries of $\Vg$ are

\begin{align}
\label{chiral rep gauge transformations}
e^{2\Vg} &\to e^\Lambdatt e^{2\Vg}e^\Lambda
&
&\Longleftrightarrow
&
\Vg &\to \frac{1}{2}\big(\Lambdat\star(2\Vg)\star\Lambda\big)
\end{align}

\noindent where $\Lambda$ and $\Lambdat$ are taken to be respectively $\gLie$-valued chiral and antichiral superfields. This is the so-called chiral representation.

\subsubsection{Wess-Zumino gauge}

As it turns out, these residual supergauge transformations are in their turn enough to set to zero the chiral and the antichiral components of $\Vg$. It follows then that we are left with

\begin{equation}
\label{Wess-Zumino gauge}
\Vg = -\theta(i\As+\sigma)\thetat + i\lambda\theta\thetat{^2} + i\lambdat\thetat\theta^2 - \frac{1}{2}D\theta^2\thetat{^2}.
\end{equation}

\noindent We hence have the following field content (which are all $\gLie$-valued):
\begin{itemize}
\item A real 1-form $A_\mu$ of dimension 1
\item A real scalar $\sigma$ of dimension 1
\item Two \textit{independent} complex spinors $\lambda$ and $\lambdat$ of dimensions $\frac{3}{2}$
\item A real scalar $D$ of dimension 2
\end{itemize}


\noindent One could now wonder wether taking the bosonic field content to be real is consistent with the supersymmetry transformations. This issue was addressed in \cite{Pestun_2012}\cite{K_ll_n_2011}. There the action is understood as analytically continued to the space of complexified fields. The path integral is then understood as integrating a holomorphic functional over a half-dimensional contour in complexified field space.

\subsection{Residual Gauge Symmetries}

We have now derived the $\gLie$-valued component fields for the vector multiplet. However, we haven't yet derived if and how their transformational properties relate to those of non-supersymmetric gauge theories.
As it turns out, these are related to the residual gauge symmetries of the Wess-Zumino gauge, which only partially fixes the gauge \cite{sohnius1985introducing}\cite{DavidTongSUSYFT}. We note that the choices of $\Lambda$ and $\Lambdat$ in equation \ref{chiral rep gauge transformations} which preserve the Wess-Zumino gauge \ref{Wess-Zumino gauge} are parametrised by a real $\gLie$-valued local parameter $\vartheta$ (\textit{not} a superfield) and given by
\begin{align}
\Lambda_\res(\vartheta) &= e^{+i\thetatt\dels\theta}(+\vartheta)
= +\vartheta - i\theta(\dels\vartheta)\thetat - \frac{1}{4}\square\vartheta\theta^2\thetat{^2}
\\
\Lambdat_\res(\vartheta) &= e^{-i\thetatt\dels\theta}(-\vartheta) = -\vartheta - i\theta(\dels\vartheta)\thetat + \frac{1}{4}\square\vartheta\theta^2\thetat{^2}
\end{align}
\noindent One finds that the prepotential $\Vg$ transforms under these residual gauge transformations as
\begin{equation}
\begin{split}
\delta_\vartheta\Vg
&= \frac{\ad\Vg}{1-e^{-2\ad\Vg}}\Lambda_\res(\vartheta) + \frac{-\ad\Vg}{1-e^{2\ad\Vg}}\Lambdat_\res(\vartheta)
\\
&\overeq{\ref{BCH infinitesimal expansion}} -\frac{1}{2}\big(\Lambdat_\res(\vartheta)+\Lambda_\res(\vartheta)\big) + \frac{1}{2}\big[\Vg, ~ \Lambdat_\res(\vartheta) - \Lambda_\res(\vartheta)\big]
\end{split}
\end{equation}

\noindent where higher order terms vanish identically due to the anticommuting nature of Grassmann numbers. Working this out we find that
\begin{equation}
\begin{split}
\delta_\vartheta\Vg
&= -\theta\Big(-i(\dels+\ad\As)\vartheta + [\vartheta,\sigma]\Big)\thetat + i[\vartheta,\lambda]\theta\thetat{^2} + i[\vartheta,\lambdat]\thetat\theta^2 - \frac{1}{2}[\vartheta,D]\theta^2\thetat{^2}
\\
&=: -\theta(i\delta_\vartheta\As + \delta_\vartheta\sigma)\thetat + i\delta_\vartheta\lambda\theta\thetat{^2} + i\delta_\vartheta\lambdat\thetat\theta^2 - \frac{1}{2}\delta_\vartheta D\theta^2\thetat{^2}
\end{split}
\end{equation}

\noindent In conclusion, we find that

\begin{tcolorbox}

\subsubsection{Residual gauge symmetries of the Wess-Zumino gauge}

\begin{align}
\delta_\vartheta A_\mu &= -\Dg_\mu\vartheta
&
\delta_\vartheta\sigma &= [\vartheta,\sigma]
&
\delta_\vartheta D &= [\vartheta,D]
\Big.\\
\delta_\vartheta\lambda &= [\vartheta,\lambda]
&
\delta_\vartheta\lambdat &= [\vartheta,\lambdat]
\Big.
\end{align}

\noindent That is, the residual gauge transformations parametrised by $\vartheta$ constitute ordinary gauge transformations, for which $A_\mu$ is a $G$-connection 1-form and $\sigma$, $\lambda$, $\lambdat$ and $D$ lie in the adjoint reperesentation of $G$.

\end{tcolorbox}

\subsection{Supersymmetries in the WZ Gauge}

We now move on to discuss supersymmetry transformations in the Wess-Zumino gauge. These are generated in the same way as for the chiral multiplet, only now we have to include compensating gauge transformations which ensure the Wess-Zumino gauge is upheld. These compensating gauge transfromations can themselves be taken to be residual gauge transformations \ref{chiral residual gauge symmetries} of the chiral representation since this condition is supersymmetric. Hence, we find that supersymmetry transformations are covariantly taken to be of the form

\begin{align}
\delta_{\zeta,\zetatt}\Vg &= \zeta Q + \zetat\Qt + \delta_{\Lambda,\Lambdatt},
&
\Lambda &= \Lambda(\zeta,\zetat,\Vg),
&
\Lambdat &= \Lambdat(\zeta,\zetat,\Vg)
\end{align}

\noindent where the gauge symmetry $\delta_{\Lambda,\Lambdatt}$ is included to ensure the preservation of the Wess-Zumino gauge.

\subsubsection{Compensating gauge transformations}

We start off by working out the compensating gauge transformations of the Wess-Zumino gauge. To this end we start off by noting that the out of gauge terms are given by
\begin{align}
(\zeta Q)\Vg\big|_{\text{out of gauge}} &= -\zeta(+i\As+\sigma)\thetat + i(\zeta\lambda)\thetat{^2}
\\
(\zetat\Qt)\Vg\big|_{\text{out of gauge}} &= -\zetat(-i\As+\sigma)\theta + i(\zetat\lambdat)\theta^2
\end{align}

\noindent From this we can read off that the parameters for the compensating gauge transformations are given by
\begin{align}
\begin{aligned}
\Lambdat &= 2e^{+i\thetatt\dels\theta}\Big[-\zeta(+i\As+\sigma)\thetat + i(\zeta\lambda)\thetat{^2}\Big]
\Big.\\
&= -2\zeta(+i\As+\sigma)\thetat + 2i(\zeta\lambda)\thetat{^2} + i\zeta(+i\As+\sigma)\delsl\theta\thetat{^2}
\Big.
\end{aligned}
\\
\begin{aligned}
\Lambda &= 2e^{-i\thetatt\dels\theta}\Big[-\zetat(-i\As+\sigma)\theta + i(\zetat\lambdat)\theta^2\Big]
\Big.\\
&= -2\zetat(-i\As+\sigma)\theta + 2i(\zetat\lambdat)\theta^2 + i\zetat(-i\As+\sigma)\delsl\thetat\theta^2
\Big.
\end{aligned}
\end{align}

\noindent In analogy to the case of the residual gauge symmetries we find that the compensating gauge transformations simplify to

\begin{equation}
\delta_{\Lambda,\Lambdatt}\Vg = -\frac{1}{2}\big(\Lambda+\Lambdat\big) + \frac{1}{2}\big[\Lambda-\Lambdat, ~ \Vg\big]
\end{equation}

\noindent again due to the anticommuting nature of Grassmann numbers.

\subsubsection{Computing the SUSY transformations}

We are now ready to compute the gauge covariant supersymmetry transformations. Inserting the aforementioned results we find (after some tedious computations) that
\begin{equation}
\begin{split}
\delta_{\zeta,\zetatt}\Vg
&= \big(\zeta Q + \zetat\Qt + \delta_{\Lambda,\Lambdatt}\big)\Vg
\\
&= 
\big(\zeta Q - \frac{1}{2}\ad\Lambdat\big)\Vg - \frac{1}{2}\Lambdat
+ 
\big(\zetat\Qt + \frac{1}{2}\ad\Lambda\big)\Vg - \frac{1}{2}\Lambda
\\
&=
\begin{aligned}[t]
&+ \theta\bigg[i\big(-\zeta\gamma_\mu\lambdat + \zetat\gamma_\mu\lambda\big) + \big(i\zeta\lambdat + i\zetat\lambda\big)\bigg]\thetat
\\
&+ i\zeta\bigg[iD - \frac{1}{2}\epsilon^{\mu\nu\rho}F_{\mu\nu}\gamma_\rho + \Dgs\sigma\bigg]\theta\thetat{^2}
+ i\zetat\bigg[iD + \frac{1}{2}\epsilon^{\mu\nu\rho}F_{\mu\nu}\gamma_\rho + \Dgs\sigma\bigg]\thetat\theta^2
\\
&- \frac{i}{2}\zeta\big(i\Dgs + \ad\sigma\big)\lambdat - \frac{i}{2}\zetat\big(i\Dgs - \ad\sigma\big)\lambda
\end{aligned}
\\
&=: +\theta\Big(i\gamma^\mu\delta_{\zeta,\zetatt}A_\mu + \delta_{\zeta,\zetatt}\sigma\Big)\thetat + i\delta_{\zeta,\zetatt}\lambda\theta\thetat{^2} + i\delta_{\zeta,\zetatt}\lambdat\thetat\theta^2 - \frac{1}{2}\delta_{\zeta,\zetatt}D\theta^2\thetat{^2}
\end{split}
\end{equation}

\noindent From this, we conclude:

\begin{tcolorbox}

\subsubsection{Supersymmetry transformations of the WZ gauge vector multiplet}

\begin{align}
\label{vector multiplet SUSY A}
\delta_{\zeta,\zetatt} A_\mu &= -\zeta\gamma_\mu\lambdat + \zetat\gamma_\mu\lambda
\bigg.\\
\delta_{\zeta,\zetatt}\sigma &= i\zeta\lambdat + i\zetat\lambda
\bigg.\\
\delta_{\zeta,\zetatt}\lambda &= \zeta\bigg[iD + \frac{1}{2}\epsilon^{\mu\nu\rho}F_{\mu\nu}\gamma_\rho + \Dgs\sigma\bigg]
\bigg.\\
\delta_{\zeta,\zetatt}\lambdat &= \zetat\bigg[iD - \frac{1}{2}\epsilon^{\mu\nu\rho}F_{\mu\nu}\gamma_\rho + \Dgs\sigma\bigg]
\bigg.\\
\label{vector multiplet SUSY D}
\delta_{\zeta,\zetatt} D &= i\zeta\big(i\Dgs + \ad\sigma\big)\lambdat + i\zetat\big(i\Dgs - \ad\sigma\big)\lambda
\bigg.
\end{align}

\noindent the gauge covariantised supersymmetry transformations, where henceforth we shall denote

\begin{equation}
\delta_{\zeta,\zetatt} = \delta_\zeta + \deltat_\zetatt
\end{equation}

\noindent as this will become convenient notation for localisations.

\end{tcolorbox} 

\subsection{Commentary on the Algebra}
\label{Commentary on the Algebra}

Let us comment a bit on this algebra. We recall that the supersymmetry algebra is characterised by the anticommutation relation
\begin{align}
\{\delta_\zeta,\deltat_\zetatt\} &= -2i(\zeta\gamma^\mu\zetat)\del_\mu
&
&\Leftrightarrow
&
\text{SUSY}^2 &= \text{transl}
\end{align}

\noindent That is, supersymmetry transformations square to translations. However, this doesn't take into account the fact that the Wess-Zumino gauge is not supersymmetric and has to be complemented by gauge transformations to stay in this gauge. Taking this into account the algebra takes on the form
\begin{gather}
\label{algebra WZ gauge vector multipley}
\{\delta_\zeta,\deltat_\zetatt\} = -2iK^\mu\del_\mu + \delta_{\text{gauge}}(-2iK^\mu A_\mu - 2\zeta\zetat\sigma)
\\
\Updownarrow\nonumber
\\
\text{SUSY}^2 = \text{transl} + \text{gauge}
\end{gather}

\noindent where we defined the Killing vector $K^\mu = \zeta\gamma^\mu\zetat$. That is, the algebra now becomes accompanied by a field dependent gauge transformation. Indeed, a direct computation yields generalisations of the nilpotency

\begin{equation}
\{\delta_\zeta,\delta_\eta\} = \{\deltat_\zetatt,\deltat_\etatt\} = 0
\end{equation}

\noindent as well as non-trivial relations
\begin{align}
\{\delta_\zeta,\deltat_\zetatt\}A_\mu &= -2iK^\nu F_{\nu\mu} + 2\zeta\zetat\Dg_\mu\sigma
\\
\{\delta_\zeta,\deltat_\zetatt\}\sigma &= -2iK^\mu\Dg_\mu\sigma
\\
\{\delta_\zeta,\deltat_\zetatt\}\lambda^\alpha &= -2iK^\mu\Dg_\mu\lambda^\alpha - 2\zeta\zetat[\sigma,\lambda^\alpha]
\\
\{\delta_\zeta,\deltat_\zetatt\}\lambdat{^\alpha} &= -2iK^\mu\Dg_\mu\lambdat{^\alpha} - 2\zeta\zetat[\sigma,\lambdat{^\alpha}]
\\
\{\delta_\zeta,\deltat_\zetatt\}D &= -2iK^\mu\Dg_\mu D - 2\zeta\zetat[\sigma,D]
\end{align}

\noindent in agreement with the algebra \ref{algebra WZ gauge vector multipley}.



\subsubsection{An aside on odd derivatives}

A source of some confusion could the question of whether to take the (Grassmann-odd!) supersymmetry transformation $\delta$ to be a left or a right derivative. As it turns out despite us using left derivatives it are right derivatives which are more fundamental \cite{ferrara2003supersymmetry}. However, the difference will be at most an overall minus sign if the object acted on isn't the sum of objects with differing Grassmann parities. Hence, for our purposes the distinction won't be relevant. A way to translate between left and right acting derivatives is through
\begin{align}
(\text{even})\deltal &= +\delta(\text{even}),
&
(\text{odd})\deltal &= -\delta(\text{odd}).
\end{align}

\section{Supersymmetric Actions and Localisation}

In this section we'll go over two kinds of supersymmetric gauge theories: super-Yang-Mills theory and super-Chern-Simons theory. Particularly, as it turns out, it is in super-Yang-Mills theory that we find the clue to localising supersymmetric gauge theories.

\subsection{Super-Yang-Mills Theory}

We start off by giving a description of super-Yang-Mills theory. We start off by noting a very interesting and relevant fact to this thesis. The action of super-Yang-Mills theory is $Q$-exact \cite{willett2017localization}\cite{kapustin2010exact}! The action of super-Yang-Mills theory is given by

\begin{equation}
\label{SYM action}
S_\SYM[\Vg]
=
\frac{1}{g^2}\int\extd^3x ~ \tr\bigg[-\frac{1}{4}F_{\mu\nu}F^{\mu\nu} - \frac{1}{2}\Dg_\mu\sigma\Dg^\mu\sigma - \frac{1}{2}D^2
- \lambdat\big(i\Dgs - \ad\sigma\big)\lambda\bigg]
\end{equation}

\noindent where $g^2$ a constant of mass dimension 1 and where we used $\Vg$ to denote the collective field content of the vector multiplet in WZ gauge. Showing this action is supersymmetric will be done through its $Q$-exactness. Particularly, we note that this action can be written in different ways as

\begin{align}
\label{SYM exact 1}
g^2S_\SYM[\Vg]
&= \frac{1}{2|\zeta|^2}\delta_\zeta\int\extd^3x ~ \tr\Big[(\delta_\zeta\lambda)^\dagger\lambda\Big]
\bigg.\\
\label{SYM exact 2}
&= \frac{1}{2|\zeta|^2}\deltat_\zeta\int\extd^3x ~ \tr\Big[(\deltat_\zeta\lambdat)^\dagger\lambdat\Big]
\bigg.\\
\label{SYM exact 3}
&= \frac{1}{4|\zeta|^2}\Qloc_\zeta\int\extd^3x ~ \tr\bigg[(\Qloc_\zeta\lambda)^\dagger\lambda + (\Qloc_\zeta\lambdat)^\dagger\lambdat\bigg]
\bigg.
\end{align}

\noindent where we defined

\begin{equation}
\Qloc_\zeta := \delta_\zeta + \deltat_\zetatt
\end{equation}

\noindent where now we take the two conjugate SUSY trasformations but with the same non-zero constant bosonic spinor $\zeta$. For conventions regarding the norm $|\bullet|^2$ and Hermitian conjugation as it relates to the index formalism for spinors we refer the reader to section \ref{Hermitian Conjugation} of the appendix. Let us verify this fact a little more explicitly:

\subsubsection{Expressions \ref{SYM exact 1} and \ref{SYM exact 2}}

We start off by taking a look at the first description, that is, expression \ref{SYM exact 1}. We find that

\begin{equation}
g^2S_\SYM[\Vg] = \frac{1}{2|\zeta|^2}\int\extd^3x ~ \tr\bigg[|\delta_\zeta\lambda|^2 + \delta_\zeta(\delta_\zeta\lambda)^\dagger\lambda\bigg]
\end{equation}

\noindent This gives us respectively bosonic and fermionic sectors
\begin{align}
g^2S_\SYM[\Vg]\big|_\bos &= \frac{1}{2|\zeta|^2}\int\extd^3x ~ \tr|\delta_\zeta\lambda|^2,
\\
g^2S_\SYM[\Vg]\big|_\fer &= \frac{1}{2|\zeta|^2}\int\extd^3x ~ \tr\delta_\zeta(\delta_\zeta\lambda)^\dagger\lambda
\end{align}

\noindent Working out the bosonic part in some closer detail will be important since due to it's exactness it will function as a localising term for which the bosonic part will be responsible for the Gaussian damping central to localisations. Noting that
\begin{align}
\delta_\zeta\lambda &= \Big[iD - \frac{1}{2}\epsilon^{\mu\nu\rho}F_{\mu\nu}\gamma_\rho - \Dgs\sigma\Big]\zeta
\\
(\delta_\zeta\lambda)^\dagger &= \zetad\Big[iD + \frac{1}{2}\epsilon^{\mu\nu\rho}F_{\mu\nu}\gamma_\rho + \Dgs\sigma\Big]
\end{align} 

\noindent we find that

\begin{equation}
\label{computation partial localisation}
\begin{split}
\tr|\delta_\zeta\lambda|^2 
&= \tr\zetad\bigg[-D^2-\Big(\frac{1}{2}\epsilon^{\mu\nu\rho}F_{\mu\nu}\gamma_\rho + \Dgs\sigma\Big)^2\bigg]\zeta
\\
&= |\zeta|^2\tr\bigg[-D^2 - \Big(\frac{1}{2}\epsilon_{\mu\rho\sigma}F^{\rho\sigma} + \Dg_\mu\sigma\Big)\Big(\frac{1}{2}\epsilon^{\mu\rho\sigma}F_{\rho\sigma} + \Dg^\mu\sigma\Big)\bigg]
\\
&= |\zeta|^2\tr\bigg[-D^2 - \frac{1}{2}F_{\mu\nu}F^{\mu\nu} - \Dg_\mu\sigma\Dg^\mu\sigma - \Dg_\mu\big(\epsilon^{\mu\rho\sigma}F_{\rho\sigma}\sigma\big)\bigg]
\end{split}
\end{equation}

\noindent where we used the Bianchi identity $\Dg F \equiv 0$ to turn the final term into a total derivative. This yields a bosonic sector

\begin{equation}
S_\SYM[\Vg]\big|_\bos
= \frac{1}{g^2}\int\extd^3x ~ \tr\bigg[-\frac{1}{4}F_{\mu\nu}F^{\mu\nu} - \frac{1}{2}\Dg_\mu\sigma\Dg^\mu\sigma - \frac{1}{2}D^2\bigg]
\end{equation}

\noindent which is indeed in agreement with super-Yang-Mills action \ref{SYM action}. Doing the fermionic sector is straightforward but tedious and not particularly insightful to any future discussion of the localisation of the Third Way Theory. Hence we will simply give the answer which is

\begin{equation}
\label{SYM action fermionic}
S_\SYM[\Vg]\big|_\fer
= \frac{1}{g^2}\int\extd^3x ~ \tr\bigg[-\lambdat\big(i\Dgs - \ad\sigma\big)\lambda\bigg]
\end{equation}

\noindent also in agreement with the super-Yang-Mills action \ref{SYM action}. This verifies that the expression \ref{SYM exact 1} agrees with the super-Yang-Mills action \ref{SYM action}. The agreement of expresion \ref{SYM exact 2} follows similarly.

\subsubsection{Expression \ref{SYM exact 3}}

\noindent As for the third expression for the super-Yang-Mills action, equation \ref{SYM exact 3}, there are a couple of interesting remarks to make which will become relevant later on when looking at the Third Way Theory. For this expression we find that

\begin{equation}
g^2S_\SYM[\Vg] = g^2S_\SYM[\Vg]\big|_\bos + g^2S_\SYM[\Vg]\big|_\fer
\end{equation}

\noindent with bosonic and fermionic sectors
\begin{align}
g^2S_\SYM[\Vg]\big|_\bos
&=
\frac{1}{4|\zeta|^2}\int\extd^3x ~ \tr\bigg[|\Qloc_\zeta\lambda|^2 + |\Qloc_\zeta\lambdat|^2\bigg]
\\
g^2S_\SYM[\Vg]\big|_\fer
&=
\frac{1}{4|\zeta|^2}\int\extd^3x ~ \tr\bigg[\Qloc_\zeta(\Qloc_\zeta\lambda)^\dagger\lambda + \Qloc_\zeta(\Qloc_\zeta\lambdat)^\dagger\lambdat\bigg]
\end{align}

\noindent In close analogy to the computations for expression \ref{SYM exact 1} we find for the bosonic sector that
\begin{align}
\label{complete localisation SYM 1}
\tr|\Qloc_\zeta\lambda|^2
&= |\zeta|^2\tr\bigg[-D^2 - \Big(\frac{1}{2}\epsilon_{\mu\rho\sigma}F^{\rho\sigma} + \Dg_\mu\sigma\Big)\Big(\frac{1}{2}\epsilon^{\mu\rho\sigma}F_{\rho\sigma} + \Dg^\mu\sigma\Big)\bigg]
\\
\label{complete localisation SYM 2}
\tr|\Qloc_\zeta\lambdat|^2
&= |\zeta|^2\tr\bigg[-D^2 - \Big(\frac{1}{2}\epsilon_{\mu\rho\sigma}F^{\rho\sigma} - \Dg_\mu\sigma\Big)\Big(\frac{1}{2}\epsilon^{\mu\rho\sigma}F_{\rho\sigma} - \Dg^\mu\sigma\Big)\bigg]
\end{align}

\noindent where the reader is encouraged to note the relative minus signs between the two expressions. Because of these relative minus signs one finds that

\begin{equation}
\label{complete localisation SYM 3}
\frac{1}{2}\tr\Big[|\Qloc_\zeta\lambda|^2 + |\Qloc_\zeta\lambdat|^2\Big]
=
\tr\bigg[-D^2 - \frac{1}{2}F_{\mu\nu}F^{\mu\nu} - \Dg_\mu\sigma\Dg^\mu\sigma\bigg]
\end{equation}

\noindent \textit{without invoking the Bianchi identities!} Now, one may wonder why this fact could possibly be useful, since it just yields the same result as previous cases modulo a total derivative. The reason that this is interesting is that the localisation of the Third Way Theory will be attempted through a deformation of the supersymmetry transformations. With these deformations it will be the case that we can't use the Bianchi identities as we did for expressions \ref{SYM exact 1} and \ref{SYM exact 2}, and hence an analogue of expression \ref{SYM exact 3} will be our starting point for localising the Third Way Theory.

To continue our discussion, we find that

\begin{equation}
S_\SYM[\Vg]\big|_\bos
= \frac{1}{g^2}\int\extd^3x ~ \tr\bigg[-\frac{1}{4}F_{\mu\nu}F^{\mu\nu} - \frac{1}{2}\Dg_\mu\sigma\Dg^\mu\sigma - \frac{1}{2}D^2\bigg]
\end{equation}

\noindent nicely in agreement with the super-Yang-Mills action \ref{SYM action} and without having to care about boundary terms. As for the fermionic sector, we start off by noting that
\begin{align}
(\delta_\zeta\lambda)^\dagger 
&= \zetad\Big[iD + \frac{1}{2}\epsilon^{\mu\nu\rho}F_{\mu\nu}\gamma_\rho + \Dgs\sigma\Big]
= \delta_\zetad\lambda
\end{align}

\noindent with in analogy following
\begin{align}
(\delta_\zeta\lambda)^\dagger &= \delta_\zetad\lambda,
&
(\deltat_\zeta\lambdat)^\dagger &= \deltat_\zetad\lambdat,
&
(\Qloc_\zeta\lambda)^\dagger &= \Qloc_\zetad\lambda,
&
(\Qloc_\zeta\lambdat)^\dagger &= \Qloc_\zetad\lambdat.
\end{align}

\noindent We thus find that
\begin{equation}
\begin{split}
&\Qloc_\zeta(\Qloc_\zeta\lambda)^\dagger\lambda + \Qloc_\zeta(\Qloc_\zeta\lambdat)^\dagger\lambdat
\\
&=
(\delta_\zeta + \deltat_\zeta)\delta_\zetad\lambda\lambda + (\delta_\zeta + \deltat_\zeta)\deltat_\zetad\lambdat\lambdat
\\
&=
\big(\delta_\zeta\delta_\zetad\lambda\big)\lambda + \big(\{\deltat_\zeta,\delta_\zetad\}\lambda\big)\lambda
+
\big(\deltat_\zeta\deltat_\zetad\lambdat\big)\lambdat + \big(\{\delta_\zeta,\deltat_\zetad\}\lambdat\big)\lambdat
\\
&= 
\underbrace{
\Big[\delta_\zeta(\delta_\zeta\lambda)^\dagger\lambda + \deltat_\zeta(\deltat_\zeta\lambdat)^\dagger\lambdat\Big]
}_{\text{compare to \ref{SYM exact 1} and \ref{SYM exact 2}}}
+
\underbrace{
\Big[\{\deltat_\zeta,\delta_\zetad\}(\lambda^2) + \{\delta_\zeta,\deltat_\zetad\}(\lambdat{^2})\Big]
}_{\text{transl + gauge}}
\end{split}
\end{equation}

\noindent Recalling the algebra as well as the results from expressions \ref{SYM exact 1} and \ref{SYM exact 2} as well as the fact that $\int\tr(\text{transl + gauge}) = 0$ we find that

\begin{equation}
S_\SYM[\Vg]\big|_\fer
= \frac{1}{g^2}\int\extd^3x ~ \tr\bigg[-\lambdat\big(i\Dgs - \ad\sigma\big)\lambda\bigg]
\end{equation}

\noindent again nicely agreeing with the super-Yang-Mills action \ref{SYM action}.

\subsubsection{Supersymmetry of SYM theory}

It now follows almost trivially that the super-Yang-Mills action is invariant under supersymmetry. Indeed, we find that
\begin{align}
g^2\delta_\zeta S_\SYM &= \frac{1}{2|\zeta|^2}\delta^2_\zeta\int\extd^3x ~ \tr\Big[(\delta_\zeta\lambda)^\dagger\lambda\Big] = 0
\\
g^2\deltat_\zeta S_\SYM &= \frac{1}{2|\zeta|^2}\deltat{^2_\zeta}\int\extd^3x ~ \tr\Big[(\deltat_\zeta\lambdat)^\dagger\lambdat\Big] = 0
\end{align}

\noindent proving that super-Yang-Mills theory is indeed supersymmetric.

\subsection{Super-Chern-Simons Theory}

We now move on to describe the superanalogue of Chern-Simons theory \cite{willett2017localization}\cite{closset2013supersymmetric}\cite{nishino1993chern}\cite{ruiz1996lectures}. Super-Chern-Simons theory at level $k$ is described through an action
\begin{equation}
\label{SCS action}
S^k_\SCS[\Vg] 
= \frac{k}{2\pi}S_\SCS[\Vg]
= \frac{k}{2\pi}S_\CS[A] + \frac{k}{2\pi}\int\extd^3x ~ \tr\Big[\lambda\lambdat - \sigma D\Big].
\end{equation}

\noindent This theory has a very interesting property which will become relevant when studying the Third Way Theory: Chern-Simons theory and its superanalogue are physically equivalent! Indeed, the additional field content $\sigma$, $\lambda$, $\lambdat$, $D$ can all be integrated out and the dynamics of the connection 1-form are left unchanged. This will be interesting in the case of the Third Way Theory because if it is physically equivalent to its conjectured superanalogue and it can be localised then we've achieved essentially a localisation of the Third Way Theory as is without having to change its physics. These kind of circumstance is exceedingly rare and being able to achieve this would prove significant theoretical progress.

\subsubsection{Supersymmetry of SCS theory}

Let us now verify that this theory is supersymmetric. This is a relatively short computation but we will do so explicitly because this is also instructive to understanding progress made towards localising the Third Way Theory.

Now to perform this computation: Varying this action yields
\begin{equation}
\label{SCS computation 1}
\delta S_\SCS[\Vg]
=
\int\extd^3x ~ \tr\bigg[\frac{1}{2}\epsilon^{\mu\nu\rho}F_{\mu\nu}\delta A_\rho + \delta\lambda\lambdat + \delta\lambdat\lambda - D\delta\sigma - \sigma\delta D\bigg]
\end{equation}

\noindent We hence find that the different contributions from supersymmetry transformations are given as follows: 
We start off by writing out the most complicated contribution to this transformation which comes from the gaugino fields $\lambda$ and $\lambdat$. Here we find that
\begin{equation}
\label{SCS computation 2}
\begin{multlined}
\int\extd^3x ~ \tr\Big[\delta\lambda\lambdat + \delta\lambdat\lambda\Big]
=
\int\extd^3x ~ \tr\bigg[\zeta\Big(iD + \frac{1}{2}\epsilon^{\mu\nu\rho}F_{\mu\nu}\gamma_\rho + \Dgs\sigma\Big)\lambdat
\\
+ \zetat\Big(iD - \frac{1}{2}\epsilon^{\mu\nu\rho}F_{\mu\nu}\gamma_\rho + \Dgs\sigma\Big)\lambda\bigg]
\end{multlined}
\end{equation}

\noindent This consists of three pairs of terms. We find that the $D$ terms cancel against

\begin{equation}
\label{SCS computation 3}
\int\extd^3x ~ \tr\Big[-D\delta\sigma\Big]
=
\int\extd^3x ~ \tr\Big[\zeta(-iD)\lambdat + \zetat(-iD)\lambda\Big].
\end{equation}

\noindent We furthermore find that the $\Dg\sigma$ terms cancel against
\begin{multline}
\label{SCS computation 4}
\int\extd^3x ~ \tr\Big[-\sigma\delta D\Big]
=
\int\extd^3x ~ \tr\Big[-i\sigma\zeta\big(i\Dgs + \ad\sigma\big)\lambdat - i\sigma\zetat\big(i\Dgs - \ad\sigma\big)\lambda\Big]
\\
=
\int\extd^3x ~ \tr\Big[\zeta(-\Dgs\sigma)\lambdat + \zetat(-\Dgs\sigma)\lambda\Big]
\end{multline}

\noindent where we used the Leibniz rule as well as the fact that $[\sigma,\sigma] = 0$. Finally we move on to the $F$ terms. Interestingly, these cancel because they are proportional to the Chern-Simons field equations. Indeed, one finds that
\begin{equation}
\label{SCS computation 5}
\int\extd^3x ~ \tr\bigg[\frac{1}{2}\epsilon^{\mu\nu\rho}F_{\mu\nu}\delta A_\rho\bigg]
=
\int\extd^3x ~ \tr\bigg[-\zeta\Big(\frac{1}{2}\epsilon^{\mu\nu\rho}F_{\mu\nu}\gamma_\rho\Big)\lambdat + \zetad\Big(\frac{1}{2}\epsilon^{\mu\nu\rho}F_{\mu\nu}\gamma_\rho\Big)\lambda\bigg]
\end{equation}

\noindent indeed cancelling against the $F$ terms in the variation of the gauginos. We this conclude that
\begin{equation}
\delta_\zeta S_\SCS[\Vg] = \deltat_\zetatt S_\SCS[\Vg] = 0.
\end{equation}

\noindent That is, super-Chern-Simons theory is indeed supersymmetric.

\subsection{Localisation of SYM and SCS Theory}
\label{Localisation of SYM and SCS Theory}

Let us now move on to describe the way supersymmetric gauge theories are localised. As already mentioned, this will be done through the super-Yang-Mills action. Particularly, through the observation that expectation value of BPS operators in a SYM theory is indepedent on the coupling strength $g^2$ of the theory. As such, we can take the limit $t := g^{-2} \to \infty$ localising the theory. Here we will discuss two differing localisation schemes and discuss their equivalence and their meaning in the context of the Third Way Theory. 

\subsubsection{$\delta$-localisation scheme}

This localisation scheme will be centered around the localising supersymmetry $\delta_\zeta$ (the $\deltat_\zeta$ case follows similarly). To rephrase in the language of section \ref{Localisation Arguments for Supersymmetric Path Integrals}, we have a fermionic functional given by

\begin{equation}
\Fg_\loc[\Vg] = \frac{1}{2|\zeta|^2}\int\extd^3x ~ \tr\Big[(\delta_\zeta\lambda)^\dagger\lambda\Big].
\end{equation}

\noindent The corresponding localising action is then given by

\begin{equation}
tS_\loc[\Vg] = t\delta_\zeta\Fg_\loc[\Vg] = S_\SYM[\Vg]\big|_{t=g^{-2}}.
\end{equation}

\noindent The BPS operators then consist of operators which are annihilated by $\delta_\zeta$ and the BPS configurations on which the path integral localises are given by

\begin{align}
F_{\mu\nu} &= 0
&
\Dg_\mu\sigma &= 0
&
D &= 0
\end{align}

\noindent all in accordance to the discussion of section \ref{Localisation Arguments for Supersymmetric Path Integrals}. For this localisation argument to make sense we need the fermionic localising functional to be annihilated by $\delta^2_\zeta$. However we recall from our discussion in section \ref{Commentary on the Algebra} that

\begin{align}
\delta_\zeta^2 &= \frac{1}{2}\{\delta_\zeta,\delta_\zeta\} \equiv 0
&
&\Longrightarrow
&
\delta^2_\zeta\Fg_\loc[\Vg] &= 0
\end{align}

\noindent thus verifying that this is indeed the case. 

\subsubsection{$\Qloc$-localisation scheme}

We now move on to the $\Qloc$-localisation scheme which is based on the localising supersymmetry transformation $\Qloc_\zeta$. In this case the fermionic localising functional is given by

\begin{equation}
\Fg_\loc[\Vg]
=
\frac{1}{4|\zeta|^2}\int\extd^3x ~ \tr\Big[(\Qloc_\zeta\lambda)^\dagger\lambda + (\Qloc_\zeta\lambdat)^\dagger\lambdat\Big]
\end{equation}

\noindent which again yields a localising action

\begin{equation}
tS_\loc[\Vg] = t\Qloc_\zeta\Fg_\loc[\Vg] = S_\SYM[\Vg]\big|_{t=g^{-2}}.
\end{equation}

\noindent In this scheme the BPS operators will consist of those annihilated by $\Qloc_\zeta$ and the localisation locus will again consist of BPS configurations

\begin{align}
F_{\mu\nu} &= 0
&
\Dg_\mu\sigma &= 0
&
D &= 0.
\end{align} 

\noindent Most seems to be the same for this localisation argument compared to the previous case. However, now the condition for this localisation argument to make sense is somewhat different. Now, we need the fermionic localising functional to be annihilated by $\Qloc_\zeta^2$. In accordance to our discussion in previous sections we recall that this is indeed the case since

\begin{equation}
\Qloc_\zeta^2 = \frac{1}{2}\{\Qloc_\zeta,\Qloc_\zeta\} = -2iK^\mu\del_\mu + \delta_{\text{gauge}}(-2iK^\mu A_\mu)
\end{equation}

\noindent where we recall that we defined the Killing vector $K^\mu = \zeta\gamma^\mu\zeta$ as a spinor bilinear from the Killing spinor $\zeta$. From this it follows that

\begin{equation}
\Qloc_\zeta^2\Fg_\loc[\Vg] = \int\extd^3x ~ \tr\Big[\Qloc_\zeta^2\big\{\dots\big\}\Big] = 0.
\end{equation}

\noindent We thus find that the $\Qloc$-localisation scheme also works.

\subsubsection{Comparison between the two schemes}

Let us now compare these two schemes. We note that

\begin{itemize}
\item The $\delta$-localisation scheme uses the Bianchi identity while the $\Qloc$-localisation scheme doesn't. If one considers different supersymmetry algebras (e.g. include a central charge) the Bianchi identity might not be sufficient anymore to seperate the $A_\mu$ and $\sigma$ loci. Furthermore if this isn't the case anymore the resulting locus might not be on-shell anymore.
\item The $\delta$-localisation scheme uses the nilpotency of $\delta_\zeta$ while the $\Qloc$-scheme uses the fact that $\Qloc_\zeta^2 = \text{transl} + \text{gauge}$. If one changes to algebra the nilpotency remains generically unaffected while the nontrivial relation becomes augmented, possibly in a way which could obstruct the localisation argument.
\end{itemize}

\noindent We hence find that both schemes in more general contexts have strong and weak points. For more general supersymmetry representations the $\delta$-localisation scheme is more likely to work but might localise to looser, possibly off-shell configurations. On the other hand, the $\Qloc$-localisation scheme usually will localise on-shell but has stronger requirements to work. In more general contexts ---particularly the Third Way Theory--- these issues will become glaringly clear.

\section{Proca-Chern-Simons Theory}

\subsection{Non-Supersymmetric Formulation}
\label{Non-Supersymmetric Formulation}

Before moving on to discussing how to localise the Third Way Theory, we will discuss a toy model, namely a model which we coin Proca-Chern-Simons theory. As the name suggests, this theory is simply given by combining a Proca term with Chern-Simons theory. That is, we consider a theory given by an action

\begin{tcolorbox}

\subsubsection{Proca-Chern-Simons Theory}

\bigskip

\begin{equation}
S_\PCS[A] = S_\CS[A] - \int\tr\Big[\frac{m}{2}A\ast A\Big]
\end{equation}

\noindent with $A$ a $G$-connection 1-form, where for simplicity we assume Euclidean signature, and where $m$ is a parameter of mass dimension 1. This theory has field equations

\begin{equation}
\label{PCS field equations}
F - m\ast A = 0.
\end{equation}

\end{tcolorbox}

\noindent As is generally the case the case for Proca terms, they break gauge symmetry. The gauge group is broken down to

\begin{equation}
G \to \{1\}.
\end{equation}

\noindent Let us now take a closer look at the field equations \ref{PCS field equations} of this theory. Taking the covariant exterior derivative of these field equations one finds

\begin{equation}
\label{Lorenz gauge}
0 = \Dg\big(F - m\ast A\big) \equiv -m\extd\ast A
\end{equation}

\noindent where we used the Bianchi identity as well as the fact that $[A,\ast A] = 0$. This is very interesting, because the Lorenz gauge $\extd\ast A = 0$ is implied by the field equations themselves. This can be regarded as a consequence of the fact that the Proca mass term breaks gauge invariance.

\subsection{Super-Proca-Chern-Simons Theory}

Let us now supersymmetrise this theory. The way we supersymmetrise this theory will be by deforming the supersymmetry transformations of super-Chern-Simons theory in such a way that we're \textit{forced} to include a mass term. That is to say, in strong analogy to the action \ref{SCS action} of super-Chern-Simons theory we have:

\begin{tcolorbox}[breakable, enhanced]

\subsubsection{Super-Proca-Chern-Simons Action}
\begin{multline}
\label{SPCS action}
S_\SPCS[\Vg] 
= S_\SCS[\Vg] - \int\tr\bigg[\frac{m}{2}A\ast A + \frac{m}{2}\sigma\ast\sigma\bigg]
\\
= S_\CS[A] + \int\extd^3x ~ \tr\bigg[\lambda\lambdat - \sigma D - \frac{m}{2}A_\mu A^\mu - \frac{m}{2}\sigma^2\bigg]
\end{multline}

\noindent where $\Vg = (A_\mu,\sigma,\lambda,\lambdat,D)$ denotes the field content of the WZ gauge vector multiplet. The field equations are given by
\begin{align}
F - m\ast A &= 0
&
\lambda &= 0
&
\lambdat &= 0
&
D + m\sigma &= 0
&
\sigma &= 0.
\end{align}

\end{tcolorbox}

\noindent The supersymmetries of this theory are deformed supersymmetry transformations of the vector multiplet $\Vg$, chosen to satisfy three purposes:

\begin{itemize}
\item supersymmetrise Proca-Chern-Simons theory,
\item yield nilpotent supersymmetry transformations, i.e. $\delta_\zeta^2 = \deltat{^2_\zetatt} = 0$,
\item localise Proca-Chern-Simons theory to on-shell field configurations.
\end{itemize}

\noindent The supersymmetry transformations we arrived at are then given by:

\begin{tcolorbox}[breakable, enhanced]

\subsubsection{Massive Supersymmetry Transformations}

\begin{align}
\label{SPCS SUSY transformation A}
\delta A_\mu &= -\zeta\gamma_\mu\lambdat + \zetat\gamma_\mu\lambda
\bigg.\\
\label{SPCS SUSY transformation sigma}
\delta\sigma &= i\zeta\lambdat + i\zetat\lambda
\bigg.\\
\label{SPCS SUSY transformation lambda}
\delta\lambda &= \zeta\bigg[iD + \frac{1}{2}\epsilon^{\mu\nu\rho}F_{\mu\nu}\gamma_\rho + \Dgs\sigma \mathcolor{blue}{- m\As + im\sigma}\bigg]
\bigg.\\
\label{SPCS SUSY transformation lambdab}
\delta\lambdat &= \zetat\bigg[iD - \frac{1}{2}\epsilon^{\mu\nu\rho}F_{\mu\nu}\gamma_\rho + \Dgs\sigma \mathcolor{blue}{+ m\As + im\sigma}\bigg]
\bigg.\\
\label{SPCS SUSY transformation D}
\delta D &= i\zeta\big(i\Dgs + \ad\sigma\big)\lambdat + i\zetat\big(i\Dgs - \ad\sigma\big)\lambda
\bigg.
\end{align}

\noindent where the highlighted terms are the `deformations' of the usual supersymmetry transformations \ref{vector multiplet SUSY A}-\ref{vector multiplet SUSY D}. Note that these are proportional to the mass parameter $m$ of Proca-Chern-Simons theory. Again we denote

\begin{equation}
\delta_{\zeta,\zetatt} = \delta_\zeta + \deltat_\zetatt
\end{equation}

\noindent from now on.

\end{tcolorbox}

\noindent The supersymmetry of the action follows in complete analogy to the computations \ref{SCS computation 1}-\ref{SCS computation 5}. To highlight why this action still is supersymmetric, we note that the new contributions to \ref{SCS computation 2} are countered by the SUSY transformations of the mass terms in the SPCS action:
\begin{equation}
\begin{split}
\delta S_\SCS[\Vg]\big|_{\textcolor{blue}{\text{new}}} 
&=
\int\extd^3x ~ \tr\Big[\delta\lambda\lambdat + \delta\lambdat\lambda\Big]\Big|_{\textcolor{blue}{\text{new}}}
\\
&= 
\int\extd^3x ~ \tr\Big[\zeta\big(-m\As+im\sigma\big)\lambdat + \zetat\big(m\As+im\sigma\big)\lambda\Big]
\\
&= -\delta\int\tr\bigg[-\frac{m}{2}A\ast A - \frac{m}{2}\sigma\ast\sigma\bigg].
\end{split}
\end{equation}

\noindent As for the nilpotency of the supersymmetry transformations, this follows from the results in section \ref{Commentary on the Algebra} as well as the observation that

\begin{equation}
\label{tussenstap 3}
\delta(i\As \pm \sigma)_\alpha{^\beta}
=
\begin{cases}
-2i\zeta_\alpha\lambdat{^\beta} + 2i\zetat{^\beta}\lambda_\alpha & \uparrow
\\
-2i\zeta^\beta\lambdat_\alpha + 2i\zetat_\alpha\lambda^\beta & \downarrow
\end{cases}
\end{equation}

\noindent which follows as a direct consequence of the Fierz identity \ref{Fierz identity}. The nilpotency of the SUSY transformations on $A_\mu$, $\sigma$ and $D$ follows directly from previous results and for the gauginos $\lambda$ and $\lambdat$ we find
\begin{align}
\delta^2_\zeta\lambda^\alpha\big|_\new 
&=
+im\zeta^\beta\delta_\zeta(i\As + \sigma)_\beta{^\alpha}
\overeq{\ref{tussenstap 3}} 2m\zeta^2\lambdat{^\alpha} = 0
\\
\deltat{^2_\zetatt}\lambdat{^\alpha}\big|_\new 
&=
-im\zetat{^\beta}\deltat_\zetatt(i\As - \sigma)_\beta{^\alpha}
\overeq{\ref{tussenstap 3}} 2m\zetat{^2}\lambda^\alpha = 0
\end{align}

\noindent proving that indeed

\begin{equation}
\label{nilpotency}
\delta^2_\zeta = \deltat{^2_\zetatt} = 0
\end{equation}

\noindent the massive SUSY transformations are nilpotent.

\subsection{The Algebra of Massive SUSY}

Let us now more closely work out the algebra associated to these massive SUSY transformations. A straightforward if not somewhat tedious computation yields

\begin{equation}
\{\delta_\zeta,\delta_\eta\} = \{\deltat_\zetatt,\deltat_\etatt\} = 0.
\end{equation}

\noindent As for the remaining anticommutators, we find
\begin{align}
\{\delta_\zeta,\deltat_\zetatt\}A_\mu &= -2iK^\nu (F_{\nu\mu} - m\epsilon_{\nu\mu\rho}A^\rho) + 2\zeta\zetat\Dg_\mu\sigma
\\
\{\delta_\zeta,\deltat_\zetatt\}\sigma &= -2iK^\mu\Dg_\mu\sigma
\\
\{\delta_\zeta,\deltat_\zetatt\}\lambda^\alpha &= -2iK^\mu\Dg_\mu\lambda^\alpha - 2\zeta\zetat[\sigma,\lambda^\alpha] - 2m\zetat{^\alpha}(\zeta\lambda)
\\
\{\delta_\zeta,\deltat_\zetatt\}\lambdat{^\alpha} &= -2iK^\mu\Dg_\mu\lambdat{^\alpha} - 2\zeta\zetat[\sigma,\lambdat{^\alpha}] - 2m\zeta^\alpha(\zetat\lambdat)
\\
\{\delta_\zeta,\deltat_\zetatt\}D &= -2iK^\mu\Dg_\mu (D + im\sigma) - 2\zeta\zetat[\sigma,D] + 2m\zeta\zetat\del_\mu A^\mu
\end{align}

\noindent Interestingly, the algebra now takes on the form

\begin{equation}
\text{SUSY}^2 = \text{transl} + \text{gauge} + \text{mass}
\end{equation}

\noindent where `mass' refers to the inclusion of terms proportional to the mass $m$. Interestingly, this \textit{does not} have the interpretation of a central charge. Originally it was attempted to realise the mass term in Proca-Chern-Simons theory as arising through a central charge. However, the relative signs between the new terms in $\delta\lambda$ and $\delta\lambdat$ in equations \ref{SPCS SUSY transformation lambda} and \ref{SPCS SUSY transformation lambdab} this can't be the case (compare to e.g. \cite{closset2013supersymmetric}). Furthermore, such relative signs to our knowledge wouldn't allow for the construction of a supersymmetric Proca-Chern-Simons action.

\subsection{Localisation of Proca-Chern-Simons Theory}

In this section we will study the localisation of Proca-Chern-Simons theory. It should be noted that due to the fact that all additional field content of SPCS being auciliary a localisation of PCS theory and SPCS theory would be equivalent. 

\subsubsection{$\delta$-Localisation Scheme}

Let us start off by studying the $\delta$-localisation scheme. In this scheme the fermionic localising function will be given by

\begin{equation}
\Fg_\loc[\Vg]
=
\frac{1}{2|\zeta|^2}\int\extd^3x ~ \tr\Big[(\delta_\zeta\lambda)^\dagger\lambda\Big].
\end{equation}

\noindent We now work out the corresponding localising action

\begin{equation}
S_\loc[\Vg] = \delta_\zeta\Fg_\loc[\Vg] = \frac{1}{2|\zeta|^2}\int\extd^3x ~ \tr\Big[|\delta_\zeta\lambda|^2 + \delta_\zeta(\delta_\zeta\lambda)^\dagger\lambda\Big].
\end{equation}

\noindent Starting off with the bosonic part we find in analogy to computation \ref{computation partial localisation} that

\begin{equation}
\begin{split}
\frac{1}{2|\zeta|^2}|\delta_\zeta\lambda|^2
&=
\tr\bigg[
\begin{aligned}[t]
&-\frac{1}{2} (D+m\sigma)^2 - \frac{1}{2}\Big(\frac{1}{2}\epsilon_{\mu\rho\sigma}F^{\rho\sigma} - mA_\mu + \Dg_\mu\sigma\Big)
\\
&\times
\Big(\frac{1}{2}\epsilon^{\mu\rho\sigma}F_{\rho\sigma} - mA^\mu + \Dg^\mu\sigma\Big)\bigg]
\end{aligned}
\\
&=\tr\bigg[
\begin{aligned}[t]
&-\frac{1}{2} (D + m\sigma)^2 - \frac{1}{4}F_{\mu\nu}F^{\mu\nu} - \frac{1}{2}\Dg_\mu\sigma\Dg^\mu\sigma 
\\
&- \frac{1}{2}\Dg_\mu\Big(\frac{1}{2}\epsilon^{\mu\rho\sigma}F_{\rho\sigma}\sigma-mA^\mu\sigma\Big)
+m\del_\mu A^\mu\sigma\bigg].
\end{aligned}
\end{split}
\end{equation}

\noindent It is here that our remarks with regards to the comparison of the $\delta$- and $\Qloc$-localisation schemes become important. Particularly, we find that an obstruction to the seperation of the two field equations in the $\delta$-localisation scheme is proportional to $\del_\mu A^\mu$. Interestingly, if we had a good reason for assuming a Lorenz gauge $\del_\mu A^\mu = 0$ this would seperate the two field equations. We hence arrive at a bosonic localising action
\begin{equation}
\begin{split}
S_\loc[\Vg]\big|_\bos
&=
\int\extd^3x ~ \tr\bigg[
\begin{aligned}[t]
&- \frac{1}{4}\Big(F_{\mu\nu} - m\epsilon_{\mu\nu\rho}(A+\Dg\sigma)^\rho\Big)
\Big(F^{\mu\nu} - m\epsilon^{\mu\nu\rho}(A+\Dg\sigma)_\rho\Big)
\\
&-\frac{1}{2} (D+m\sigma)^2\bigg]
\end{aligned}
\\
&= \int\extd^3x ~ \tr\bigg[
\begin{aligned}[t]
&-\frac{1}{2}(D+m\sigma)^2 - \frac{1}{4}\big(F_{\mu\nu} - m\epsilon_{\mu\nu\rho}A^\rho\big)\big(F^{\mu\nu} - m\epsilon_{\mu\nu\rho}A^\rho\big)
\\
&- \frac{1}{2}\Dg_\mu\sigma\Dg^\mu\sigma + m\del_\mu A^\mu\sigma\bigg]
\end{aligned}
\end{split}
\end{equation}

\noindent As for the fermionic part of the localising action, we find new contributions
\begin{equation}
\tr\Big[\delta_\zeta(\delta_\zeta\lambda)^\dagger\lambda\Big]\Big|_\new 
= \tr\bigg[im\zetad{^\beta}\delta_\zeta(i\As + \sigma)_\beta{^\alpha}\lambda_\alpha\bigg]
\overeq{\ref{tussenstap 3}} \tr\Big[2m|\zeta|^2\lambdat\lambda\Big]
\end{equation}

\noindent Comparing this to equation \ref{SYM action fermionic} we hence find find that the fermionic part of the localising functional is given by

\begin{equation}
S_\loc[\Vg]\big|_\fer = \int\extd^3x ~ \tr\bigg[-\lambdat\big(i\Dgs - \ad\sigma - m\big)\lambda\bigg].
\end{equation}

\noindent In conclusion, we find that the localising action is given by

\begin{tcolorbox}[breakable, enhanced]

\subsubsection{$\delta$-Localisation of Proca-Chern-Simons Theory}

\begin{equation}
S_\loc[\Vg] 
=
\int\extd^3x ~ \tr\bigg[
\begin{aligned}[t]
&-\frac{1}{4}\Big(F_{\mu\nu} - m\epsilon_{\mu\nu\rho}A^\rho + \epsilon_{\mu\nu\rho}\Dg^\rho\sigma)\Big)
\\
&\times\Big(F^{\mu\nu} - m\epsilon^{\mu\nu\rho}A_\rho + \epsilon^{\mu\nu\rho}\Dg_\rho\sigma)\Big)
\\
&-\frac{1}{2}(D + m\sigma)^2 - \lambdat\big(i\Dgs - \ad\sigma - m\big)\lambda\bigg]
\end{aligned}
\end{equation}

\noindent which yields a localisation locus
\begin{align}
D+m\sigma &= 0,
&
F - m\ast A &= -\ast\Dg\sigma.
\end{align}

\noindent localising the action on \textit{sourced} Proca-Chern-Simons $A_\mu$ field equations as well as the $\sigma$ field equations.

\end{tcolorbox}

\noindent We note that for the covariant flatness condition $\Dg\sigma = 0$ to seperate from the $A_\mu$ field equations and hence become its own localisation locus the condition $\del_\mu A^\mu = 0$ has to be satisfied.
A perhaps meaningful way of looking at this is that this would be the case if Noether's second theorem were to hold somehow. However, due to the fact that PCS theory violates gauge symmetry it doesn't have to. That is to say,
\begin{align}
\delta_\theta S_\PCS[A] &~\slashed{\equiv}~ 0,
&
\Dg_\mu\fder{S_\PCS}{A_\mu} &~\slashed{\equiv}~ 0
&
&\Leftrightarrow
&
\del_\mu A^\mu &~\slashed{\equiv}~ 0.
\end{align}

\noindent What we note though is that if the gauge field $A_\mu$ is taken to be on-shell, it does hold:
\begin{align}
\fder{S_\PCS}{A_\mu} &= 0
&
&\Leftrightarrow
&
F_{\mu\nu} - m\epsilon_{\mu\nu\rho}A^\rho &= 0
&
&\Longrightarrow
&
\del_\mu A^\mu &= 0.
\end{align}

\noindent This is an interesting fact, since terms appearing in supersymmetry which vanish on-shell are often symptomatic of the elimination of auxiliary field content. Perhaps this suggests that we need to introduce additional field content for the Proca-Chern-Simons localisation to make sense off-shell. 

\subsubsection{$\Qloc$-Localisation Scheme}

As an attempt to get rid of the terms which appear in the $\delta$-localisation scheme, we will now perform the $\Qloc$-localisation scheme, which isn't equivalent to the former with these supersymmetry transformations.
In analogy to section \ref{Localisation of SYM and SCS Theory} we have a fermionic localising functional

\begin{equation}
\Fg_\loc[\Vg] = \frac{1}{4|\zeta|^2}\int\extd^3x ~ \tr\Big[(\Qloc_\zeta\lambda)^\dagger\lambda + (\Qloc_\zeta\lambdat)^\dagger\lambdat\Big]
\end{equation}

\noindent In complete analogy to the computations \ref{complete localisation SYM 1}-\ref{complete localisation SYM 3} we find
\begin{equation}
\begin{split}
S_\loc[\Vg]\big|_\bos 
&= \frac{1}{4|\zeta|^2}\int\extd^3x ~ \tr\Big[|\Qloc_\zeta\lambda|^2 + |\Qloc_\zeta\lambdat|^2\Big]
\\
&= \int\extd^3x ~ \tr\bigg[
\begin{aligned}[t]
&-\frac{1}{4}\big(F_{\mu\nu} - m\epsilon_{\mu\nu\rho}A^\rho\big)\big(F^{\mu\nu} - m\epsilon^{\mu\nu\rho}A_\rho\big)
\\
&-\frac{1}{2}\Dg_\mu\sigma\Dg^\mu\sigma - \frac{1}{2}(D + m\sigma)^2\bigg]
\end{aligned}
\end{split}
\end{equation}

\noindent which does \textit{not} involve the term proportional to $\del_\mu A^\mu$. If the total action is then supersymmetric, that is to say, $\Qloc_\zeta S_\loc = 0$, then the localisation argument makes sense. However, before moving on to this let us discuss the fermionic part of this localising action. In this case the fermionic part of the action \textit{won't} agree with that of tha $\delta$-localisation scheme. We find after some brief computations that we arrive at a fermionic sector
\begin{equation}
\begin{split}
S_\loc[\Vg]\big|_\fer 
&= \frac{1}{4|\zeta|^2}\int\extd^3x ~ \tr\Big[\Qloc_\zeta(\Qloc_\zeta\lambda)^\dagger\lambda + \Qloc_\zeta(\Qloc_\zeta\lambdat)^\dagger\lambdat\Big]
\\
&= \int\extd^3x ~ \tr\bigg[-\lambdat\big(i\Dgs - \ad\sigma - m\big)\lambda - \frac{m}{2}\big(\lambda^2 + \lambdat{^2}\big)\bigg].
\end{split}
\end{equation}

\noindent This action interestingly is augmented by non-standard mass terms. We thus arrive at a total localising action



\begin{equation}
S_\loc[\Vg]
=
\int\extd^3x ~ \tr\bigg[
\begin{aligned}[t]
&-\frac{1}{4}\big(F_{\mu\nu} - m\epsilon_{\mu\nu\rho}A^\rho\big)\big(F^{\mu\nu} - m\epsilon^{\mu\nu\rho}A_\rho\big)
\\
&-\frac{1}{2}\Dg_\mu\sigma\Dg^\mu\sigma - \frac{1}{2}(D + m\sigma)^2
\\
&-\lambdat\big(i\Dgs - \ad\sigma - m\big)\lambda - \frac{m}{2}\big(\lambda^2 + \lambdat{^2}\big)\bigg].
\end{aligned}
\end{equation}

\noindent \textit{If} this action is supersymmetric, $\Qloc_\zeta S_\loc[\Vg] = 0$, this localises on the localisation locus
\begin{align}
D+m\sigma &= 0,
&
F_{\mu\nu} - m\epsilon_{\mu\nu\rho}A^\rho &= 0,
&
\Dg_\mu\sigma &= 0.
\end{align}

\noindent consisting of the $\sigma$ field equation, $A_\mu$ field equation and the $\sigma$ field being covariantly constant. However, as it turns out, this action is \textit{not} supersymmetric under $\Qloc_\zeta$:

\begin{equation}
\Qloc_\zeta S_\loc[\Vg] \neq 0.
\end{equation}


\noindent In conclusion, we find that the $\Qloc$-localisation scheme, which aimed at removing the problems which involve using the Bianchi identities in the derivation of the $\delta$-localisation, does not work for localising Proca-Chern-Simons theory, despite its initial appeal. Having set up this toy case we are now finally ready to move on to the supersymmetrisation and localisation of the Third Way Theory.


\section{Localisation of the Third Way Theory}

\subsection{Motivation}

Having treated Proca-Chern-Simons theory as a toy model we are now ready to move on to the localisation of the Third Way Theory. Let us motivate again why this localisation ---if successful--- is so interesting: 

\smallskip

\noindent As we saw in previous chapters, the Third Way Theory originated as a new kind of consistent sourcing of the Yang-Mills field equations, with a source which is conserved under the remarkable condition that the gauge fields are taken to be on-shell. Due to this rather peculiar fact it followed that no gauge invariant action could yield the Third Way field equations. The way to solve this issue was then to introduce auxiliary gauge fields which transform diagonally with the original gauge fields and which can't be integrated out of the action, but for which the combined field equations yield the Third Way field equation. This action, rather than taking on the form of some augmented TMYM action, took on the form of two Chern-Simons actions connected with each other through a peculiar gauge invariant mass term. 

\smallskip

\noindent This then leads us to our results in previous sections on localisation: One of the remarkable facts of the supersymmetrisation of Chern-Simons theory is that the field content added to supersymmetrise the theory is all auxiliary. This means that Chern-Simons theory and its supersymmetrisation are physically equivalent. This then in turn means that results derived from localising super-Chern-Simons theory yield results in Chern-Simons theory. Due to the Chern-Simons-like action of the Third Way Theory, this then leads to the hypothesis that this theory can be supersymmetrised, and that its supersymmetrisation is such that in analogy to Chern-Simons theory it is physically equivalent to the original theory. The results from a conjectured localisation of the super-Third Way Theory would then lead to exact results in the Third Way Theory. That is to say, if we can localise the Third Way Theory what we have localised is a deformation of topologically massive Yang-Mills theory with a highly non-trivial source term and which by its own formulation isn't inherently supersymmetric.

\subsection{Arvanitakis $\delta$-Localisation Scheme}

We start off by briefly describing a localisation scheme originally suggested by Arvanitakis. This scheme does not closely resemble the localisation scheme developed for Proca-Chern-Simons Theory which was developed throughout this thesis. For example, rather than forcing the mass turn upon us through deformed supersymmetry transformations it seeks to eliminate it through parity properties of the transformations. It further also assumes that $\mu = 0 \Leftrightarrow m = \mb \Leftrightarrow k = \kb$. Instead of trying to localise on the $A$ and $\Ab$ field equations this scheme tries to localise on the diagonal and antidiagonal field equations. That is,

\begin{align}
&
\begin{cases}
F - m\ast(\Ab-A) = 0
\\
\Fb - m\ast(\Ab-A) = 0
\end{cases}
&
&\Leftrightarrow
&
&
\begin{cases}
F - \Fb = 0
\\
F + \Fb = 2m\ast(\Ab-A)
\end{cases}
\end{align}

\subsubsection{Supersymmetric Action and Supersymmetries}

In this scheme we have a supersymmetric action which depends on field content $(A,\Ab,\sigma,\lambda,\lambdat,D)$. That is, the field content of a vector multiplet $\Vg$ alongside the `auxiliary' gauge field $\Ab$. This is a little strange because one would typically expect that supersymmetric theories have matching fermionic and bosonic degrees of freedom. However, this isn't a problem since in terms of field content this isn't a hard rule, as we can always just add auxiliary fields to a theory indefinitely. The corresponding action is now given by

\begin{equation}
S[\Vg,\Ab] 
= \frac{k}{2\pi}S_\CS[\Ab] - \frac{k}{2\pi}S_\SCS[\Vg] - \frac{k}{2\pi}S_\mass[\Ab-A]
\end{equation}

\noindent with the Third Way mass term given by

\begin{equation}
S_\mass[\Ab-A] = \int\tr\bigg[\frac{1}{2\ell}(\Ab-A)\ast(\Ab-A)\bigg].
\end{equation}

\noindent As is the case for Chern-Simons theory the additional field content $(\sigma,\lambda,\lambdat,D)$ is all auxiliary. The proposed supersymmetries of this action satisfy the property $\delta A = \delta\Ab$, thus automatically annihilating the mass term. These are given by
\begin{align}
\delta A_\mu &= \delta\Ab_\mu = -\zeta\gamma_\mu\lambdat + \zetat\gamma_\mu\lambda
\bigg.\\
\delta\sigma &= i\zeta\lambdat + i\zetat\lambda
\bigg.\\
\delta\lambda &= \zeta\bigg[iD + \frac{1}{2}\epsilon^{\mu\nu\rho}(\Fb-F)_{\mu\nu}\gamma_\rho + [\Abs-\As,\sigma]\bigg]
\bigg.\\
\delta\lambdat &= \zetat\bigg[iD - \frac{1}{2}\epsilon^{\mu\nu\rho}(\Fb-F)_{\mu\nu}\gamma_\rho + [\Abs-\As,\sigma]\bigg]
\bigg.\\
\delta D &= -\zeta[\Abs-\As,\lambdat] - \zetat[\Abs-\As,\lambda]
\bigg.
\end{align}

\noindent Originally these only included the $\zeta$ components, since this scheme only cared about a $\delta$-localisation (for good reasons). 
It is worth noting that these transformations don't resemble standard supersymmetry transformations as much anymore, due to the fact that they don't contain derivatives anymore except for the field strengths.

\medskip

\noindent Let us verify whether the action is eliminated by these supersymmetry transformations. First and foremost we note that
\begin{align}
\delta A &= \delta\Ab
&
&\Longrightarrow
&
\delta S_\mass[\Ab-A] &= 0.
\end{align}

\noindent We further also note that in close analogy to computations \ref{SCS computation 1}-\ref{SCS computation 5} one finds
\begin{equation}
\begin{split}
\delta\Big\{S_\CS[\Ab] - S_\CS[A]\Big\}
&=
\int\extd^3x ~ \tr\bigg[
\begin{aligned}[t]
&- \frac{1}{2}\epsilon^{\mu\nu\rho}(\Fb-F)_{\mu\nu}\zeta\gamma_\rho\lambdat
\\
&+ \frac{1}{2}\epsilon^{\mu\nu\rho}(\Fb-F)_{\mu\nu}\zetat\gamma_\rho\lambda\bigg]
\end{aligned}
\\
&= - \delta\int\extd^3x ~ \tr\Big[\lambda\lambdat - \sigma D\Big]
\end{split}
\end{equation}

\noindent We hence conclude that this action is supersymmetric.

\subsubsection{Localisation}

The localisation scheme for this procedure is the $\delta$-localisation scheme. The reason for this is that due to the unconventional form of the supersymmetries the $\Qloc$-localisation scheme is hopeless in yielding any results. The fermionic localising functional is again given by

\begin{equation}
\Fg_\loc[\Vg,\Ab] = \frac{1}{2|\zeta|^2}\int\extd^3x ~ \tr\Big[(\delta_\zeta\lambda)^\dagger\lambda\Big].
\end{equation}

\noindent Working out the $\delta$-localisation scheme we find in analogy to computation \ref{computation partial localisation} that
\begin{multline}
S_\loc[\Vg,\Ab]
=
\int\extd^3x ~ \tr\bigg[
-\frac{1}{2}D^2 - \frac{1}{4}\Big(\Fb_{\mu\nu} - F_{\mu\nu} + \epsilon_{\mu\nu\rho}[(\Ab-A)^\rho,\sigma]\Big)
\\
\times
\Big(\Fb{^{\mu\nu}} - F^{\mu\nu} + \epsilon^{\mu\nu\rho}[(\Ab-A)_\rho,\sigma]\Big)\bigg]
\end{multline}

\noindent It is clear now that we don't have any Bianchi identities at our disposal to make meaningful simplifications. As such we arrive at a localisation locus

\begin{align}
\Fb_{\mu\nu} - F_{\mu\nu} + \epsilon_{\mu\nu\rho}[(\Ab-A)^\rho,\sigma] &= 0,
&
D &= 0.
\end{align}

\noindent This localises the action to the off-shell sourced diagonal field equations and the $\sigma$ field equation.

\subsection{The Super-Third Way Theory}

We now formulate the super-Third Way Theory. The construction of this theory follows in close analogy to super-Proca-Chern-Simons theory. Again we deform the supersymmetry transformations in such a way that a mass term is forced upon us. In strong analogy to the super-Proca-Chern-Simons action we find

\begin{tcolorbox}[breakable, enhanced]

\subsubsection{The Super-Third Way Action}

\medskip

\begin{equation}
S\SuperThirdWay[\Vg,\Vgb] = 
\frac{k}{2\pi}S_\SCS[\Vgb] - \frac{\kb}{2\pi}S_\SCS[\Vg]
- \frac{k\kb}{4\pi^2}S_\mass[\Vgb-\Vg]
\end{equation}

\noindent the supersymmetrisation of the Third Way Action. Its field content consists of two vector multiplets $\Vg = (A,\sigma,\lambda,\lambdat,D)$ and $\Vgb = (\Ab,\sigmab,\lambdab,\lambdabt,\Db)$. Its super-Chern-Simons component actions are given by \ref{SCS action} and the mass term is in this case given by
\begin{equation}
S_\mass[\Vgb-\Vg] = \int\tr\bigg[\frac{1}{2\ell}(\Ab-A)\ast(\Ab-A)+ \frac{1}{2\ell}(\sigmab-\sigma)\ast(\sigmab-\sigma)\bigg]
\end{equation}

\noindent in close analogy to the super-Proca-Chern-Simons case \ref{SPCS action}. Its field equations are given by
\begin{align}
F + \frac{k}{2\pi\ell}\ast(\Ab-A) &= 0
&
\Fb + \frac{\kb}{2\pi\ell}\ast(\Ab-A) &= 0
\bigg.\\
D + \frac{k}{2\pi\ell}(\sigmab - \sigma) &= 0
&
\Db + \frac{\kb}{2\pi\ell}(\sigmab - \sigma) &= 0
\bigg.\\
\sigma = \lambda = \lambdat &= 0
&
\sigmab = \lambda = \lambdabt &= 0
\bigg.
\end{align}

\end{tcolorbox}

\noindent Its supersymmetries are also derived in close analogy to super-Proca-Chern-Simons theory. We recall that the criteria which were used to derive these supersymmetry transformations were given by

\begin{itemize}
\item Supersymmetrising the Third Way Theory.
\item Yielding nilpotent supersymmetry transformations.
\item Ideally localising the action on on-shell field configuration.
\end{itemize}

\noindent These conditions yield the following supersymmetry transformations:

\newpage

\begin{tcolorbox}

\subsubsection{Third Way Supersymmetry Transformations}

We denote the Third Way supersymmetry transformation by

\begin{equation}
\delta_{\zeta,\zetatt,\zetab,\zetabtt} 
= 
\delta_\zeta + \deltat_\zetatt + \deltab_\zetab + \deltabt_\zetabtt
\end{equation}

\noindent this time parametrised by \textit{four} independent constant 2-spinors $\zeta$, $\zetat$, $\zetab$, $\zetabt$. The supersymmetry transformations read
\begin{align}
\delta A_\mu &= -\zeta\gamma_\mu\lambdat + \zetat\gamma_\mu\lambda
\bigg.\\
\delta\Ab_\mu &= -\zetab\gamma_\mu\lambdabt + \zetabt\gamma_\mu\lambdab
\bigg.\\
\delta\sigma &= i\zeta\lambdat + i\zetat\lambda
\bigg.\\
\delta\sigmab &= i\zetab\lambdabt + i\zetabt\lambdab
\bigg.\\
\delta\lambda &= \zeta\bigg[iD + \frac{1}{2}\epsilon^{\mu\nu\rho}F_{\mu\nu}\gamma_\rho + \Dgs\sigma \mathcolor{blue}{- \frac{k}{2\pi\ell}(\Abs-\As) + \frac{ik}{2\pi\ell}(\sigmab-\sigma)}\bigg]
\bigg.\\
\delta\lambdab &= \zetab\bigg[i\Db + \frac{1}{2}\epsilon^{\mu\nu\rho}\Fb_{\mu\nu}\gamma_\rho + \Dgbs\sigmab \mathcolor{blue}{- \frac{\kb}{2\pi\ell}(\Abs-\As) + \frac{i\kb}{2\pi\ell}(\sigmab-\sigma)}\bigg]
\bigg.\\
\delta\lambdat &= \zetat\bigg[iD - \frac{1}{2}\epsilon^{\mu\nu\rho}F_{\mu\nu}\gamma_\rho + \Dgs\sigma \mathcolor{blue}{+ \frac{k}{2\pi\ell}(\Abs-\As) + \frac{ik}{2\pi\ell}(\sigmab-\sigma)}\bigg]
\bigg.\\
\delta\lambdabt &= \zetabt\bigg[i\Db - \frac{1}{2}\epsilon^{\mu\nu\rho}\Fb_{\mu\nu}\gamma_\rho + \Dgbs\sigmab \mathcolor{blue}{+ \frac{\kb}{2\pi\ell}(\Abs-\As) + \frac{i\kb}{2\pi\ell}(\sigmab-\sigma)}\bigg]
\bigg.\\
\delta D &= i\zeta\big(i\Dgs + \ad\sigma\big)\lambdat + i\zetat\big(i\Dgs - \ad\sigma\big)\lambda
\bigg.\\
\delta\Db &= i\zetab\big(i\Dgbs + \ad\sigmab\big)\lambdabt + i\zetabt\big(i\Dgbs - \ad\sigmab\big)\lambdab
\bigg.
\end{align}

\noindent where the highlighted terms are the new terms which diverge from the standard supersymmetry transformations of the vector multiplet. Note that these terms indeed vanish as $\ell\to\infty$.

\end{tcolorbox}

\newpage

\noindent Let us now verify that these indeed leave the super-Third Way action invariant. We find that the new contributions are given by
\begin{align}
\frac{k}{2\pi}\delta S_\SCS[\Vgb]\Big|_\new &= \frac{k\kb}{4\pi^2\ell}\int\extd^3x ~ \tr\bigg[
\begin{aligned}[t]
&\zetab\Big(-(\Abs-\As) + i(\sigmab-\sigma)\Big)\lambdabt
\\
+~&\zetabt\Big(+(\Abs-\As) + i(\sigmab-\sigma)\Big)\lambdab\bigg]
\end{aligned}
\\
-\frac{\kb}{2\pi}\delta S_\SCS[\Vg]\Big|_\new &= \frac{k\kb}{4\pi^2\ell}\int\extd^3x ~ \tr\bigg[
\begin{aligned}[t]
&\zeta\Big(+(\Abs-\As) - i(\sigmab-\sigma)\Big)\lambdat
\\
+~&\zetat\Big(-(\Abs-\As) - i(\sigmab-\sigma)\Big)\lambda\bigg]
\end{aligned}
\end{align}

\noindent These cancel precisely against the supersymmetry transformations of the mass term:

\begin{equation}
-\frac{k\kb}{4\pi^2}\delta S_\mass[\Vgb-\Vg]
= -\frac{k}{2\pi}\delta S_\SCS[\Vgb]\Big|_\new + \frac{\kb}{2\pi}\delta S_\SCS[\Vg]\Big|_\new
\end{equation}

\noindent hence proving that the super-Third Way Theory is supersymmetric

\begin{equation}
\delta_{\zeta,\zetatt,\zetab,\zetabtt}S\SuperThirdWay[\Vg,\Vgb] = 0.
\end{equation}

\noindent As for the nilpotency of these transformations, these follow from precisely the same steps as equations \ref{tussenstap 3}-\ref{nilpotency}, thus resulting in

\begin{equation}
\delta^2_\zeta = \deltat{^2_\zetatt} = \deltab{^2_\zetab} = \deltab{^2_\zetabtt} = 0.
\end{equation}

\noindent That is, the Third Way supersymmetries are nilpotent.

\subsection{The Third Way Supersymmetry Algebra}

\subsubsection{Four Killing Spinors}

Let us now take a closer look at the supersymmetry algebra of the super-Third Way Theory. This algebra will take on a significantly more complicated form, since now we have split it up into four different supersymmetries. We split these up into three different cathegories:

\begin{itemize}
\item $\delta^2$, $\deltab{^2}$, $\deltat{^2}$, $\deltabt{^2}$: These all vanish generalising the nilpotency of the transformations as
\begin{equation}
\{\delta_\zeta,\delta_\eta\} = \{\deltat_\zetatt,\deltat_\etatt\} = \{\deltab_\zetab,\deltab_\etab\} = \{\deltabt_\zetabtt,\deltabt_\etabtt\} = 0
\end{equation}
\item $\delta\deltab$, $\deltat\deltab$, $\delta\deltabt$, $\deltat\deltabt$: These interestingly ---or perhaps alarmingly--- show some non-vanishing results:
\begin{align}
\label{unusual 1}
\{\delta_\zeta,\deltab_\zetab\}\lambda^\alpha &= \frac{k}{\pi\ell}(\zeta\zetab)\lambdabt{^\alpha}
\\
\label{unusual 2}
\{\delta_\zeta,\deltab_\zetab\}\lambdab{^\alpha} &= \frac{\kb}{\pi\ell}(\zeta\zetab)\lambdat{^\alpha}
&
\{\delta_\zeta,\deltab_\zetab\}\big|_\other &= 0
\\
\{\deltat_\zetatt,\deltab_\zetab\}\lambdat{^\alpha} &= -\frac{k}{\pi\ell}\zetab{^\alpha}(\zetat\lambdabt)
\\
\{\deltat_\zetatt,\deltab_\zetab\}\lambdab{^\alpha} &= \frac{\kb}{\pi\ell}\zetat{^\alpha}(\zetab\lambda)
&
\{\deltat_\zetatt,\deltab_\zetab\}\big|_\other &= 0
\\
\{\delta_\zeta,\deltabt_\zetabtt\}\lambda^\alpha &= -\frac{k}{\pi\ell}\zetabt{^\alpha}(\zeta\lambdab)
\\
\{\delta_\zeta,\deltabt_\zetabtt\}\lambdabt{^\alpha} &= \frac{\kb}{\pi\ell}\zeta^\alpha(\zetabt\lambdat)
&
\{\delta_\zeta,\deltabt_\zetabtt\}\big|_\other &= 0
\\
\label{unusual 3}
\{\deltat_\zetatt,\deltabt_\zetabtt\}\lambdat{^\alpha} &= -\frac{k}{\pi\ell}(\zetat\zetabt)\lambdab{^\alpha}
\\
\label{unusual 4}
\{\deltat_\zetatt,\deltabt_\zetabtt\}\lambdabt{^\alpha} &= - \frac{\kb}{\pi\ell}(\zetat\zetabt)\lambda^\alpha
&
\{\deltat_\zetatt,\deltabt_\zetabtt\}\big|_\other &= 0
\end{align}
\item $\delta\deltat$, $\deltab\deltabt$: These are the Third Way analogues of the non-trivial relations:
\begin{align}
\{\delta_\zeta,\deltat_\zetatt\}A_\mu &= -2iK^\nu\Big(F_{\nu\mu} - \frac{k}{2\pi\ell}\epsilon_{\nu\mu\rho}(\Ab-A)^\rho\Big) + 2\zeta\zetat\Dg_\mu\sigma
\\
\{\delta_\zeta,\deltat_\zetatt\}\sigma &= -2iK^\mu\Dg_\mu\sigma
\\
\{\delta_\zeta,\deltat_\zetatt\}\lambda^\alpha &= -2iK^\mu\Dg_\mu\lambda^\alpha - 2\zeta\zetat[\sigma,\lambda^\alpha] + \frac{k}{\pi\ell}\zetat{^\alpha}(\zeta\lambda)
\\
\{\delta_\zeta,\deltat_\zetatt\}\lambdat{^\alpha} &= -2iK^\mu\Dg_\mu\lambdat{^\alpha} - 2\zeta\zetat[\sigma,\lambdat{^\alpha}] + \frac{k}{\pi\ell}\zeta^\alpha(\zetat\lambdat)
\\
\{\delta_\zeta,\deltat_\zetatt\}D &= 
\begin{aligned}[t]
&-2iK^\mu\Dg_\mu\Big(D + \frac{k}{2\pi\ell}(\sigmab-\sigma)\Big) - 2\zeta\zetat[\sigma,D]
\\ 
&+ \frac{k}{\pi\ell}\Dg_\mu(\Ab-A)^\mu\zeta\zetat
\end{aligned}
\\
\{\delta_\zeta,\deltat_\zetatt\}\Vgb &= 0
\end{align}
and similarly

\begin{align}
\{\deltab_\zetab,\deltabt_\zetabtt\}\Ab_\mu &= -2i\Kb{^\nu}\Big(\Fb_{\nu\mu} - \frac{\kb}{2\pi\ell}\epsilon_{\nu\mu\rho}(\Ab-A)^\rho\Big) + 2\zetab\zetabt\Dgb_\mu\sigmab
\\
\{\deltab_\zetab,\deltabt_\zetabtt\}\sigmab &= -2i\Kb{^\mu}\Dgb_\mu\sigmab
\\
\{\deltab_\zetab,\deltabt_\zetabtt\}\lambdab{^\alpha} &= -2i\Kb{^\mu}\Dgb_\mu\lambdab{^\alpha} - 2\zetab\zetabt[\sigmab,\lambdab{^\alpha}] - \frac{\kb}{\pi\ell}\zetabt{^\alpha}(\zetab\lambdab)
\\
\{\deltab_\zetab,\deltabt_\zetabtt\}\lambdabt{^\alpha} &= -2i\Kb{^\mu}\Dgb_\mu\lambdabt{^\alpha} - 2\zetab\zetabt[\sigmab,\lambdabt{^\alpha}] - \frac{\kb}{\pi\ell}\zetab{^\alpha}(\zetabt\lambdabt)
\\
\{\deltab_\zetab,\deltabt_\zetabtt\}\Db &= 
\begin{aligned}[t]
&-2i\Kb{^\mu}\Dgb_\mu\Big(\Db + \frac{\kb}{2\pi\ell}(\sigmab-\sigma)\Big) - 2\zetab\zetabt[\sigmab,\Db] 
\\
&+ \frac{\kb}{\pi\ell}\Dgb_\mu(\Ab-A)^\mu\zetab\zetabt
\end{aligned}
\\
\{\deltab_\zetab,\deltabt_\zetabtt\}\Vg &= 0
\end{align}
where by $\Vg$ and $\Vgb$ we denoted the collective field content of the respective vector multiplets and where we defined the Killing vectors
\begin{align}
K^\mu &= \zeta\gamma^\mu\zetat,
&
\Kb{^\mu} &= \zetab\gamma^\mu\zetabt.
\end{align}
\end{itemize}

\subsubsection{Two Killing Spinors}

\noindent We note that due to the relations \ref{unusual 1}-\ref{unusual 4}, and in particular equations \ref{unusual 1}, \ref{unusual 2}, \ref{unusual 3} and \ref{unusual 4} the algebra takes on quite an unusual form. However, we note that these are all proportional to either $\zeta\zetab$ or $\zetat\zetabt$. Hence this suggests that to make these vanish we could take $\zeta = \zetab$ and $\zetat = \zetabt$ and redefine
\begin{align}
\label{redefinition}
\delta_\zeta,\deltab_\zetab &\to \delta_\zeta := \delta_{\zeta,0,\zeta,0}
&
\deltat_\zetatt,\deltabt_\zetabtt &\to \deltat_\zetatt := \delta_{0,\zetatt,0,\zetatt}.
\end{align}

\noindent The algebra then takes on the more familiar form
\begin{align}
\{\delta_\zeta,\delta_\eta\} &= 0
&
\{\deltat_\zetatt,\deltat_\etatt\} &= 0
\end{align}

\noindent with non-trivial anticommutators

\begin{align}
\{\delta_\zeta,\deltat_\zetatt\}A_\mu &= -2iK^\nu\Big(F_{\nu\mu} - \frac{k}{2\pi\ell}\epsilon_{\nu\mu\rho}(\Ab-A)^\rho\Big) + 2\zeta\zetat\Dg_\mu\sigma
\bigg.\\
\{\delta_\zeta,\deltat_\zetatt\}\Ab_\mu &= -2iK^\nu\Big(\Fb_{\nu\mu} - \frac{\kb}{2\pi\ell}\epsilon_{\nu\mu\rho}(\Ab-A)^\rho\Big) + 2\zeta\zetat\Dgb_\mu\sigmab
\bigg.\\
\{\delta_\zeta,\deltat_\zetatt\}\sigma &= -2iK^\mu\Dg_\mu\sigma
\bigg.\\
\{\delta_\zeta,\deltat_\zetatt\}\sigmab &= -2iK^\mu\Dgb_\mu\sigmab
\bigg.\\
\{\delta_\zeta,\deltat_\zetatt\}\lambda^\alpha &= -2iK^\mu\Dg_\mu\lambda^\alpha - 2\zeta\zetat[\sigma,\lambda^\alpha] - \frac{k}{\pi\ell}\zetat{^\alpha}\zeta(\lambdab - \lambda)
\bigg.\\
\{\delta_\zeta,\deltat_\zetatt\}\lambdab{^\alpha} &= -2iK^\mu\Dgb_\mu\lambdab{^\alpha} - 2\zeta\zetat[\sigmab,\lambdab{^\alpha}] - \frac{\kb}{\pi\ell}\zetat{^\alpha}\zeta(\lambdab - \lambda)
\bigg.\\
\{\delta_\zeta,\deltat_\zetatt\}\lambdat{^\alpha} &= -2iK^\mu\Dg_\mu\lambdat{^\alpha} - 2\zeta\zetat[\sigma,\lambdat{^\alpha}] - \frac{k}{\pi\ell}\zeta^\alpha\zetat(\lambdabt - \lambdat)
\bigg.\\
\{\delta_\zeta,\deltat_\zetatt\}\lambdabt{^\alpha} &= -2iK^\mu\Dgb_\mu\lambdabt{^\alpha} - 2\zeta\zetat[\sigmab,\lambdabt{^\alpha}] - \frac{\kb}{\pi\ell}\zeta^\alpha\zetat(\lambdabt - \lambdat)
\bigg.\\
\{\delta_\zeta,\deltat_\zetatt\}D &= 
\begin{aligned}[t]
&-2iK^\mu\Dg_\mu\Big(D + \frac{k}{2\pi\ell}(\sigmab-\sigma)\Big) - 2\zeta\zetat[\sigma,D] 
\\
&+ \frac{k}{\pi\ell}\Dg_\mu(\Ab-A)^\mu\zeta\zetat
\end{aligned}
\bigg.\\
\{\delta_\zeta,\deltat_\zetatt\}\Db &= 
\begin{aligned}[t]
&-2iK^\mu\Dgb_\mu\Big(\Db + \frac{\kb}{2\pi\ell}(\sigmab-\sigma)\Big) - 2\zeta\zetat[\sigmab,\Db] 
\\
&+ \frac{\kb}{\pi\ell}\Dgb_\mu(\Ab-A)^\mu\zeta\zetat
\end{aligned}
\bigg.
\end{align}

\noindent where we now only defined a single Killing vector

\begin{equation}
K^\mu = \zeta\gamma^\mu\zetat.
\end{equation}

\noindent In analogy to the Proca-Chern-Simons case this transformation takes on the form

\begin{equation}
\text{SUSY}^2 = \text{transl} + \text{gauge} + \text{mass}.
\end{equation}

\noindent However, an aspect in which it is different from Proca-Chern-Simons theory is that the field dependent gauge transformation are different between the two multiplets $\Vg$ and $\Vgb$. Aside from the issues we incountered in Proca-Chern-Simons theory for the $\Qloc$-localisation argument this also will hinder it.

\subsection{Localising the Super-Third Way Theory}

Finally, we move on to perform the localisation procedure of the super-Third Way Theory. This will follow in close analogy to the Proca-Chern-Simons case.

\subsubsection{$\Qloc$-Localisation Scheme}

Let us now start with the $\Qloc$-localisation scheme instead of the $\delta$-localisation scheme, because this is the less promising one for similar reasons as the Proca-Chern-Simons case. In this case the fermionic localising functional is given by

\begin{equation}
\Fg_\loc[\Vg,\Vgb] = \frac{1}{4|\zeta|^2}\int\extd^3x ~ \tr\bigg[(\Qloc_\zeta\lambda)^\dagger\lambda + (\Qloc_\zeta\lambdat)^\dagger\lambdat + (\Qloc_\zeta\lambdab)^\dagger\lambdab + (\Qloc_\zeta\lambdabt)^\dagger\lambdabt\bigg]
\end{equation}

\noindent where we defined a localising supersymmetry

\begin{equation}
\Qloc_\zeta = \delta_\zeta + \deltat_\zeta = \delta_{\zeta,\zeta,\zeta,\zeta}
\end{equation}

\noindent again following the definition \ref{redefinition}. However, for reasons similar to the Proca-Chern-Simons case it sadly doesn't work, since

\begin{equation}
\Qloc_\zeta S_\loc[\Vg,\Vgb] \neq 0.
\end{equation}

\noindent The explicit expression isn't given because as one might suspect it's quite a mess, and it doesn't seem to vanish on-shell, suggesting that adding auxiliary field content won't obviously help. However, this doesn't stop one from trying and multiple ways of adjusting the supersymmetries were tried out, though all being unsuccessful. Most notably a way to come up with the `right' supersymmetries was starting from superspace, and breaking the group of supergauge transformations. This would lead to supersymmetries somewhat resembling the Third Way supersymmetries but which didn't seem fit for constructing a supersymmetric action.

\subsubsection{$\delta$-localisation scheme}

Now, we treat the more promising $\delta$-localisation scheme. Here the fermionic localising functional is given by

\begin{equation}
\Fg_\loc[\Vg,\Vgb] = \frac{1}{2|\zeta|^2}\int\extd^3x ~ \tr\bigg[(\delta_\zeta\lambda)^\dagger\lambda + (\delta_\zeta\lambdab)^\dagger\lambdab\bigg]
\end{equation}

\noindent where we defined the supersymmetry transformation $\delta_\zeta = \delta_{\zeta,0,\zeta,0}$ as in equation \ref{redefinition}. The systematics of the computations are identitcal to Proca-Chern-Simons theory but the equations are over twice as long. Hence, we will only present the reader with the results, these are given by:

\begin{tcolorbox}[enhanced,breakable]

\subsubsection{$\delta$-Localisation of the Super-Third Way Theory}

\smallskip

\begin{equation}
\begin{split}
&S_\loc[\Vg,\Vgb] = \delta_\zeta\Fg_\loc[\Vg,\Vgb]
\\
&=
\int\extd^3x ~ \tr\bigg[
\begin{aligned}[t]
&- \frac{1}{4}\Big(F_{\mu\nu} - \frac{k}{2\pi\ell}\epsilon_{\mu\nu\rho}(\Ab-A)^\rho + \epsilon_{\mu\nu\rho}\Dg^\rho\sigma\Big)
\\
&\times\Big(F^{\mu\nu} - \frac{k}{2\pi\ell}\epsilon^{\mu\nu\rho}(\Ab-A)_\rho + \epsilon^{\mu\nu\rho}\Dg_\rho\sigma\Big)
\\
&- \frac{1}{4}\Big(\Fb_{\mu\nu} - \frac{\kb}{2\pi\ell}\epsilon_{\mu\nu\rho}(\Ab-A)^\rho + \epsilon_{\mu\nu\rho}\Dgb{^\rho}\sigmab\Big)
\\
&\times\Big(\Fb{^{\mu\nu}} - \frac{\kb}{2\pi\ell}\epsilon^{\mu\nu\rho}(\Ab-A)_\rho + \epsilon^{\mu\nu\rho}\Dgb_\rho\sigmab\Big)
\\
&-\frac{1}{2}\Big(D + \frac{k}{2\pi\ell}(\sigmab-\sigma)\Big)^2
-\frac{1}{2}\Big(\Db + \frac{\kb}{2\pi\ell}(\sigmab-\sigma)\Big)^2
\\
&- \lambdat\Big(i\Dgs - \ad\sigma + \frac{k}{2\pi\ell}\Big)\lambda + \frac{k}{2\pi\ell}\lambdabt\lambda
\\
&- \lambdabt\big(i\Dgbs - \ad\sigmab - \frac{\kb}{2\pi\ell}\big)\lambdab - \frac{\kb}{2\pi\ell}\lambdat\lambdab\bigg]
\end{aligned}
\end{split}
\end{equation}

\noindent which yields a localisation locus
\begin{align}
F - \frac{k}{2\pi\ell}\ast(\Ab-A) &= -\ast\Dg\sigma
&
D + \frac{k}{2\pi\ell}(\sigmab-\sigma) &= 0
\\
\Fb - \frac{\kb}{2\pi\ell}\ast(\Ab-A) &= -\ast\Dgb\sigmab
&
\Db + \frac{\kb}{2\pi\ell}(\sigmab-\sigma) &= 0
\end{align}

\noindent where the Third Way field equations become sourced by the scalar fields. 

\end{tcolorbox}

\noindent In analogy to the Proca-Chern-Simons case, we note that for the covariant flatness $\Dg\sigma = \Dgb\sigmab = 0$ to seperate from the $A$, $\Ab$ field equations the condition $\Dg_\mu(\Ab-A)^\mu \equiv \Dgb_\mu(\Ab-A)^\mu = 0$ would have to be satisfied. The deeper reason why this isn't the case can again be traced back to the violation of Noether's second theorem.
Due to the fact that the action isn't invariant under a full $G\times G$ but rather the diagonal subgroup $G$, Noether's second theorem doesn't have to hold anymore:
\begin{equation}
\delta_{\theta,\thetab}S\ThirdWay ~\slashed{\equiv}~ 0
\end{equation}

\noindent where we now have
\begin{align}
\Dg_\mu\fder{S\ThirdWay}{A_\mu} &~\slashed{\equiv}~ 0
&
&\Leftrightarrow
&
\Dgb_\mu\fder{S\ThirdWay}{\Ab_\mu} &~\slashed{\equiv}~ 0
&
&\Leftrightarrow
&
\Dg_\mu(\Ab-A)^\mu \equiv \Dgb_\mu(\Ab-A)^\mu &~\slashed{\equiv}~ 0.
\end{align}

\noindent However, if the fields are taken to be on-shell, this requirement is satisfied:
\begin{align}
\fder{S\ThirdWay}{A_\mu} &= 0
&
&\Longrightarrow
&
\Dg_\mu(\Ab-A)^\mu &= 0
&
&\Longleftarrow
&
\fder{S\ThirdWay}{\Ab_\mu} &= 0
\end{align}

\noindent Again it is worth noting that the appearance of terms which vanish on-shell could be an indicator for auxiliary fields which have been integrated away. Perhaps there exists some addition of auciliary fields, which could after integrating them away impose the condition $\Dg_\mu(\Ab-A)^\mu \equiv \Dgb_\mu(\Ab-A)^\mu = 0$. Attempts have been made to do this consistently, and while the possibility of this isn't excluded, no such addition of additional auxiliary fields is currently known.







\chapter*{Conclusion}
\addcontentsline{toc}{chapter}{Conclusion}

Let us now conclude this thesis. The main goal of this thesis was to localise the Third Way Theory. However, a lot of interesting original computations also emerged from our treatment of the Witten index in chapter 2.

\bigskip

\noindent For the Witten index, a number of original computations have been performed. These derived already established results which ---to the best of our knowledge--- haven't been applied elsewhere in the literature. For starters, the $\beta$-independence of the Witten index is typically only derived using the Hilbert space formalism. Instead in this thesis we used the path integral formalism to derive this result. We arrived at this result by using the superspace formalism to note that the action is $Q$-exact. Similarly, the `kinetic' and `potential' rescalings were shown to be symmetries of the Witten index, again making use of superspace methods. These methods had the advantage of showing that a term is manifestly $Q$-exact, rather than forcing us to engage in educated guess work. Finally, the computations were overall more thorough than those in the source material (particularly \cite{DavidTong}).



\bigskip

\noindent As for the Third Way Theory, let us remind the reader of the significance of localising this theory again: On-shell, the Third Way Theory is a continuous deformation of Yang-Mills theory. However, the action which describes this theory consists of two independent Chern-Simons theories over the same gauge group and a mass term which breaks the two copies of the gauge group to its diagonal subgroup. Combined with the observation that $\NSUSY = 0$ Chern-Simons theory can be localised due to its equivalence with $\NSUSY = 2$ Chern-Simons theory, one is lead to hypothesise that the $\NSUSY = 0$ and $\NSUSY = 2$ Third Way theories are equivalent. As a corollary to this, it would then follow that localising the $\NSUSY = 2$ Third Way Theory is equivalent to localising the $\NSUSY = 0$ Third Way Theory, thus achieving a method of doing non-perturbative computations in a highly non-trivial deformation of $d=3$ Yang-Mills theory.



\bigskip

\noindent Progress to this end has been made in various ways: Significant progress has been made into understanding Proca-Chern-Simons theory, the toy model which leads to the Third Way localisation. A new kind of supersymmetry transformation has been introduced to supersymmetrise this theory, which we coined `massive supersymmetry'. Using this supersymmetry, Proca-Chern-Simons theory was then superymmetrised, such that its superanalogue is physically equivalent to the original theory. After this, the theory was localised on its field equations, sourced by auxiliary fields. Then, a proposed localisation scheme by Arvanitakis was reviewed which localised the Third Way Theory on its diagonal field equation, sourced by auxiliary fields. Finally, the results of Proca-Chern-Simons theory were generalised to the Third Way Theory, leading to a supersymmetrisation of the Third Way Theory which is physically equivalent to the original theory. Using this supersymmetrisation, the Third Way Theory was then localised on both its field equations, sourced by auxiliary fields.

\bigskip

\noindent In conclusion, significant progress has been made towards localising the Third Way Theory. We found a localising action which localises the theory on sourced field equations. One assuption which has been made throughout the thesis though is that the path integral measure is supersymmetric. This is at this moment the only barrier between definitively being able to say that the theory is localised.

\appendix

\chapter*{Appendix}
\refstepcounter{chapter}
\addcontentsline{toc}{chapter}{Appendix}

\section{Differential Geometry}
\label{Differential Geometry}

\subsection{Riemannian Geometry}
\label{Riemannian Geometry}

We denote the flat space metric as

\begin{equation}
\eta_{ab} = \diag(-1,\dots,-1,+1,\dots,+1).
\end{equation}

\noindent The curved space metric is given by

\begin{equation}
g_{\mu\nu} = e_\mu{^a}e_\nu{^b}\eta_{ab}
\end{equation}

\noindent where $e_\mu{^a}$ the vielbeins with inverses $e^\mu{_a}$ which we take such that
\begin{align}
\eta &:= \det (\eta_{ab}) = (-)^{\text{\# time directions}}
\\
e &:= \det (e_\mu{^a}) > 0
\\
g &:= \det (g_{\mu\nu}) = \eta e^2
\end{align}

\noindent We denote the Christoffel symbol by $\Gamma_{\mu\sigma}{^\rho}$ and the spin connection by $\omega_\mu{^{ab}}$ \cite{nakahara2018geometry}\cite{ortin2004gravity}.

\subsection{Lie algebras and Connections}

\subsubsection{Lie algebras}

We consider a simple compact Lie group $G$ with Lie algebra $\gLie$. The Lie algebra $\gLie$ is taken to be generated by $T_I$ with structure constants $f_{IJ}{^K}$ and Killing form $\kappa_{IJ}$ defined through
\begin{align}
[T_I,T_J] &= f_{IJ}{^K}T_K
&
\kappa_{IJ} &= \tr\big[T_IT_J\big]
\end{align}

\noindent in our conventions $f_{IJ}{^K}$ are real and $\kappa_{IJ}$ non-degenerate and negative semi-definite. Let $\omega_p$ and $\eta_q$ be $\gLie$-valued forms. Their commutator is then defined as

\begin{align}
[\omega_p,\eta_q] &:= \omega_p\eta_q - (-)^{pq}\eta_q\omega_p
&
&\Leftrightarrow
&
[\omega_p,\eta_q]^I &:= f_{JK}{^I}\omega_p^J\eta_q^K.
\end{align}

\noindent The $\mathbb{Z}_2$-graded Jacobi identity here takes on the form

\begin{equation}
\label{Z2 graded Jacobi identity}
[\alpha_r,[\omega_p,\eta_q]] = [[\alpha_r,\omega_p],\eta_q] + (-)^{rp}[\omega_p,[\alpha_r,\eta_q]].
\end{equation}

\noindent One should think of this as analogous to the super-Jacobi identities for Lie superalgebras \cite{fuchs2003symmetries}\cite{nakahara2018geometry}.

\subsubsection{Connections}

Connections of this group are $\gLie$-valued 1-forms $A$ which transform as
\begin{align}
A &\to e^\theta(A+\extd)e^{-\theta}
&
\delta_\theta A &= -\extd \theta - [A,\theta] =: -\Dg\theta
\end{align}

\noindent and field strengths are denoted as
\begin{align}
F &:= \extd A + A^2
&
F &\to e^\theta Fe^{-\theta}
&
\delta_\theta F &= [\theta,F]
\end{align}

\noindent where $\theta = \theta^IT_I$ a $\gLie$-valued local gauge parameters. Field strengths satisfy the Bianchi identity

\begin{align}
\Dg F &\equiv 0
&
&\Leftrightarrow
&
\Dg_{[\mu}F_{\rho\sigma]} &\equiv 0.
\end{align}

\noindent where we defined $\Dg = \extd + \ad A$ \cite{nakahara2018geometry}.

\subsection{Levi-Civita Symbol and Hodge Dualities}

\subsubsection{Levi-Civita symbol and tensor}

The Levi-Civita \textit{symbol} is characterised by
\begin{align}
\varepsilon_{[a_1\dots a_d]} &= \varepsilon_{a_1\dots a_d}
&
\varepsilon_{0\dots(d-1)} &= 1
\\
\varepsilon^{[a_1\dots a_d]} &= \varepsilon^{a_1\dots a_d}
&
\varepsilon^{0\dots(d-1)} &= 1
\end{align}

\noindent Similarly, we define the Levi-Civita \textit{tensor} for flat indices as
\begin{align}
\epsilon_{[a_1\dots a_d]} &= \epsilon_{a_1\dots a_d}
&
\epsilon_{0\dots(d-1)} &= 1
\\
\epsilon^{[a_1\dots a_d]} &= \epsilon^{a_1\dots a_d}
&
\epsilon^{0\dots(d-1)} &= \eta
\end{align}

\noindent and the Levi-Civita tensor for curved indices is obtained as

\begin{equation}
\epsilon_{\mu_1\dots\mu_d} := e_{\mu_1}{^{a_1}}\dots e_{\mu_d}{^{a_d}}\varepsilon_{a_1\dots a_d} = e\varepsilon_{\mu_1\dots\mu_d}.
\end{equation}

\noindent Contractions between Levi-Civita symbols are computed as
\begin{equation}
\varepsilon^{a_1\dots a_pc_{p+1}\dots c_d}\varepsilon_{b_1\dots b_pc_{p+1}\dots c_d} = p!(d-p)!\delta^{a_1\dots a_p}_{b_1\dots b_p}
\end{equation}

\noindent where the generalised Kronecker delta symbol is defined as

\begin{equation}
\delta^{a_1\dots a_p}_{b_1\dots b_p} = \delta^{[a_1}{_{b_1}}\dots\delta^{a_d]}{_{b_d}}.
\end{equation}

\noindent For the Levi-Civita tensor this gives

\begin{align}
\epsilon^{a_1\dots a_pc_{p+1}\dots c_d}\epsilon_{b_1\dots b_pc_{p+1}\dots c_d} &= p!(d-p)!\cdot\eta\cdot\delta^{a_1\dots a_p}_{b_1\dots b_p}
\\
\epsilon^{\mu_1\dots\mu_p\rho_{p+1}\dots\mu_d}\epsilon_{\nu_1\dots\nu_p\rho_{p+1}\dots\mu_d}
&=
p!(d-p)!\cdot\eta\cdot\delta^{\mu_1\dots\mu_p}_{\nu_1\dots\nu_p}.
\end{align}

\noindent with additional factors $\eta$ \cite{nakahara2018geometry}\cite{ortin2004gravity}.

\subsubsection{Hodge dualities}

We define the Hodge star operator $\ast:\Omega^p(M)\to\Omega^{d-p}(M)$ as

\begin{equation}
\ast\omega_{\mu_{p+1}\dots\mu_d} := \frac{1}{p!}\epsilon_{\mu_1\dots\mu_d}\omega^{\mu_1\dots\mu_p}
\end{equation}

\noindent The Hodge star satisfies

\begin{align}
\ast_{d-p}\ast_p &= \eta (-)^{p(d-p)}
&
&\Longrightarrow
&
\ast_p^{-1} &= \eta(-)^{p(d-p)}\ast_{d-p}
\end{align}

\noindent giving us an expression for inverse Hodge operators \cite{nakahara2018geometry}\cite{ortin2004gravity}.

\subsubsection{Inner products between forms}

Making use of the Hodge star we can construct inner products

\begin{align}
(\bullet|\bullet) &: \Omega^p(M)^{\otimes 2}\to\mathbb{R}
&
(\bullet|\bullet) &: (\Omega^p(M)\otimes\gLie)^{\otimes 2}\to\mathbb{R}
\end{align}

\noindent between ($\gLie$-valued) forms as

\begin{equation}
(\omega|\eta) := \int\tr\big[\omega\ast\eta\big] = \int\extd^dx ~ \frac{e}{p!}\kappa_{IJ}\omega^I_{\mu_1\dots\mu_p}\eta^{J\mu_1\dots\mu_p}.
\end{equation}

\noindent The non-$\gLie$-valued case is simply obtained by not taking the trace as it isn't defined. This product is symmetric and non-degenerate. On Riemannian manifolds it's also negative semi-definite in the $\gLie$-valued case and positive semi-definite in the non-$\gLie$-valued case \cite{nakahara2018geometry}.

\subsubsection{Adjoint exterior derivatives}

Using the aforementioned inner product we can now define adjoints of exterior derivatives. Let $\omega_p$ and $\eta_{p+1}$ be respectively $p$- and $(p+1)$-forms. We define the adjoint $\extdd_{p+1} : \Omega^{p+1}(M)\to\Omega^p(M)$ of the exterior derivative \cite{nakahara2018geometry}\cite{ortin2004gravity} through the relation
\begin{gather}
(\extd_p\omega_p|\eta_{p+1}) = (\omega_p|\extdd_{p+1}\eta_{p+1})
\\
\Downarrow\nonumber
\\
\extdd = -\eta(-)^{p(d-p)}\ast\extd\ast = -(-)^{d+p}\ast\extd\invast = (-)^p\invast\extd\ast = -\extd\inn
\end{gather}

\noindent where we defined the contraction between forms as

\begin{gather}
\eta_p\inn\omega_{p+q} = \omega_{p+q}\innr\eta_p := (-)^{r(d-r+s)}\invast(\eta_s\ast\omega_{r+s})\Big.
\\
\mathcal{m}\Big.\nonumber
\\
(\eta_p\inn\omega_{p+q})_{\mu_1\dots\mu_p} = (\omega_{p+q}\innr\eta_p)_{\mu_1\dots\mu_p} := \frac{1}{q!}\eta^{\nu_1\dots\nu_q}\omega_{\nu_1\dots\nu_q\mu_1\dots\mu_p}
\end{gather}

\noindent Similarly it follows that

\begin{align}
\Dgd &= -\Dg\inn
&
(\ad A)^\dagger &= -\ad A\inn
\end{align}



\subsection{Matrix Exponentiation and BCH Formula}
\label{Matrix Exponentiation and BCH Formula}

Let us mention some standard formulas related to matrix exponentiation. Let $X,Y$ be arbitrary matrices. One could verify that
\begin{align}
e^XYe^{-X} &= e^{\ad X}Y = Y + [X,Y] + \frac{1}{2}[X,[X,Y]] + \dots
\\
e^X\delta e^{-X} &= \frac{1 - e^{-\ad X}}{\ad X}\delta X = \delta X - \frac{1}{2}[X,\delta X] + \frac{1}{6}[X,[X,\delta X]] - \dots
\end{align}

\noindent where $\delta$ is an arbitrary derivative with respect to matrix multiplication \cite{Pope}.

\subsubsection{Baker-Campbell-Hausdorff (BCH) formula}

For matrices which don't commute the identity $e^xe^y = e^{x+y}$ doesn't generally hold anymore. Instead, it is replaced by the Baker-Campbell-Hausdorff (BCH) formula \cite{gilmore1974baker}

\begin{equation}
e^Xe^Y = e^{X\star Y}
\end{equation}

\noindent where $X\star Y$ is formally expressed as

\begin{equation}
\label{BCH formula}
X\star Y = \ln(e^Xe^Y) = Y + \int_0^1\extd t ~ g\big(e^{t\ad X}e^{\ad Y}\big)X = -(-Y)\star(-X)
\end{equation}

\noindent with $g(z)$ given by

\begin{align}
g(z) &= \frac{\ln z}{1-z} = \sum_{n=0}^\infty\frac{(1-z)^n}{n+1},
&
|1-z| &< 1.
\end{align}

\noindent Some of the leading order terms are given by
\begin{equation}
X\star Y
=
X+Y + \frac{1}{2}[X,Y] + \frac{1}{12}[[X,Y],Y] + \frac{1}{12}[X,[X,Y]] + \Og(X^pY^q)_{p+q\geq 4}
\end{equation}

\noindent Another interesting way to expand $X\star Y$ is as

\begin{equation}
\label{BCH formula infinitesimal}
\begin{split}
X\star Y 
&= X + \frac{\ad X}{1 - e^{-\ad X}}Y + \Og(Y^2)
\\
&= Y - \frac{\ad Y}{1 - e^{+\ad Y}}X + \Og(X^2)
\end{split}
\end{equation}

\noindent to leading orders in respectively $Y$ and $X$ \cite{fuchs2003symmetries}. Interestingly, it is the case that

\begin{equation}
\label{BCH infinitesimal expansion}
\frac{z}{e^z-1} = \sum_{n=0}^\infty \frac{B_n}{n!}z^n = 1 - \frac{1}{2}z + \frac{1}{12}z^2 + \Og(z^4)
\end{equation}

\noindent where $B_n$ the \textit{Bernoulli numbers} \cite{carlitz1968bernoulli} recursively defined as

\begin{align}
B_0 &= 1,
&
B_{n+1} &= \sum_{k=0}^{n+1}\binom{n+1}{k}B_k
&
&\Leftrightarrow
&
B_n &= \frac{-1}{n+1}\sum_{k=0}^{n-1}\binom{n+1}{k}B_k.
\end{align} 

\noindent Some of the leading values are given by

\begin{align}
B_0 &= 1,
&
B_1 &= -\frac{1}{2},
&
B_2 &= \frac{1}{6},
&
B_3 &= 0,
&
B_4 &= -\frac{1}{30},
&
&\dots
\end{align}

\newpage


\section{$d = 0 + 3$, $\NSUSY = 2$ Supersymmetry}

\subsection{Spinors and Gamma Matrices}
\label{Spinors and Gamma Matrices}

The gamma matrices are $2\times 2$ matrices given by

\begin{equation}
\gamma^\mu = (\gamma^\mu{_\alpha}{^\beta}) = (\sigma^3,~-\sigma^1,~-\sigma^2)
\end{equation}

\noindent with $\mu = 1,2,3$ and $\alpha,\beta = 1,2$, where lower spinorial indices are column indices and upper spinorial indices are row indices, and $\sigma^\mu$ are the Pauli matrices \cite{closset2013supersymmetric}\cite{ruiz1996lectures}. These satisfy an algebra

\begin{equation}
\label{Clifford algebra}
\gamma^\mu\gamma^\nu = \eta^{\mu\nu}\idop + i\epsilon^{\mu\nu\rho}\gamma_\rho
\end{equation}

\noindent where we normalised the $x$-space Levi-Civita symbol as $\epsilon_{012} = 1$ and the metric is simply given by $\eta_{\mu\nu} = \delta_{\mu\nu}$.
Indices are raised and lowered using the charge conjugation matrix which in this case is just the Levi-Civita symbol $\varepsilon_{\alpha\beta}$ in spinor space, where we follow NW-SE conventions:
\begin{align}
\psi^\alpha &= \varepsilon^{\alpha\beta}\psi_\beta,
&
\psi_\alpha &= \psi^\beta\varepsilon_{\beta\alpha}.
\end{align}

\noindent where we normalised $\varepsilon_{12} = \varepsilon^{12} = 1$. Let us name a few interesting properties of the tensors $\varepsilon_{\alpha\beta}$ and $\gamma^\mu{_\alpha}{^\beta}$:

\begin{align}
\varepsilon_\alpha{^\beta} &= -\varepsilon^\beta{_\alpha} = \delta_\alpha{^\beta}
&
\gamma^\mu{_\alpha}{^\beta} &= \gamma^{\mu\beta}{_\alpha}
&
&\text{etc.}
\end{align}

\noindent Similarly, we have for example

\begin{align}
\psi\chi &:= \psi^\alpha\chi_\alpha = (-)^{|\psi||\chi|+1}\chi\psi
\\
\psi\gamma^\mu\chi &:= \psi^\alpha\gamma^\mu{_\alpha}{^\beta}\chi_\beta = (-)^{|\psi||\chi|}\chi\gamma^\mu\psi
\end{align}

\noindent where $|\bullet|$ denotes the Grassmann parity of an object. In case there is no(t much) room for confusion, contracted indices will usually be dropped, with always an assumption of NW-SE contraction. Some more advanced identities of the gamma matrices include the Fierz identity

\begin{align}
\label{Fierz identity}
M &= \frac{1}{2}\idop \tr M + \frac{1}{2}\gamma^\mu \tr(M\gamma_\mu)
&
&\Leftrightarrow
&
M_\alpha{^\beta} &= \frac{1}{2}M_\gamma{^\gamma}\varepsilon_\alpha{^\beta} + \frac{1}{2}M_\gamma{^\delta}\gamma_{\mu\delta}{^\gamma}\gamma^\mu{_\alpha}{^\beta}
\end{align}

\noindent as well as the following few lemmas \cite{closset2013supersymmetric}\cite{ruiz1996lectures}:

\subsection{Some identities}

\begin{tcolorbox}

\noindent \textbf{Lemma:}

\begin{equation}
\label{Lemma symmetric}
\gamma^\mu{_{\alpha\beta}}\gamma_\mu{^{\gamma\delta}} = -2\varepsilon_{(\alpha}{^\gamma}\varepsilon_{\beta)}{^\delta}
\end{equation}

\end{tcolorbox}

\noindent \textbf{Proof:} We start off by noting that

\begin{equation}
\label{tussenstap}
\begin{aligned}
\gamma^\mu\gamma_\mu &\overeq{\ref{Clifford algebra}} \delta^\mu{_\mu}\idop = 3\idop,
\\
\gamma^\nu\gamma^\mu\gamma_\nu 
&\overeq{\ref{Clifford algebra}}
\gamma^\mu + i\epsilon^{\nu\mu\rho}\gamma_\rho\gamma_\nu
\overeq{\ref{Clifford algebra}}
\gamma^\mu - \epsilon^{\nu\mu\rho}\epsilon_{\rho\nu\sigma}\gamma^\sigma = -\gamma^\mu.
\end{aligned}
\end{equation}

\noindent Applying the Fierz identity to respectively $_\alpha{^\gamma}$ and $_\beta{^\gamma}$ we now find that
\begin{equation}
\begin{split}
\gamma^\mu{_{\alpha\beta}}\gamma_\mu{^{\gamma\delta}}
&\overeq{\ref{Fierz identity}} -\frac{1}{2}\gamma^\mu\gamma_{\mu(\beta}{^\delta}\varepsilon_{\alpha)}{^\gamma} + \frac{1}{2}\gamma^\mu\gamma_\nu\gamma_{\mu(\beta}{^\delta}\gamma^\nu{_{\alpha)}}{^\gamma}
\\
&\overeq{\ref{tussenstap}} -\frac{3}{2}\varepsilon_{(\alpha}{^\gamma}\varepsilon_{\beta)}{^\delta} - \frac{1}{2}\gamma_{\mu(\alpha}{^\gamma}\gamma^\mu{_{\beta)}}{^\delta}
\\
&\overeq{\ref{Fierz identity}} -\frac{3}{2}\varepsilon_{(\alpha}{^\gamma}\varepsilon_{\beta)}{^\delta} - \frac{1}{4}\gamma^\mu\gamma^\nu\gamma_{\mu\gamma\delta}\gamma_\nu{^{\alpha\beta}}
\\
&\overeq{\ref{tussenstap}} -\frac{3}{2}\varepsilon_{(\alpha}{^\gamma}\varepsilon_{\beta)}{^\delta} + \frac{1}{4}\gamma^\mu{_{\alpha\beta}}\gamma_\mu{^{\gamma\delta}}
\end{split}
\end{equation}

\noindent which is clearly equivalent to the desired identity.\qed

\begin{tcolorbox}

\noindent \textbf{Lemma:} 

\begin{equation}
\label{Lemma antisymmetric}
\epsilon^{\mu\rho\sigma}\gamma_{\rho\alpha}{^\beta}\gamma_{\sigma\gamma}{^\delta} = i\gamma^\mu{_\alpha}{^\delta}\varepsilon_\gamma{^\beta} - i\varepsilon_\alpha{^\delta}\gamma^\mu{_\gamma}{^\beta}
\end{equation}

\end{tcolorbox}

\noindent \textbf{Proof:} Application of the Fierz identity on $^\beta{_\gamma}$ yields

\begin{equation}
\epsilon^{\mu\rho\sigma}\gamma_{\rho\alpha}{^\beta}\gamma_{\sigma\gamma}{^\delta}
\overeq{\ref{Fierz identity}} \frac{1}{2}\epsilon^{\mu\rho\sigma}\Big(\gamma_\rho\gamma_{\sigma\alpha}{^\delta}\varepsilon_\gamma{^\beta} + \gamma_\rho\gamma_\nu\gamma_{\sigma\alpha}{^\delta}\gamma^\nu{_\gamma}{^\beta}\Big).
\end{equation}

\noindent For the first term we find

\begin{equation}
\frac{1}{2}\epsilon^{\mu\rho\sigma}\gamma_\rho\gamma_{\sigma\alpha}{^\delta} 
\overset{\text{\ref{Clifford algebra}}}{=} \frac{i}{2}\epsilon^{\mu\rho\sigma}\epsilon_{\rho\sigma\nu}\gamma^\nu{_\alpha}{^\delta} = i\gamma^\mu{_\alpha}{^\delta}.
\end{equation}

\noindent As for the second term we find
\begin{equation}
\begin{split}
\frac{1}{2}\epsilon^{\mu\rho\sigma}\gamma_\rho\gamma_\nu\gamma_{\sigma\alpha}{^\delta}
&\overeq{\ref{Clifford algebra}}
\frac{1}{2}\epsilon^{\mu\rho\sigma}\big(\eta_{\rho\nu} + i\epsilon_{\rho\nu\tau}\gamma^\tau\big)\gamma_{\sigma\alpha}{^\delta}
\\
&\overeq{\ref{Clifford algebra}}
\frac{1}{2}\epsilon^\mu{_{\nu\sigma}}\gamma^\sigma{_\alpha}{^\delta} - i\delta^{\mu\sigma}_{\nu\tau}\big(\delta^\tau{_\sigma}\varepsilon_\alpha{^\delta} + i\epsilon^\tau{_{\sigma\kappa}}\gamma^\kappa{_\alpha}{^\delta}\big)
\\
&= \frac{1}{2}\epsilon^\mu{_{\nu\sigma}}\gamma^\sigma{_\alpha}{^\delta} - i\delta^\mu{_\nu}\varepsilon_\alpha{^\beta} - \frac{1}{2}\epsilon^\mu{_{\nu\sigma}}\gamma^\sigma{_\alpha}{^\delta}
\\
&= -i\delta^\mu{_\nu}\varepsilon_\alpha{^\beta}
\end{split}
\end{equation}

\noindent Combining these results we find
\begin{equation}
\epsilon^{\mu\rho\sigma}\gamma_{\rho\alpha}{^\beta}\gamma_{\sigma\gamma}{^\delta}
= \frac{1}{2}\epsilon^{\mu\rho\sigma}\Big(\gamma_\rho\gamma_{\sigma\alpha}{^\delta}\varepsilon_\gamma{^\beta} + \gamma_\rho\gamma_\nu\gamma_{\sigma\alpha}{^\delta}\gamma^\nu{_\gamma}{^\beta}\Big)
= i\gamma^\mu{_\alpha}{^\delta}\varepsilon_\gamma{^\beta} - i\varepsilon_\alpha{^\delta}\gamma^\mu{_\gamma}{^\beta}
\end{equation}
thus proving our lemma.\qed

\noindent Finally, an interesting identity which could prove useful is

\begin{tcolorbox}

\noindent \textbf{Lemma:} Let $\Og_{\alpha\beta\gamma}$ be any spinor space tensor. Then

\begin{equation}
\Og^\beta{_{\alpha\beta}}
= \Og_\alpha{^\beta}{_\beta} + \Og^\beta{_{\beta\alpha}}.
\end{equation}

\end{tcolorbox}

\noindent \textbf{Proof:} Since antisymmetric 3-tensors in spinor space always vanish, we find that

\begin{equation}
0 \equiv 3\Og_{[\alpha\beta\gamma]} = \Og_{\alpha[\beta\gamma]} + \Og_{[\gamma|\alpha|\beta]} + \Og_{[\beta\gamma]\alpha}.
\end{equation}

\noindent Contracting $\beta$ and $\gamma$ then yields the desired identity.\qed

\subsection{Hermitian Conjugation}
\label{Hermitian Conjugation}

Let us now make a few comments on Hermitian conjugation. Throughout this thesis we will repeatedly make use of norms of vectors, which arise from Hilbert space inner products, as opposed to the contractions of indices through the charge conjugation matrix $\varepsilon_{\alpha\beta}$. The former is more easy to understand using column matrices and the other in index notation. In this section of the appendix we wish to shed some light on this potentially confusing tension between these two notations.

To be specific, we note that

\begin{align}
\zeta\eta &:= \zeta^\alpha\eta_\alpha = \zeta_\alpha\eta_\beta\varepsilon^{\beta\alpha}
&
(\zeta|\eta) &:= (\zeta_\alpha)^\ast\eta_\alpha
&
|\zeta|^2 &:= (\zeta_\alpha)^\ast\zeta_\alpha
\end{align}

\noindent These are covariant expressions, however, the way the indices are contracted may be confusing to readers. Now looking at the column vector notation we find that

\begin{align}
(\zeta|\eta) &:= \zeta^\dagger\eta
&
|\zeta|^2 &:= \zeta^\dagger\zeta
\end{align}

\noindent interpreting now $\zeta$ as a column matrix and $\zeta^\dagger$ as a row matrix. As it might be regarded as rather confusing that there are two ways to look at these inner products we harmonise the two by introducing the spinor notation

\begin{equation}
\zeta^{\dagger\alpha} := (\zeta_\alpha)^\ast
\end{equation}

\noindent in terms of representations this also makes sense, since if one is in the defining representation of SU(2) the other will be in the conjugate representation, which makes sense index-wise. We can now harmoniously write

\begin{align}
(\zeta|\eta) &= \zeta^\dagger\eta = \zeta^{\dagger\alpha}\eta_\alpha
&
|\zeta|^2 &= \zeta^\dagger\zeta = \zeta^{\dagger\alpha}\zeta_\alpha
\end{align}

\noindent Let us now move on to the gamma matrices. We note that

\begin{align}
(\gamma^\mu)^\dagger &= \gamma^\mu 
&
&\Leftrightarrow
&
(\gamma^\mu{_\alpha}{^\beta})^\ast &= \gamma^\mu{_\beta}{^\alpha}
\end{align}

\noindent As an example we find that

\begin{equation}
(\gamma^\mu\zeta)^{\dagger\alpha} = (\gamma^\mu{_\alpha}{^\beta}\zeta_\beta)^\ast = (\zeta_\beta)^\ast(\gamma^\mu{_\alpha}{^\beta})^\ast = \zeta^{\dagger\beta}\gamma^\mu{_\beta}{^\alpha} = \zeta^\dagger\gamma^{\mu\alpha}
\end{equation}

\noindent which is nicely in agreement with the results one would expect from using matrices instead of index notation \cite{closset2013supersymmetric}.



\backmatter%

\bibliographystyle{ieeetr}
\bibliography{referenties}

\end{document}